\documentclass[journal]{IEEEtran}

\usepackage{caption}

\usepackage{array}
\newcolumntype{L}[1]{>{\raggedright\let\newline\\\arraybackslash\hspace{0pt}}m{#1}}
\newcolumntype{C}[1]{>{\centering\let\newline\\\arraybackslash\hspace{0pt}}m{#1}}
\newcolumntype{R}[1]{>{\raggedleft\let\newline\\\arraybackslash\hspace{0pt}}m{#1}}

\usepackage[dvipsnames]{xcolor}
\usepackage{amsmath}
\usepackage{amssymb}
\usepackage{amsthm}
\newtheorem{definition}{Definition}
\usepackage{adjustbox}

\usepackage{graphicx}
\usepackage{tikz}
\usepackage{pgf-pie}
\usetikzlibrary{arrows,automata}
\usetikzlibrary{shapes.geometric}
\usetikzlibrary{positioning,calc}
\usetikzlibrary{shadows,arrows.meta}
\tikzset{parent/.style={align=center,text width=2cm,fill=green!20,rounded corners=2pt},
    child/.style={align=center,text width=2.8cm,fill=green!50,rounded corners=6pt},
    grandchild/.style={fill=pink!50,text width=2.3cm}
}
\usepackage{forest}
\usetikzlibrary{arrows.meta,shadows}

\usepackage{fontawesome}
\usepackage{balance}
\usepackage{verbatim}
\usepackage{smartdiagram}
\usepackage{mathtools}

\usepackage{pifont}
\usepackage{wasysym}
\usepackage{amssymb}
\usepackage{multirow}
\usepackage{hyperref}
\usepackage{colortbl}
\newcommand{\lb}{\left\lbrace}

\usepackage[vlined,ruled,linesnumbered]{algorithm}

\graphicspath{{./figures},{./}}
\def\headline#1{\hbox to \hsize{\hrulefill\quad\lower.3em\hbox{#1}\quad\hrulefill}}
\ifCLASSINFOpdf
\else
\fi
\begin{document}

\pgfkeys{/forest,
  rect/.append style   = {rectangle, rounded corners = 2pt,
                         inner color = col6in, outer color = col6out},
  ellip/.append style  = {ellipse, inner color = col5in,
                          outer color = col5out},
  orect/.append style  = {rect, font = \sffamily\bfseries\LARGE,
                         text width = 325pt, text centered,
                         minimum height = 10pt, outer color = col7out,
                         inner color=col7in},
  oellip/.append style = {ellip, inner color = col8in, outer color = col8out,
                          font = \sffamily\bfseries\large, text centered}}
                          
	\title{A Survey of Moving Target Defenses for\\Network Security}
	
	% --- For IEEE --- %
% 	\author{Sailik~Sengupta\textsuperscript{*} \quad\qquad \and
% 		Ankur~Chowdhary\textsuperscript{*} \quad\qquad \and
% 		Abdulhakim~Sabur \quad\qquad \and 
% 		Adel~Alshamrani\quad\qquad \and
% 		~~~~Dijiang~Huang  \quad\qquad \and
% 		Subbarao~Kambhampati
    \author{
        Sailik~Sengupta\textsuperscript{*} \qquad 
	    Ankur~Chowdhary\textsuperscript{*} \qquad
	    Abdulhakim~Sabur \qquad 
		Adel~Alshamrani  \qquad \\
		Dijiang~Huang \qquad
		Subbarao~Kambhampati
		\thanks{\textsuperscript{*}These two authors have contributed equally to this work. }
		\thanks{The authors Sailik Sengupta, Ankur Chowdhary, Abdulhakim Sabur, Dijiang Huang, and Subbarao Kambhampati are with the School of Computing, Informatics, and Decision Systems Engineering (CIDSE), Arizona State University, Tempe, AZ, USA (e-mail: ssengu15@asu.edu, achaud16@asu.edu, asabur@asu.edu, dijiang@asu.edu, rao@asu.edu). Abdulhakim Sabur is also with Taibah University, Madinah, Saudi Arabia. Adel Alshamrani is with University of Jeddah, Jeddah, Saudi Arabia (e-mail: asalshamrani@uj.edu.sa)}}
	
	% --- For ArXiv --- %
% 	\author{
% 	    Sailik Sengupta\textsuperscript{*}$^\dagger$, Ankur Chowdhary\textsuperscript{*}$^\dagger$, Abdulhakim Sabur$^\dagger$, Dijiang Huang$^\dagger$,\\
% 	    Adel Alshamrani$^\ddagger$, Subbarao Kambhampati$^\dagger$\\
% 	    $^\dagger$Arizona State University, $^\ddagger$University of Jeddah\\
% 	    $^\dagger$\{ssengu15, achaud16, asabur, dijiang, rao\}@asu.edu, $^\ddagger$asalshamrani@uj.edu.sa\\
	    
% 		\thanks{\textsuperscript{*}The first two authors contributed equally to this work.}
% 		\thanks{This work has been submitted to the IEEE for possible publication. Copyright may be transferred without notice, after which this version may no longer be accessible.}
% 	}
	
	%\markboth{}%
	%{Shell \MakeLowercase{\textit{et al.}}: A Survey of Advanced Persistent Threats: Techniques, Solutions, Challenges, and Research Opportunities}



	\maketitle

	\begin{abstract}
Network defenses based on traditional tools, techniques, and procedures (TTP) fail to account for the attacker's inherent advantage present due to the static nature of network services and configurations. To take away this asymmetric advantage, Moving Target Defense (MTD) continuously shifts the configuration of the underlying system, in turn reducing the success rate of cyberattacks. In this survey, we analyze the recent advancements made in the development of MTDs and highlight (1) how these defenses can be defined using common terminology, (2) can be made more effective with the use of artificial intelligence techniques for decision making, (3) be implemented in practice and (4) evaluated.
We first define an MTD using a simple and yet general notation that captures the key aspects of such defenses. We then categorize these defenses into different sub-classes depending on {\em what} they move, {\em when} they move and {\em how} they move. In trying to answer the latter question, we showcase the use of domain knowledge and game-theoretic modeling can help the defender come up with effective and efficient movement strategies. Second, to understand the practicality of these defense methods, we discuss how various MTDs have been implemented and find that networking technologies such as Software Defined Networking and Network Function Virtualization act as key enablers for implementing these dynamic defenses.
We then briefly highlight MTD test-beds and case-studies to aid readers who want to examine or deploy existing MTD techniques. Third, our survey categorizes proposed MTDs based on the qualitative and quantitative metrics they utilize to evaluate their effectiveness in terms of security and performance. We use well-defined metrics such as risk analysis and performance costs for qualitative evaluation and metrics based on Confidentiality, Integrity, Availability (CIA), attack representation, QoS impact, and targeted threat models for quantitative evaluation. Finally, we show that our categorization of MTDs is effective in identifying novel research areas and highlight directions for future research. 
    
    %(i) Intelligent Cyberdeception; The adaptation problem associated with MTD - how and what to mutate. We consider the timing considerations for performing environment changes (when to perform MTD?). The game theore  (ii) MTD implementation which is focused on deployment aspects of MTD. Use of programmable network frameworks such as SDN/NFV for cyberdeception are analyzed in this section of the survey. 

    %We explore the recent advancements in the field of MTD with the advent of programmable network platforms like SDN/NFV, Artificial Intelligence (AI), and Game Theory for cost-benefit analysis of MTD. We evaluate the effectiveness of the current MTD technqiues, how likely they are to be deployed on a production grade network based on adoption and experience stories in the Industry.  
	\end{abstract}
	

	\IEEEpeerreviewmaketitle

% \texthl{Keep this at present for a quick overview of the survey.}
% \tableofcontents

\section{Introduction}

%\Comment{Image of overall framework goes here}
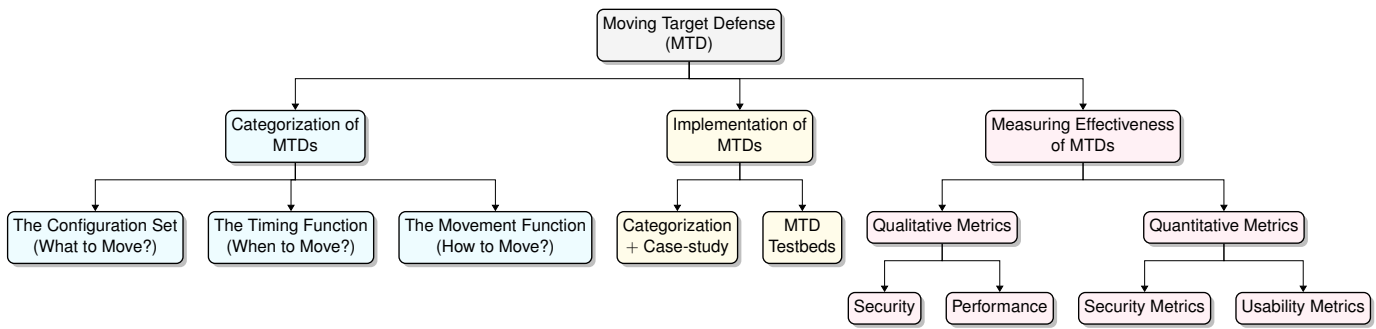
\begin{figure*}[t]
\centering
\begin{adjustbox}{width=\textwidth}
    \begin{forest}
        for tree={
            thick,
            draw,
            rounded corners,
            drop shadow,
            font=\sffamily,
            inner color=Gray!9,
            outer color=Gray!9,
            align=center,
            child anchor=north,
            parent anchor=south,
            l sep=1cm,
            s sep=0.5cm,
            edge path={
                \noexpand\path[
                    >={Triangle[]},
                    ->,
                    \forestoption{edge}]
                (!u.parent anchor) -- +(0pt,-10pt) -|
                (.child anchor)\forestoption{edge label};
            }
        }
        [Moving Target Defense\\(MTD),
            [Categorization of\\MTDs
              , inner color=CornflowerBlue!10!white
              , outer color=CornflowerBlue!10!white
              [The Configuration Set\\(What to Move?)
              , inner color=CornflowerBlue!10!white
              , outer color=CornflowerBlue!10!white
              ]
              [The Timing Function\\(When to Move?)
              , inner color=CornflowerBlue!10!white
              , outer color=CornflowerBlue!10!white
              ]
              [The Movement Function\\(How to Move?)
              , inner color=CornflowerBlue!10!white
              , outer color=CornflowerBlue!10!white
              ]
            ]
            [Implementation of\\MTDs
              , inner color=Goldenrod!10!white
              , outer color=Goldenrod!10!white
              [Categorization\\$+$ Case-study
              , inner color=Goldenrod!10!white
              , outer color=Goldenrod!10!white
              ]
              [MTD\\Testbeds
              , inner color=Goldenrod!10!white
              , outer color=Goldenrod!10!white
              ]
            ]
            [Measuring Effectiveness\\of MTDs
            , inner color=Salmon!10!white
            , outer color=Salmon!10!white
              [Qualitative Metrics
              , inner color=Salmon!10!white
              , outer color=Salmon!10!white
                [Security
                , inner color=Salmon!10!white
                , outer color=Salmon!10!white
                ][Performance
                , inner color=Salmon!10!white
                , outer color=Salmon!10!white
                ]][Quantitative Metrics
              , inner color=Salmon!10!white
              , outer color=Salmon!10!white
                  [Security Metrics
                  , inner color=Salmon!10!white
                  , outer color=Salmon!10!white
                    %[Network Downtime\\Rule-Conflict]
                  ][Usability Metrics
                  , inner color=Salmon!10!white
                  , outer color=Salmon!10!white
                  ]]]]
    \end{forest}
\end{adjustbox}
\caption{We survey various defense techniques based on the paradigm of Moving Target Defense (MTD) and showcase that a common terminology and directions for future work naturally emerges due to the categorization in our survey.}
\label{fig:workflowedge}
\end{figure*}

\IEEEPARstart{T}{he} network and cloud infrastructures have become both ubiquitous and more complex in the past few years. Gartner predicts that by the year 2025, 80\% of the entire IT infrastructure which includes deployed applications, technologies and services will be cloud-based \cite{cloudTrends}. 
While the performance aspects with regards to storage capacity, networking efficiency, and hardware have received due attention and evolved with business demands, aspects that govern the security of cloud infrastructure are still managed using traditional means. Given that security breaches can lead to loss of customer trust and worsen business reputation, a key question is \emph{how effective are these traditional security approaches?} Is monitoring network traffic for malicious patterns, routinely patching known vulnerabilities, and relying on the network and perimeter defense such as Firewalls, Intrusion Detection Systems, Anti-malware tools, {\em etc.} enough to detect and thwart attacks?

There exist multiple shortcomings in these traditional defense mechanisms. First, the attacker, with time on their side, can spend reconnaissance effort in modeling the cloud system (the defenses in place) and then, carefully plan their attacks. Second, implementations of these defenses in practice are often far from ideal, thereby allowing attackers even more opportunities to exploit the system. A report from 2016 predicts that by the end of 2020, 99\% of the vulnerabilities exploited will be known to security and IT professionals since a year ago \cite{gartnerReport}. A major reason for this is the time and complexity associated with the process of routinely patching vulnerabilities. Often, the fear of degradation in the Quality of Service (QoS) provided to customers deters Cloud Service Providers (CSPs) from making changes to the existing system configuration. Third, {\em zero-day} attacks developed using the information the attacker gathers during the reconnaissance phase can render traditional defenses useless.

To address the shortcomings of existing defense methods, \emph{Moving Target Defense} (MTD) has emerged as a solution that provides \emph{proactive defense} against adaptive adversaries. The goal of MTD is to constantly move between multiple configurations in a cyber-system (such as changing the open network ports, network configuration, software, {\em etc.}.) thereby increasing the uncertainty for the attacker; in effect, diminishing the advantage of reconnaissance that an attacker inherently has against traditional defense mechanisms. The advantages of MTDs go away if the shifting mechanism is deterministic because the attacker, with time on their side, will eventually be able to predict this movement and design attacks accordingly. Thus, for MTDs to be effective, they need to have implicit randomness built into them.  This survey categorizes MTDs based on what they shift, when they shift and how they shift. 
  
The dynamic aspect of MTD adds an extra layer of complexity in implementing these defenses. To address this, one can leverage advances in networking technology. First, \emph{Network Functions Virtualization} (NFV)~\cite{han2015network} has emerged as a technology to provide a virtualized implementation of hardware-based equipment such as firewalls, routers, {\em etc.}. through \emph{Virtual Machines} (VMs) or containers running on top of a physical server in cloud computing environments. Given MTDs, often, need more hardware than conventional software systems, such virtualization helps to reduce the cost of implementation.
Second, \emph{Software-Defined Networking} (SDN)~\cite{kreutz2015software}, which serves as an enabling technology for NFV, provides a centralized security policy enforcer. The programmable interface afforded by SDN can help administrators implement optimal movement strategies for MTDs.
Futhermore, enabling MTD implementation at scale will help us evaluate the effectiveness of these defenses in practical settings.
  
The key contributions of this survey are: (1) provides an umbrella under which we can categorize the array of MTD techniques proposed for network environments, (2) introduces a common language that can be utilized to describe (and understand) the assumptions and threat models of various MTDs, (3) gives an overview of how these defenses are implemented by researchers, highlighting testbeds and case studies to guide its large-scale deployment and (4) discusses how MTDs have been evaluated from a qualitative and a quantitative standpoint, in effect, shedding light on how effective these tools and techniques are with regards to security and performance. Figure \ref{fig:workflowedge} gives a a quick overview of this survey.

The rest of the survey is organized in the following manner. In Section \ref{sec:background}, we introduce the reader to some background knowledge about the various stages of an attack in cloud systems, popular for detections and defenses against malicious traffic, and formal frameworks to capture knowledge about attacks in networking systems. In Section~\ref{sec:mtd}, we propose a universal notation that captures the key aspect of all the MTDs proposed and use it to investigate and categorize the defenses depending on how they answer the questions (1) \emph{what to move}, (2) \emph{when to move} and (3) \emph{how to move}. In this regard, we also highlight how the different cyber surfaces - discussed under {\em what to move}-- relate to the various stages of an attack described in Section \ref{sec:background}. In Section \ref{sec:mtd_impl}, we discuss how the various MTD works have been implemented in practice, with special emphasis on the role of (and the effectiveness) of SDN and NFV in enabling them. We showcase examples of existing MTD testbeds and perform case-studies that can help security personnel in the adoption of MTD solutions for production-grade networks.  In Section~\ref{effect}, we elaborate on various qualitative and quantitative metrics that have been presented in the literature; this can help a cloud administrator decide if a defense mechanism is secure enough. In Section~\ref{research}, we highlight areas, in terms of our categorization, that have received less attention and discuss an array of future research directions. Finally, we conclude the survey in Section~\ref{concl}.
\begin{table*}[t]
\centering
\begin{tabular}{|c|c|c|c|c|c|c|c|c|}
    \hline
    \multirow{2}{*}{Research Work} & \multicolumn{3}{c|}{Cyber Surface Shift Analysis} & \multirow{2}{*}{Relation to APTs} & \multirow{2}{*}{MTD Implementation} & \multicolumn{2}{c|}{MTD Evaluation} & \multirow{2}{*}{Research Directions}  \\\cline{2-4}\cline{7-8} 
    %\multicolumn{2}{c}{\multirow{2}{*}{Multi-col-row}}&X\\
    %\multicolumn{2}{c}{}&X\\
    
     & What? & When? & How? &  &  & Qualitative  & Quantitative  & \\ 
     \hline
     ~\cite{roy2010survey} & {\color{Red} \ding{51}} & {\color{Red} \ding{51}} & {\color{Gray} \ding{55}} & {\color{Gray} \ding{55}} & {\color{Gray} \ding{55}} & {\color{Gray} \ding{55}} & {\color{Gray} \ding{55}} & {\color{Red} \ding{51}} \\
    \hline
    ~\cite{okhravi2013survey}  &{\color{Red} \ding{51}} & {\color{Gray} \ding{55}} &{\color{Red} \ding{51}} & {\color{Gray} \ding{55}} & {\color{Gray} \ding{55}} & {\color{Gray} \ding{55}} & {\color{Gray} \ding{55}} & {\color{Gray} \ding{55}} \\
    \hline 
    ~\cite{okhravi2014finding}  &{\color{Red} \ding{51}} & {\color{Gray} \ding{55}} & {\color{Gray} \ding{55}} & {\color{Gray} \ding{55}} & {\color{Gray} \ding{55}} & {\color{Red} \ding{51}} & {\color{Gray} \ding{55}} & {\color{Red} \ding{51}}\\
    \hline 
    ~\cite{farris2015quantification}  & {\color{Red} \ding{51}} & {\color{Gray} \ding{55}} & {\color{Gray} \ding{55}} & {\color{Gray} \ding{55}} & {\color{Gray} \ding{55}} & {\color{Gray} \ding{55}} &{\color{Red} \ding{51}} & {\color{Gray} \ding{55}} \\
    \hline 
    ~\cite{cai2016moving} & {\color{Red} \ding{51}} & {\color{Gray} \ding{55}} & {\color{Red} \ding{51}} & {\color{Gray} \ding{55}} & {\color{Gray} \ding{55}} & {\color{Red} \ding{51}} & {\color{Red} \ding{51}} & {\color{Red} \ding{51}} \\
    \hline 
    
     ~\cite{lei2018moving} &{\color{Red} \ding{51}} & {\color{Gray} \ding{55}} & {\color{Gray} \ding{55}} & {\color{Gray} \ding{55}} & {\color{Gray} \ding{55}} & {\color{Red} \ding{51}} & {\color{Gray} \ding{55}} & {\color{Red} \ding{51}} \\
     \hline
    Our Survey & {\color{Red} \ding{51}} & {\color{Red} \ding{51}} & {\color{Red} \ding{51}} & {\color{Red} \ding{51}} & {\color{Red} \ding{51}} & {\color{Red} \ding{51}} & {\color{Red} \ding{51}} & {\color{Red} \ding{51}} \\
    \hline
\end{tabular}
\caption{Our categorization of MTDs is a super-set of the categorization considered in earlier works. The analysis of MTDs in the context of Advanced Persistent Threats (APTs), a unified view to analyse the what, when and hows of MTDs and case study of their implementation are entirely new.}
\label{tab:7}
% \vspace{-1.6em}
\end{table*}

\section{Background and Related Work}
\label{sec:background}

In this section, we first discuss related surveys in the area of Moving Target Defense (MTD) that look at the aspects of characterizing and evaluating proactive defenses. This helps us to situate our survey and highlight several aspects of MTDs that existing surveys ignore. Second, we describe the various stages of an attack that establishes a realistic threat model against a cyber-system. This helps us understand which step(s) of an attack a particular MTD technique seeks to disarm.  Third, we highlight the traditional defense methods that are presently used to detect or reduce the impact of cyberattacks. As seen later, MTD mechanisms often leverage these traditional defenses, adding movement to the way they are deployed; this makes it harder for an attacker to fool the overall defense. Finally, we provide an overview of existing databases (NVD, OVSDB, CVSS)~\cite{ovsdb, cve2012exposures} that are used to obtain domain knowledge about known vulnerabilities and popular attack representation methods such as attack graphs and attack trees.

\subsection{Related Works and the Need for this Survey}

We present a comparison of our survey to existing surveys in Table~\ref{tab:7}. Firstly, we observe that most existing surveys~\cite{okhravi2013survey,cai2016moving} provide only partial coverage of topics relating to \emph{what, when, and how} to move the elements of the network. Section \ref{sec:mtd} provides a more holistic view of various Moving Target Defenses (MTDs). Moreover, the techniques surveyed do not talk about modeling Advanced Persistent Threat (APT) scenarios \cite{alshamrani2019survey,chen2014study}. Our survey, on the other hand, provides an overview of APT and its relation to the attack and defense surfaces, thereby helping us to highlight how a particular MTD may be effective against both known attacks as well as unknown attacks.

We provide an in-depth analysis of how MTDs are implemented and the role of networking technologies such as SDN and NFV in enabling them. We categorize the MTDs based on the maturity level of their implementation-- ranging from simulation-based analysis to research test-beds and production-level industrial products. Table~\ref{tab:adel} summarizes various MTDs highlighting their use of centralized networking paradigms such as SDN/NFV (yes/no), and the level of maturity at which they have been implemented (analytic/simulation/emulation/commercial). This categorization has not been considered in previous surveys.

Some existing surveys~\cite{farris2015quantification, cai2016moving} have taken a look at categorizing the evaluation metrics for understanding the effectiveness of MTDs but these works do not talk about the different components that contribute to the evaluation of MTDs. In our analysis, we consider both security and performance metrics associated with each system configuration and with the ensemble, enabling us to highlight directions that can help improve existing MTDs solutions and mechanisms.

Beyond simply providing a categorization of existing work, the goal of our survey is to establish a common language for researchers who develop MTDs. This will help in making evident the aspects that have been considered and those that have been assumed away in the development of a particular MTD in the future. Our categorization also helps to identify promising directions for future research such as prevention, and hybrid surface shifting, improving the modeling of APTs, {\em etc.} (see Figure \ref{fig:fw}).

%The different phases of MTD lifecycle in Figure~\ref{fig:9} show that we need a situation-aware model, which can provide guidelines for attack information collection, MTD decision, and network state analysis post-MTD. We map different phases of MTD lifecycle to the layers of cloud infrastructure.

%\subsection{Attacks on the Cloud System}\label{acs}
\subsection{Attack Modeling Techniques}\label{acs}
Organizations utilize advanced infrastructure management tools and follow best practices such as software patching, hardening, analysis of the system's log for reducing the attack surface. Yet, skilled adversaries manage to compromise the network assets by utilizing zero-day attacks, customized malware, {\em etc.}, that are often difficult to detect or prevent using intrusion detection systems and anti-virus tools. For the effective deployment of intelligent cyber-defenses, it is crucial to collect information (called the threat model) about the attack process followed by an adversary.

An intelligence-driven approach focused on studying particular threats from an attacker's perspective is key for the detection and mitigation of sophisticated attacks in networks, which are characterized as \emph{Advanced Persistent Threats} (APTs) \cite{chen2014study, alshamrani2019survey}. To understand the collection, correlation and categorization of data related to a cyberattack, Lockheed Martin defines the \emph{Cyber Kill Chain} (CKC)~\cite{cyberkillchain}. The evidence-based knowledge derived from their study can help us in understanding and  deploying appropriate defense measures. Thus, we will first describe the different phases of the CKC followed by a brief description of APTs and how they can be viewed through the lens of CKC. This setup will later help us understand how MTDs can be effective against the different phases of an APT.

\subsubsection{Reconnaissance}
The attacker gathers information about the target environment in this phase. For example, the attacker can perform passive monitoring using automated tools such as trace-route and Nmap to perform network probes.

\subsubsection{Weaponization}
The attacker, based on information obtained in the reconnaissance phase, utilizes tools and techniques such as a phishing e-mail, a malware-infected document, {\em etc.} to create a targeted attack payload against the victim.

\subsubsection{Delivery}
The transmission of infected payload occurs during this stage. For example, the attacker may leave a malware-infected USB on the victim site or send an email to the Chief Financial Officer (CFO) of the company. This step requires the attacker to evade the authentication and thus, the people (and not the technology) become a more important target during this phase. Thus, a trained workforce can help in reducing the attack surface area.

\subsubsection{Exploitation}
The attack detonation takes place during this stage. This phase involves the exploitation of a vulnerability and the attacker gaining elevated privileges on the victim's resources by utilizing specially crafted payload that exploits a known (or zero-day) vulnerability.

\subsubsection{Installation}
Once the attacker gains elevated privileges in the exploitation stage, they may install malware on the victim's machine or choose to harvest useful information in the victim's database. Tools that can analyze abnormal behavior such as anti-malware, \emph{host-based IDS} (HIDS), {\em etc.} become quite important in attack detection during this stage.

\subsubsection{Command and Control ($C\&C$)}
After the installation phase is complete, the attacker contacts the $C\&C$ to maintain remote control over the victim machine. Tools such as \emph{network-based IDS} (NIDS) and outbound \emph{firewall rules} are quite useful in detecting and blocking malicious outbound connections requests.

\subsubsection{Actions on Objectives}
During this phase, the attacker executes the actions to achieve the attack goals, such as data-exfiltration, service disruption, {\em etc.} Two other important behaviors often observed in this attack-step are \emph{pivoting}, which involves identifying similar target nodes that have already been exploited, and \emph{lateral movement}, which involves identifying new systems that can be exploited in the network.

\subsection{Advanced Persistent Threats (APTs)}
\label{aptdef}

\emph{Advanced Persistent Threats} (APTs) refers to a distinct set of attacks against a high-value target organization that differs from normal cyber attacks in several ways \cite{alshamrani2019survey}. First, APTs are achieved by a group of highly-skilled attackers who are well-funded. According to \emph{Mandiant Report}~\cite{center2013apt1}, APTs such as \emph{Operation Aurora, Shady Rat, and Red October} have been used on a global scale for industrial espionage \cite{friedberg2015combating, moon2014mlds}. Oftentimes, the attackers mounting APT work closely with government organizations. Second, an APT attacker is extremely persistent; they (1) pursue the objectives repeatedly, often over an extended period, (2) can adapt to a defender's efforts in trying to resist the attack, and (3) determined to maintain a level of access within the defender's system required to achieve their objectives. Third, the key objective of APTs is to either exfiltrate information or undermining the key services in the network using multiple attack vectors \cite{kissel2011glossary}.

\paragraph{Relation between APTs and CKC}
An APT can be broken down into five phases-- reconnaissance, foothold establishment, lateral movement, data exfiltration, and post exfiltration \cite{alshamrani2019survey}. These can be mapped to different phases of the Cyber Kill Chain.
The {\em reconnaissance} phase in APTs maps directly to the {\em reconnaissance} phase in CKCs, which was described earlier. The \textit{foothold establishment} phase comprises of the \emph{weaponization} and the \emph{delivery} phases of CKC. The attackers use information gathered from the reconnaissance phase in order to prepare an attack plan and eventually exploit vulnerabilities uncovered in the target organization's web application, databases, and other software.

Once the attacker has gained a foothold into the target environment, they can try to move laterally in the target environment and gain access to other sensitive hosts and critical information in the organizational network in the lateral movement phase of APT. In this setting, the attacker needs to continuously explore (perform reconnaissance) and exploit the various components of the defender's system, mapping to the exploitation phase of the CKC.

In the data ex-filtration phase of APT, the attacker tries to exfiltrate the collected data to their \emph{command and control} $(C\&C)$ center. Most of the intrusion detection and prevention systems do ingress filtering and not egress filtering, thus, data exfiltration can often go undetected. 
The goal of the attacker in \emph{post ex-filtration} phase is to maintain persistence in the system for a long duration of time until the funding for attack campaign is finished or the goals of attacking the organization/government are fulfilled.

\subsection{Defense Methods}\label{defense-mech}

In this section, we provide a brief overview of the various defense techniques and highlight how each of these defense mechanisms is effective in various parts of a Cyber Kill Chain (CKC) in Table \ref{tab:2}. This discussion will help the reader better understand some of the MTD techniques that use these traditional defenses as a building block.

\begin{table}[t]
\centering
\begin{tabular}{p{1.8cm} | p{1.06cm} p{0.6cm}p{0.7cm}p{0.8cm}p{0.9cm}}
\hline
\textbf{Phase}&\multicolumn{5}{c}{\textbf{Network Defense Techniques}} \\

& \textbf{Detect} & \textbf{Deny} & \textbf{Disrupt} &\textbf{Degrade} & \textbf{Deceive}\\
\hline
\hline
Reconnaissance & Web Analytics & & & \\
\hline
Weaponization & NIDS & NIPS & & & \\
\hline
Delivery & Vigilant user & Proxy filter & AV & Queuing & \\
\hline
Exploitation & HIDS & Patch & DEP & & \\
\hline
Installation & HIDS & chroot & AV & & \\
\hline
C2 & NIDS & ACL & NIPS & Tarpit & DNS redirect \\
\hline
Actions on Objectives & Audit log & & & QoS & Honeypot \\
\hline
\end{tabular}
\caption{Highlights how existing defense mechanisms are related to the different phases of the Cyber Kill Chain.}
\label{tab:2}
% \vspace{-1.6em}
\end{table}

\subsubsection{Web Analytics} A huge amount of security-related information is present on the web-- in \emph{social engineering} websites, phishing sites, and dark-web forums-- including discussion about a particular target product or company. As discussed by Glass \textit{et. al.}~\cite{glass2011web}, the problem of exploring and analyzing the web for information should provide (1) Security relevant information discovery capabilities (2) \emph{Situation awareness} by performing real-time inference over the available information, and (3) predictive analysis to provide early warning for any likely security attacks. 
%One such data collection and analytic system is CACTUS~\cite{ackley2009cactus} that provides automated analytic capabilities that can be leveraged by prediction frameworks.

\subsubsection{Intrusion Detection Systems (IDS)}
There are two types of IDS agents that can be deployed in the network under attack, i.e., Network-based IDS (NIDS) and Host-based IDS (HIDS). NIDS such as \textit{Bro}~\cite{mehra2012brief} and \emph{Snort}~\cite{rehman2003intrusion}, use signature match techniques based on known attack patterns and can flag incoming network traffic as malicious or benign. HIDS such as \textit{auditd}~\cite{deshpande2018hids} or \textit{tripwire}~\cite{kim1994design}, on the other hand check the Indicators of Compromise (IOCs) on the physical server or VM, in order to identify malicious activity, e.g., privilege escalation attempt by normal user.

\subsubsection{Intrusion Prevention Systems (IPS)}
A network security threat prevention technology such as IPS~\cite{zhang2004intrusion} examines network traffic flow to detect and prevent vulnerability exploits. The exploits come in the form of malicious inputs to the target application or service. The IPS is often  deployed behind the firewall and provides a complementary layer of analysis. Compared to IDS, the IPS system takes automated action based on traffic pattern match such as (1) dropping malicious traffic packets, (2) blocking requests from a source, (3) resetting connection, and (4) sending an alarm to the administrator.

\subsubsection{Proxy Filtering}
A (reverse) proxy server such as \emph{nginx}, can act as an intermediary between the attacker located on the public network and private network. A proxy can help in shielding the real network from the attacker.

\subsubsection{Access Control List (ACL)} ACL can be applied at different enforcement points in a network. ACLs, such as \emph{netfilter} \cite{welte2000netfilter}, investigate network traffic and based on properties such as source/destination IP address, port number, protocol {\em etc.} decide either to \emph{permit} or \emph{deny} it.

\subsubsection{Data Execution Prevention (DEP)}
DEP is a security feature that can be deployed in the form of hardware or software to prevent malicious code from running on the host. They \emph{monitor programs} run on the host and ensure it uses memory in an expected (or safe) manner. 

\subsubsection{Anti-Virus (AV)}
A software program designed to protect the hosts from malicious programs such as a virus, spyware, worms, rootkits, key-loggers, {\em etc.} The AVs can be classified into two types-- \emph{signature-based AV} and \emph{behavior-based AV}. The signature-based AVs check the malicious program against the database of known malicious programs. On the other hand, the behavior-based AVs check the program behavior by running it in a sandbox environment. 

\subsubsection{Tarpit}
This defense technique involves purposeful introduction of delay when responding to queries. The key idea behind this defense mechanism is that adversaries will give up on a target if it takes too long to achieve the defined goal.

\subsubsection{QoS}
The \emph{network traffic} can be segmented on the service type and importance. The segmented traffic can be analyzed for bottlenecks and threats. The malicious traffic can be slowed down in order to increase the \emph{Cost of Attack (COA)} or selectively dropped.

\subsubsection{DNS Redirect}
The malicious connection requests can be served a different response than was expected. The attacker may seek to connect with command and control center using a page with malware, but DNS redirect will kill this chance. 

\subsubsection{Honeypot}
A security mechanism, which can be used to detect, deceive or in some cases counter a malicious user trying to gain access to key network services~\cite{pouget2004honeypot}. Honeypots such as \emph{Cowrie}~\cite{oosterhof2014cowrie} can help in better understanding the attacker's tools and tactics. The connection requests from the attacker are directed to a decoy service, which mimics the behavior of the normal service and logs the attacker's activity. 

%\texthl{Owners to be decided}\vspace{7pt}

Although there exists a large set of defense mechanisms, attackers often use clever techniques to evade detection or prevention. SANS \cite{idsevade} discusses methods like \emph{obfuscation}, \emph{fragmentation}, \emph{encryption} and \emph{Denial of Service} (DoS) attacks against signature-based detection tools.
Detection based on HIDS can also be evaded by a skilled attacker using techniques such as file location manipulation (using directories or files white-listed by IDS), application hijacking, {\em etc.} Furthermore, machine learning models that can help overcome some of the shortcomings of signature-based detection tools can themselves be fooled by adversarial attacks \cite{szegedy2013intriguing,al2018adversarial}. On similar lines, deception techniques such as DNS redirect and Honeypot can help in deceiving the attacker temporarily, but over a longer period of time, the attacker will eventually change their attack methodologies thereby reducing the effectiveness of these defenses. Stojanovski \textit{et. al.} \cite{stojanovski2007bypassing} performed experimental analysis on how to bypass \emph{DEP protection} on \emph{Windows XP} systems.
Thus, there is a need to perform intelligent manipulation to assure the attacker's likelihood of reaching the desired goal is limited, even if they can evade the traditional detection methods. Additionally, the defense mechanisms discussed in the Section~\ref{defense-mech} target known attacks often with easy to detect signature patterns.

\paragraph{Proactive Defenses against APTs}

The differences between APTs and regular cyberattacks, discussed in \ref{aptdef}, make it arduous to use traditional defenses and pre-specified threat models to address APTs as a whole. In this regard, proactive defenses can prove to be effective against APTs. While MTDs, as we will see later, makes it difficult for APT attackers by dynamically shuffling various system components (see Fig \ref{fig:mtd-apt}), other proactive defenses such as cyber-deception can prove to be effective in gathering threat-model information. For example, Shu \textit{et al.} propose a cyber deception to protect \emph{FTP} services against APT attackers \cite{shu2018ensuring}. In their research work, a defender reroutes attack traffic to a host, which may be a honeypot, and the defender ensures that an attacker is not able to notice a connection difference between the real IP address and the honeypot trap. By observing attacker's behavior on the honeypot, the defender updates the threat-model and in turn, hardens their FTP services. A key aspect of this work is to make the attackers continuously believe that they are interacting with the original environment as opposed to a honeypot.
% Next, the attacker attempt is transferred  to a \textit{deceptive environment} \cite{shu2018ensuring}. \textit{Bftpd} and \textit{ProFTPd} services were set up in order to attract the APT attacker to the deceptive area. The author calls the process of transforming the attacker from the target to the deceptive area a context switch. It is vital to ensure the exact communication setup remains during the context switch process, otherwise, the attacker will know they are being trapped.

\begin{figure*}[t]
    \centering
    \includegraphics[width=0.8\textwidth]{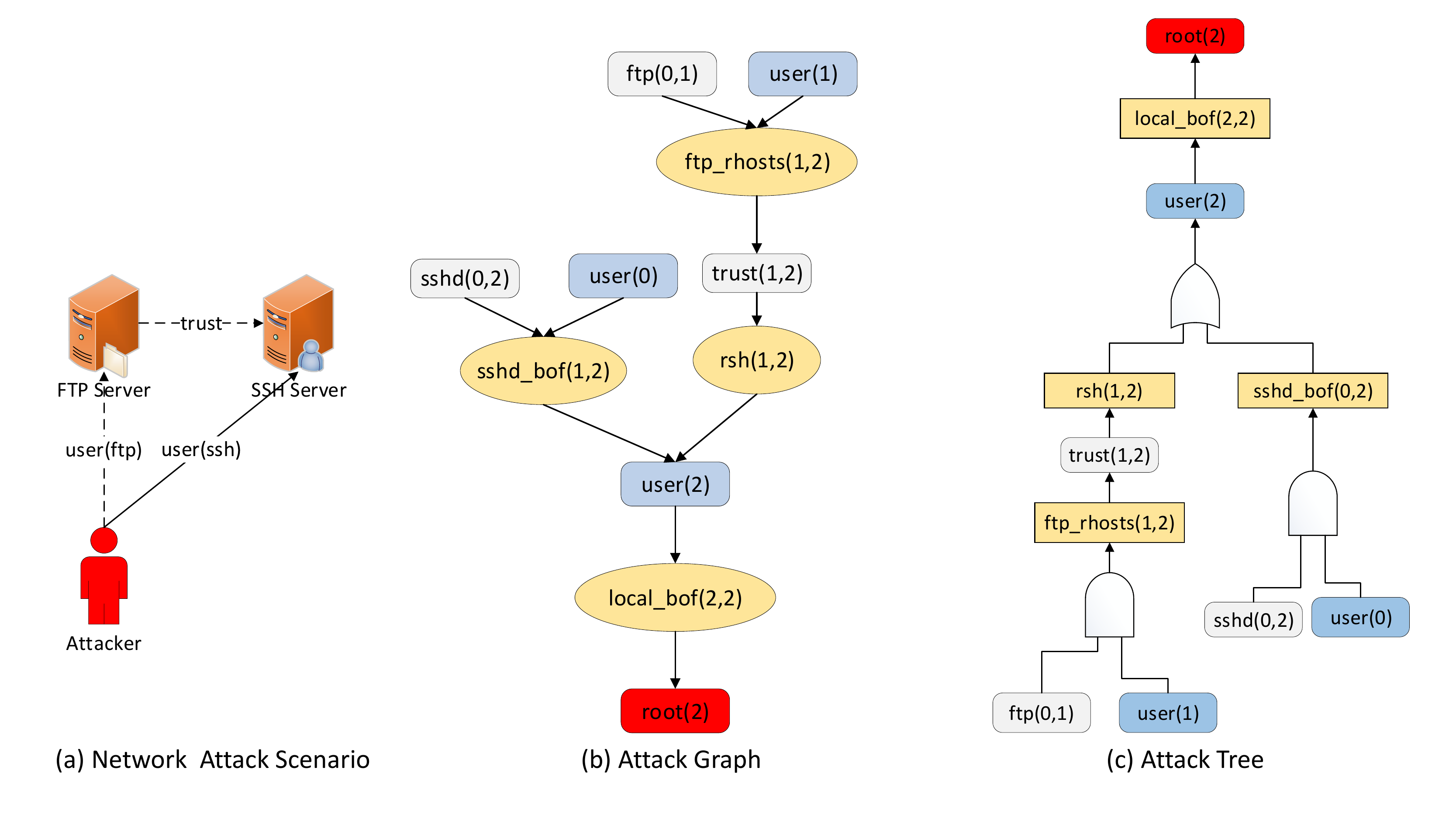}
    \caption{Attack representations such as an attack graph (b) and an attack tree (c) for a simple network attack scenario (a) in a small-size cloud system with known vulnerabilities. Creation of these representations from network vulnerability and reachability information often suffers from scalability issues for real-world networks (see Table \ref{tab:4}).}
    \label{fig:4a}
\end{figure*}

% == Commenting out this for now == %
\iffalse
\texthl{Ankur: Pointwise or tabular desc of issues with current defenses would be better.\\Sailik: I agree. But there are two reasons for my deterrence here-- (1) The write up had only described weaknesses against some of the traditional defense mechanisms (as opposed to all of them), which is necessary to make a pretty table, and (2) if we do that here and later use some of these defenses in our MTD, we will have to have another table describing why (and how) is MTD not vulnerable to any of these while each attack by themselves are.}
\begin{enumerate}
    \item Obfuscation: Use a variant of traffic signature to evade IDS signature match.
    \item Fragmentation: Breaking the attack data into multiple fragments.
    \item Encryption: Most IDS tools are configured to operate on raw traffic, so encrypted traffic evades intrusion detection. 
    \item Denial of Service: Flooding the detection system with multiple attack packets, creating alarms originating from multiple source IP addresses. This will reduce the chances of IDS detecting the actual attacker considerably.
\end{enumerate}
\fi
% === % 

\subsection{Domain Information for Modeling Cyber-attacks}

Defenders almost always have information about the system they want to protect. This knowledge can help in enhancing the threat-model and, in turn, improve the effectiveness of MTD techniques. Such information may range from knowledge of the network architecture, the capacity of the individual machines, known vulnerabilities (and an idea of how dangerous they are), {\em etc.} We discuss here a few popular models and data sources that have been leveraged by researchers.

\subsubsection{Common Vulnerabilities and Exposures (CVEs)}
are publicly known vulnerabilities and exposures that are categorized and can be referenced uniquely in the \emph{National Vulnerability Database} (NVD) using a common vulnerability enumeration identifier (CVE-ID). This system is maintained and updated by the \emph{Mitre} corporation regularly to make defenders and administrators aware of existing and new vulnerabilities.

\subsubsection{Common Vulnerability Scoring System (CVSS)}
The use of the Common Vulnerability Scoring System (CVSS) for rating attacks is well studied in security \cite{houmb2010quantifying}. For (most) CVEs listed in the NVD database, we have a six-dimensional \emph{CVSS v2} vector~\cite{cve2012exposures}, which can be decomposed into multiple components that represent Access Complexity (AC), i.e. how difficult it is to exploit a vulnerability, and the impact on Confidentiality, Integrity, and Availability (CIA) gained by exploiting a vulnerability. The values of AC are categorical \{EASY, MEDIUM, HIGH\}, while CIA values are in the range $[0, 10]$. These scores are also known as the Exploitability Score (ES) and the Impact Score (IS).

Although a defender may be aware of the known vulnerabilities present in their system (by being aware of the published CVEs that affect them), the knowledge of how they affect their system, in particular, may help them in making better decisions for security. The attack representation can be useful to quantify the attack and defense surface for MTD. To this extent, we define two heavily used representation methods that can represent known attacks present in a system-- the \emph{attack tree} and the \emph{attack graph} as shown in the Figure~\ref{fig:4a}. 

The Figure~\ref{fig:4a}, shows a network attack scenario, where an attacker has access to FTP and SSH server over the network. This example illustrates different methods for representing multi-stage attacks, i.e., attack graph and attack tree which we defined above. The FTP server consists of a vulnerability that allows the attacker to obtain remote shell (rsh) on the system. The SSH server, on the other hand, consists of buffer overflow ($sshd\_bof$) vulnerability. The goal of the attacker is to obtain root privilege on the SSH server, i.e., \textit{root(2)}. The attack progression using attack graph and attack tree has been presented in Figure~\ref{fig:4a} (b), (c) respectively.

\textbf{Attack Tree}~\cite{schneier1999attack} as shown in Figure~\ref{fig:4a}(c) is another method of representing system security. The Attack Tree represents the network attacks. Attack Tree represents a monotonic path taken by an attacker starting from a leaf node to reach a goal node. Attack Tree usually consists of set of \emph{AND} nodes (sshd(0,2), user(0) - Figure~\ref{fig:4a}(c)) and \emph{OR} nodes (rsh(1,2), $sshd\_bof$(0,2)). The \emph{OR} nodes represent one or more ways in which a goal node can be reached, whereas \emph{AND} nodes represent different conditions that must be fulfilled to achieve a goal node. Children of the node are refinements of this goal, and the attacks that can no longer be refined are represented by leaf nodes~\cite{mauw2005foundations}.

\begin{definition}
An Attack Tree~\cite{mauw2005foundations} can be defined by three tuples $(N, \rightarrow, N_r)$
\begin{itemize}
    \item N is all possible nodes in the tree;
    \item $S^+(N)$ is \emph{multi-set} of all possible subsets of nodes N;
    \item $\rightarrow \subseteq N \times S^+(N)$ denotes transition relation;
    \item $N_R$ represents the \emph{goal node} of the attack tree.
\end{itemize}    
\end{definition}

\begin{table*}[t]
    \centering
    \setlength{\extrarowheight}{3pt}
    \begin{tabular}{|p{40mm}|p{60mm}|p{60mm}|}
        \hline
        \textbf{Category} & \textbf{Details} & \textbf{Complexity}\\
        \hline
        Automated Attack Analysis ~\cite{ingols2006practical} & Multi-prerequisite graph based on vulnerability and reachability information & O(E+NlgN); N is attack graph nodes and E is graph edges \\
        \hline
        Attack Cost Modeling~\cite{albanese2012time} &
        Time Efficient Cost Effective hardening of network using Attack Graph &  $O(n^{\frac{d}{2}})$; d represents the depth of the attack graph \\
        \hline
        Attack Cost Modeling~\cite{jha2002two} & Model checking based attack graph generation using Markov Decision Process (MDP) &  Approximation algorithm $\rho(n)$ = $H(max_{a \in A}$
        $\{\mu_G(a)\})$, where A is Attacks, $\mu$ is maximization function.\\
        \hline
        Scalable Attack Graph~\cite{ou2006scalable} & Scalable attack graph using logical dependencies. & $O(N^2) - O(N^3)$, where N is number of nodes in attack graph. \\
        \hline
        
        Attack Graph based Risk Analysis~\cite{lee2009scalable} & Scalable attack graph for risk assessment using divide and conquer approach &  $O(r(n+c)^k)$, where r is small coefficient. \\
        \hline
        Attack Cost Modeling~\cite{homer2008attack} & Attack Graph cost reduction and security problem solving framework Min. Cost SAT Solving. & NP-Hard problem, SAT solving methods employed. \\
        \hline
        Ranking Attack Graphs~\cite{sawilla2008identifying} & Asset Ranking algorithm for ranking attack graphs to identify attacks. \emph{Page Rank} based algorithm & Similar to complexity of page rank algorithm. \\
        \hline
        Attack Cost Modeling~\cite{huang2011distilling} & Identifying critical portions of attack graph. Min. Cost SAT solving, Counter-example guided abstraction refinement (CEGAR) & NP-Hard problem, SAT solving methods used. \\
        \hline
    \end{tabular}
    \caption{Complexity of the various Attack Representation methods.}
    \label{tab:4}
    % \vspace{-1.6em}
\end{table*}

\textbf{Attack Graph} is a data structure used to represent attack propagation in a network with vulnerabilities as shown in Figure~\ref{fig:4a}(b). The start state of the attack graph represents the current privileges of the attacker. The goal state of the attack graph represents a state in which the intruder has successfully achieved his attack goal, e.g., data-exfiltration, root privileges on a web server, {\em etc.} Security analysts use attack graph for attack detection, network defense, and forensics~\cite{sheyner2003tools}. We formally define the attack graph as follows:

\theoremstyle{definition} 
\begin{definition}\label{definition:AG} Attack Graph (AG)
An attack graph is represented as a graph $G=\{V,E\}$, where \textit{V} is set of nodes and \textit{E} is set of edges of the graph \textit{G}, where 
\begin{enumerate}
    \item $V=N_C \cup N_D \cup N_R$, where $N_C$ denotes the set of conjunctive or exploit nodes, $N_D$ is a set of disjunctive nodes or result of an exploit, and $N_R$ is the set of a starting nodes of an attack graph, i.e. root nodes.
    \item  $E=E_{pre} \cup E_{post}$ are sets of directed edges, such that $e \in E_{pre} \subseteq N_D \times N_C$, i.e., $N_C$ must be satisfied to obtain $N_D$. An edge $e \in E_{post} \subseteq N_C \times N_D$ means that condition $N_C$ leads to the consequence $N_D$. $E_{pre}$ represents the attack graph pre-conditions (ftp(0,1) and user(1) in the Figure~\ref{fig:4a}(b)) necessary for vulnerability exploitation and $E_{post}$ are the post-conditions (rsh(1,2) in the Figure~\ref{fig:4a}(b)) obtained as a result of exploit. 
\end{enumerate}
\end{definition}
The time taken to construct attack representation methods (ARMs) grows exponentially with an increase in the number of hosts or the number of known vulnerabilities in the network \cite{hong2013scalable}. Authors in \cite{hong2017survey} survey various research works that try to address this challenge. Amman \textit{et al.} \cite{ammann2002scalable} present a scalable solution that assumes monotonicity; it allowed them to achieve scalability of $O(N^6)$.
To mitigate the state explosion problem, most of the existing solutions try to reduce the dependency among vulnerabilities by using some logical representation \cite{ou2005mulval} or by imposing a hierarchical structure to reduce the computing and analysis complexity of constructing and using ARMs \cite{hong2013performance}. 
In the latter work, the authors proposed a two-layer AG generation approach where the goal was to develop a faster method by considering network reachability and vulnerability information at different layers. The graphs constructed have vulnerability information of the individual hosts in the lower graph while the topological information of the network is in the upper layer. Unfortunately, the effectiveness of these methods on real-world networks is uncertain.

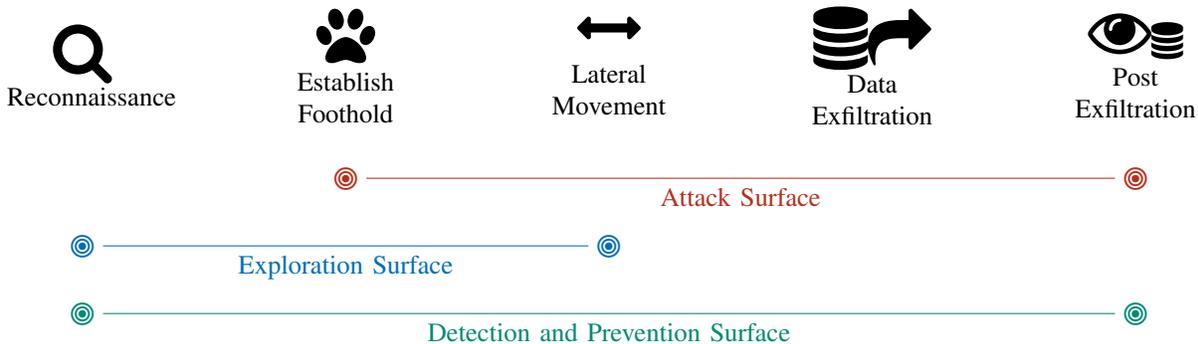
\begin{figure*}
    \centering

    \begin{tikzpicture}[
            state/.style={rectangle},
            node distance=3.5cm,
            ]
            \centering
            \node[state] (recon) [text width=2cm, align=center] {{\Huge \faSearch}\\{Reconnaissance}};
            
            \node[state] (ef) [right of=recon, text width=2cm, align=center] {{\Huge \faPaw}\\{Establish Foothold}};
            
            \node[state] (lm) [right of=ef, text width=2cm, align=center] {{\Huge \faArrowsH}\\{Lateral Movement}};

            \node[state] (de) [right of=lm, text width=2cm, align=center] {{\Huge \faDatabase \faMailForward}\\{Data Exfiltration}};
            
            \node[state] (pe) [right of=de, text width=2cm, align=center] {{\Huge \faEye}{\Large \faDatabase}\\{Post Exfiltration}};

            \node[state] (as_s) [below of=ef, yshift=2cm, text width=0.3cm, align=center] {\color{BrickRed} \faBullseye};
            \node[state] (as_e) [below of=pe, yshift=2cm, text width=0.3cm, align=center] {\color{BrickRed} \faBullseye};
            
            \node[state] (es_s) [below of=recon, yshift=1.1cm, text width=0.3cm, align=center] {\color{NavyBlue} \faBullseye};
            \node[state] (es_e) [below of=lm, yshift=1.1cm, text width=0.3cm, align=center] {\color{NavyBlue} \faBullseye};
            
            \node[state] (dps_s) [below of=recon, yshift=0.2cm, text width=0.3cm, align=center] {\color{PineGreen} \faBullseye};
            \node[state] (dps_e) [below of=pe, yshift=0.2cm, text width=0.3cm, align=center] {\color{PineGreen} \faBullseye};

            \path[BrickRed, -]
                (as_s) edge node[below] {Attack Surface} (as_e);
                
            \path[NavyBlue, -]
                (es_s) edge node[below] {Exploration Surface} (es_e);
                
            \path[PineGreen, -]
                (dps_s) edge node[below] {Detection and Prevention Surface} (dps_e);
                
    \end{tikzpicture}
    
    \caption{A mapping between the different phases of Advanced Persistent Threat (APT) and various surfaces of a cyber system that various MTDs seek to move. Shifting of the exploration surface and the attack surface are effective only against some phases on an APT, whereas shifting the detection and the prevention surface is effective throughout the APT life-cycle.}
    \label{fig:mtd-apt}
    % \vspace{-1.6em}
\end{figure*}

Table~\ref{tab:4} highlights the complexity of creating various attack graph and attack tree-based methods. As can be see, scalability is a major concern when coming up with a good attack representation, thereby impacting the effectiveness of MTD techniques. In \cite{chowdhary2016sdn}, the authors present a scalable solution for approximating the attack graph of a large scale data-centric network by using distributed attack graph computation followed by semantic clustering. We discuss attack representation methods as one of the \emph{Quantitative metrics} which can be used for measuring effectiveness of MTD in Section~\ref{effect}.

\section{Categorizing Moving Target Defenses}
\label{sec:mtd}

The goal of \emph{Moving Target Defense} (MTD) is to continually move the components of an underlying system in a randomized fashion; this ensures the information gathered by the attacker in the \emph{reconnaissance phase} becomes stale during the \emph{attack phase} given the defender has moved to a new configuration within that time. This increases the uncertainty for the attacker, making it difficult (or rather, more expensive) for them to successfully exploit the system.

An MTD can be described using a three tuple $\langle M, T, C \rangle$ where $M$ represents the movement strategy that directs the system {\em how to move}, $T$ denotes the timing function that represents {\em when} a switch action occurs in the MTD system and $C$ represents the configuration space or the set of configurations between which the system switches, answering the question of {\em what to switch}. Answers to these three questions can help us define a useful categorization for the various MTDs. We believe that this categorization (1) gives a holistic view of the various MTD systems, providing a heuristic sense of what modeling aspects, when done carefully, make a particular defense effective, and (2) highlights opportunities on how modeling of a particular component of an MTD can either be improved or amalgamated with other MTDs.

As we will see, the configuration set $C$ and the timing function $T$ are often a design choice made the system administrator depending on the threat model. However, the movement function $M: H \rightarrow C$ needs to be a stochastic function that given the history of system configurations $H$ as input, produces the next configuration the system should switch to. A stochastic function is needed because a deterministic function can easily be learned by an attacker over time and thus, does not help the defender in terms of security.

%\texthl{Incorporating the papers in the other two sections on implementation and evaluation into this section.}

\subsection{The Configuration Set $C$ -- What to Switch?}

At an abstract level, from the perspective of an attacker, every software system can be categorized into four surfaces:

\begin{itemize}
    \item Exploration Surface
    \item Attack Surface
    \item Detection Surface
    \item Prevention Surface
\end{itemize}

In this section, we use these surfaces as a basis for categorizing what the different MTDs shift.
In the context of multi-stage attacks like APTs, an adversary needs to exploit all the different surfaces, but not necessarily in a predefined order (see Fig. \ref{fig:mtd-apt}). For example, an adversary may first try to explore the target network and try to figure out its topology, bandwidth, software deployed on the different nodes, {\em etc.}, but it may also need to perform reconnaissance at an advanced stage, say, to verify if it has actually gained access and established a foothold, i.e. a new vantage point, within the system. The knowledge gained in the exploration phase helps them to execute attacks that exploit the system, to move to different points in the network, and exfiltrate important information.
As both exploration and attack traffic are malicious in nature, they tend to differ from legitimate user traffic. At that point, the detection and prevention surfaces come into play.

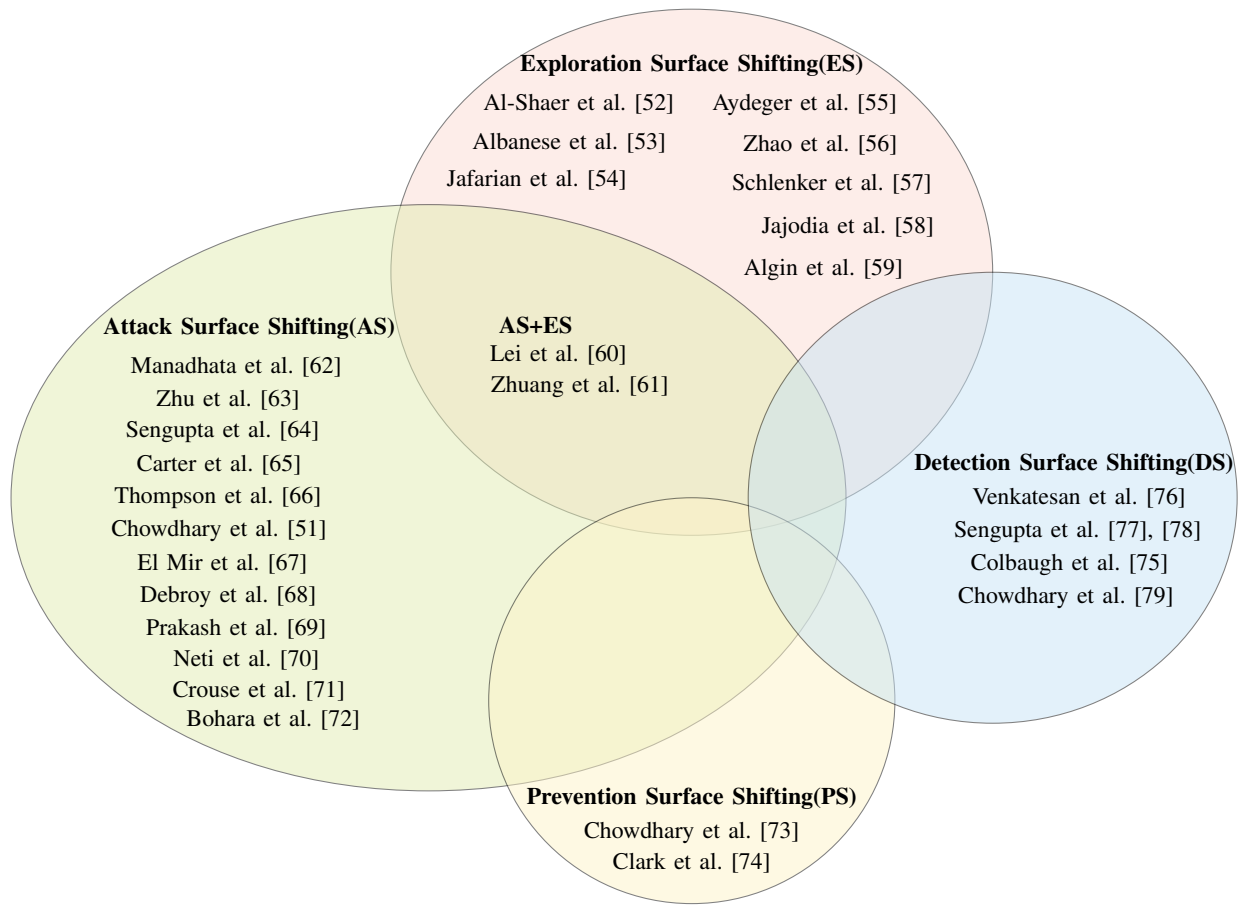
\begin{figure*}
\small
\centering
\begin{tikzpicture}
\begin{scope}[opacity=0.5]
    % everything is polar co-ordinates
    % (angle from origin [0-360], distance)
    \fill[Salmon!30!white,draw=black] (90:3) ellipse (4 and 3.5);
    \fill[SpringGreen!50!white,draw=black] (180:3.5) ellipse (5.55 and 3.9);
    \fill[Goldenrod!30!white,draw=black] (270:2.7) circle (2.7);
    \fill[CornflowerBlue!30!white,draw=black]  (360:4) circle (3.25 and 3);
    
 \end{scope}    
    % Exp Surface
    \node at (90:5.75) {\textbf{Exploration Surface Shifting(ES)}};
    \node at (106:5.45) {Al-Shaer et al.~\cite{al2012random}};
    \node at (109:5) {Albanese et al.~\cite{albanese2013moving}};
    \node at (116:4.7) {Jafarian et al.~\cite{jafarian2012openflow}};
    \node at (74:5.45) {Aydeger et al.~\cite{aydeger2016mitigating}};
    \node at (70:5) {Zhao et al.~\cite{zhao2017sdn}};
    \node at (66:4.58) {Schlenker et al.~\cite{schlenker2018deceiving}};
    \node at (60:4.15) {Jajodia et al.~\cite{jajodia2018share}};
    \node at (60:3.5) {Algin et al.~\cite{algin2017mitigating}};
     
    % Attack + Exp Surface
    \node at (132:3.1) {\textbf{AS+ES}};
    \node at (133:2.6) {Lei et al.~\cite{lei2017optimal}};
    \node at (135:2.1) {Zhuang et al.~\cite{zhuang2014towards}};
   
    % Attack Surface
    \node at (159:6.3) {\textbf{Attack Surface Shifting(AS)}};
    \node at (164:6.3) {Manadhata et al.~\cite{manadhata2013game}};
    \node at (168:6.3) {Zhu et al.~\cite{zhu2013game}};
    \node at (172:6.3) {Sengupta et al.~\cite{sengupta2017game}};
    \node at (176:6.3) {Carter et al.~\cite{carter2014game}};
    \node at (180:6.3) {Thompson et al.~\cite{thompson2014multiple}};
    \node at (184:6.3) {Chowdhary et al.~\cite{chowdhary2016sdn}};
    \node at (188:6.3) {El Mir et al.~\cite{el2016softwarel}};
    \node at (192:6.3) {Debroy et al.~\cite{debroy2016frequency}};
    \node at (196:6.3) {Prakash et al.~\cite{prakash2015empirical}};
    \node at (200:6.3) {Neti et al.~\cite{neti2012software}};
    \node at (204:6.3) {Crouse et al.~\cite{crouse2012improving}};
    \node at (208:6.3) {Bohara et al.~\cite{bohara2017moving}};
    
    % Prevention Surface
    \node at (270:4)    {\textbf{Prevention Surface Shifting(PS)}};
    \node at (270:4.45)    {Chowdhary et al.~\cite{chowdhary2017dynamic}};
    \node at (270:4.85)    {Clark et al.~\cite{clark2015game}};
  
    % Detection Surface Shifting
    \node at (350:5.1) {Colbaugh et al.~\cite{colbaugh2012predictability}};
    \node at (360:5.15) {Venkatesan et al.~\cite{venkatesan2016moving}};
    \node at (355:5.15) {Sengupta et al.~\cite{sengupta2018moving,senguptageneral}};
    \node at (345:5.15) {Chowdhary et al.~\cite{chowdhary2018markov}};
    \node at (5:5.1)    {\textbf{Detection Surface Shifting(DS)}};
  
  %\node [font=\Large] {\LaTeX};
  \end{tikzpicture}
    
%\end{comment}

\caption{The different surfaces that can be moved by a particular Moving Target Defense (MTD). Moving the exploration surface makes it harder for the attacker to figure out the exact system configuration by making the sensing actions noisy or erroneous. Moving the attack surface makes a particular attack inapplicable. Moving the detection surface, similar to `patrolling methods', helps in providing efficient detection in budget-constrained cyber environments. Moving the prevention surface makes it harder for an attacker to ex-filtrate data even when they have a strong foot-hold inside the system.}
\label{fig:what}
% \vspace{-1.6em}
\end{figure*}

%We now describe the works in context to the surface(s) they shift, followed by a discussion about works that shifts more than one surface and thus, fall at the intersection of two or more individual areas.
In Fig. \ref{fig:what}, we show a Venn diagram that categorizes existing MTDs based on the surface they shift. Although this categorization is at a level of abstraction, we discuss various works and show how different MTDs define the elements of the set $C$. Most MTDs shift one surface at a time, rarely considering scenarios where other surfaces can be shifted in conjunction. After discussing the various MTDs, we briefly mention some unexplored research areas that can lead to the development of new defenses that shift multiple surfaces.

\subsubsection{\textbf{Exploration Surface Shifting}}\label{exp}

The goal of shifting the exploration surface is to ensure that the model of a system that an attacker can gather by exploration actions such as by scanning for open-ports, sending non-malicious traffic to uncover system topology, discover vulnerabilities, {\em etc.}, are noisy or inaccurate. Thus, an adversary, with this faulty information from the reconnaissance phase, is left with no other choice but to work with faulty view of the attack surface.
%shoot arrows (attacks) in the dark.

In \cite{al2012random,albanese2013moving}, the authors propose the concept of Random Host Mutation (RHM) -- moving target hosts are assigned random virtual IP addresses (vIP) in an unpredictable and decentralized fashion. Formally, $C$ represents the set of bipartite graphs, where each configuration represents a mapping from the set of vIP addresses in the Virtual Address Range (VAR) to the set of hosts in the network. Every switch action changes the one-to-one mapping of hosts in the system to VARs. Another work \cite{jafarian2012openflow} tries to implement an MTD on the same set of configurations using a centralized approach based on \emph{(SDN)} technologies like \emph{OpenFlow}.

% In \cite{jafarian2012openflow}, the authors seek to implement the same defense method using a centralized approach that uses \emph{software-defined networking} (SDN) technologies like \emph{OpenFlow}.

Another line of work focuses on reducing the quality of information that an attacker can gather through exploration. In \cite{aydeger2016mitigating}, the authors use a centralized approach to obfuscate the network links so that the topology that an attacker can retrieve via crossfire attacks is noisy and unreliable. In this work, each state is a possible path from the source to the destination of the crossfire attack. Thus, the configuration space $C$ represents the set of all paths from a source point to a destination point in the network. The MTD paradigm comes into play because the administrator chooses a path, in a randomized fashion, so that the attacker is not able to get reliable path information between any any two points in the network. Given attacks may only succeed if the packet is sent over a particular path, attackers are forced to use a lot of attack traffic in order to succeed in exploiting the system or even obtain a reasonable estimate of the network topology. Such behavior increases their chances of getting caught or dealing with an inaccurate (and hopefully, not useful) model of the system. On similar lines, there have been works that either try to move the fingerprint of a protected host \cite{zhao2017sdn} or randomly alter the time schedule that guides when a host transmits information, reducing the effectiveness of selective jamming attacks against smart meters \cite{algin2017mitigating}. In \cite{schlenker2018deceiving}, the authors propose to send back incorrect information to an attacker trying to query information about the hosts on the network. Although they come up with deterministic strategies for responses, the possibility of sending back different lies opens up when the underlying attack surface uses an MTD. These works are termed as {\em cyber-deception} because the defender is trying to deceive the adversary by feeding them false information about the system. In such cases, the goal of the defender is to ensure that the adversary's model of the underlying network is incorrect as opposed to just being incomplete or noisy.

In \cite{jajodia2018share}, Jajodia et. al. looks at how they can deploy \emph{honey-nets} in a strategic fashion and thereby, minimize the expense of deploying all possible honey-nets at once. They show that deploying honey-nets introduces deception in the network against \emph{noisy-rich} (NR) cyber-attackers (i.e. adversaries who try to exploit all the vulnerabilities present in order to compromise the target network). In this case, if we represent the set of all honey-nets that can be placed in a network as $X$, then the configuration space $C$ consists of all budget-limited subsets of honey-nets that can be deployed by the administrator.

\subsubsection{\textbf{Attack Surface Shifting}}\label{att}

Most of the work from the research community has focused on movement of the attack surface. The main aim of switching between attack surfaces is to render invalid an attack action that the attacker chooses after some exportation. For example, an attack to exploit a \emph{Linux-based OS} will be useless if it is executed on a machine running a \emph{Windows OS}. We first discuss some MTD methods that are, for a networking environment, defined at an abstract level. We will then, focus on ones that consider moving more specific elements of the network.

In one of the earlier works \cite{manadhata2013game}, the authors have a set of systems, which forms their space of MTD configurations $C$, and each configuration $c \in C$ can be represented by an attack surface characterized by three properties-- (1) the set of entry and exit points into the system, (2) the set of \emph{channels}, and (3) the set of un-trusted items (data/system/network link). The defender aims to switch between the different configurations to minimize the similarity of the attack surface in consecutive configurations and at the same time, have a minimum impact on performance. As we will soon see, this multi-objective trade-off is a common theme in many of the works on MTD.

In \cite{zhu2013game}, Zhu and Bashar break down a full-stack system into several layers (denoted by $l$). For each layer, they have a set of technologies that can be used to provide the functionality that the layer is responsible for. Now, when all layers (or stacks) come together to form the full-stack system, all possible combinations of technologies cannot be used to provide proper functionality to the end-user. Thus, from all these possible combinations, they handpick a few combinations of technologies that meet the use-case demands for the full-stack software among which they switch, which defines the configuration space $C$.
On similar lines, Sengupta et al. \cite{sengupta2017game} also assume a full-stack web-application and thus, has similar action sets where the technologies for each layer are relevant to web-application development. The two papers differ in the way they decide {\em how to switch} (to be discussed later).

Carter et al. design an MTD system that switches between different Operating Systems (OSs) (which are called `platforms') \cite{carter2014game}. In their case, the configuration set $C$ is the set of all OSs that they can shift between. They mention a notion of similarity (or diversity) between two configurations $\in C$ based on code syntax and semantics of the functionality provided. On similar lines, authors in \cite{thompson2014multiple} move away from MTD systems that consider a set of abstract configurations and implement an MTD that can perform the OS rotation for machines deployed in the systems hosted on a network using a centralized mechanism. We now shift our attention to MTDs that move elements solely relating to networks.

In \cite{chowdhary2016sdn}, Chowdhary et al. describe an MTD that leverages port hopping to thwart known attacks on a cloud-based system. In their work, the states of the system are composed of variables, each of which indicates if a certain vulnerability in the system has been exploited (or not) and based on it, decides when and how to move. This fits well with our earlier discussion on how various surfaces are inter-related -- the attack surface shifting comes into play when an attacker can successfully evade the detection surface. Along similar lines, authors in \cite{el2016softwarel} move a deployed Virtual Machine (VM) to a different physical server if the impact of known vulnerabilities (measured using heuristic measures) on the physical server exceeds a certain threshold. In \cite{debroy2016frequency}, the authors implement MTD at a different layer of the system abstraction where they move services deployed on a particular VM to another VM. A natural extension could be to use both the layers for designing a hybrid MTD that shifts (1) software deployed on a VM and (2) individual VMs deployed on actual physical servers in the cloud network. This is similar to the concept of multi-layer MTD, as discussed in \cite{sengupta2017game,zhu2013game}.

In \cite{prakash2015empirical}, authors talk about a range of configurations (precisely, $72$ of them) and analyze the effect of using various game-theoretic strategies for switching between them. They shed some light on the amount of security gain that can be achieved in various settings. In \cite{neti2012software}, the authors point out a fundamental assumption that is often inherent in cybersecurity in general and MTD systems in particular-- the different configurations of the MTD are not vulnerable to the same attack (similar to the notion of diversity described in \cite{carter2014game}). They create a bipartite graph that maps hosts to vulnerabilities and show that MTD is more effective when the diversity of the configuration $\in C$ is higher. Similar conclusions have also been drawn in regards to an ensemble of Deep Neural Networks (DNN) where higher {\em differential immunity} to attacks leads to higher gains in robustness \cite{sengupta2018mtdeep}.

In rare cases, the number of configurations in $C$ may become large. To address scalability challenges, a solution could be to first reduce the configuration set $C$ by partitioning them into disjoint subsets \cite{senguptageneral} or choose a reasonably-sized subset and perform MTD on them separately. In \cite{crouse2012improving}, the authors take the latter route and use \textit{genetic algorithm} search methods where the fitness function measures the diversity of the subset. In \cite{bohara2017moving}, constraining the movement of software between containers both reduces the size of the set $C$ and helps to minimize performance impact.

\subsubsection{\textbf{Detection Surface Shifting}}\label{dec}

The concept of detecting attacks by inspecting traffic on the wire or the behavior of a request on a host machine has been the cornerstone of cybersecurity. Unfortunately, placing all possible Intrusion Detection Systems (IDS) on a system, especially in the case of a large network system, can lead to a degradation in the performance of the system. Thus, to minimize the impact on performance while keeping the attacker guessing about whether their attack will be detected or not, researchers have looked at intelligent ways in which a limited number of detection systems can be placed on the underlying network. These methods are similar to patrolling methods that try to identify security breaches in physical security domains like airports, wildlife sanctuaries, {\em etc.} with limited resources \cite{sinha2015physical}.

In \cite{venkatesan2016moving} and \cite{sengupta2018moving}, authors show that when faced with stealthy botnets and external adversaries, who are powerful enough to attack any internal node of a deployed system, shifting the detection surface helps to maintain system performance, while being effective in detecting attacks.  In both of these cases, they have a set of nodes $N$ that are deployed in the network. The configuration set $C$ is composed of all $k$-sized subsets of $N$ (where $k (< |N|)\}$). In \cite{chowdhary2018markov,senguptageneral}, the authors argue that in many real-world multi-layered networks, the assumption that an attacker can attack from any place in the network is too strong and split the configuration set $C$ into disjoint sets based their position in on an attack graph.

In contrast to MTD for detection surface shifting, whose goal is to maximize security while adhering to performance constraints of the underlying network, some works investigate detection surface shifting for the sole purpose of enhancing security. In \cite{colbaugh2012predictability}, the authors use an ensemble of classifiers that can distinguish a mail as spam and switch between them to make it more difficult for the attacker to fool the system. In general, the use of machine learning for anomaly detection \cite{salman2017machine} coupled with the paradigm of MTD for stochastic ensembles \cite{vorobeychik2014optimal,sengupta2018mtdeep} can lead to interesting future research, especially the introduction of a new attack surface (authors in \cite{bhamare2016feasibility} highlight several challenges in the use of supervised machine learning in cloud network settings). In these cases, each classifier is a configuration in $C$.

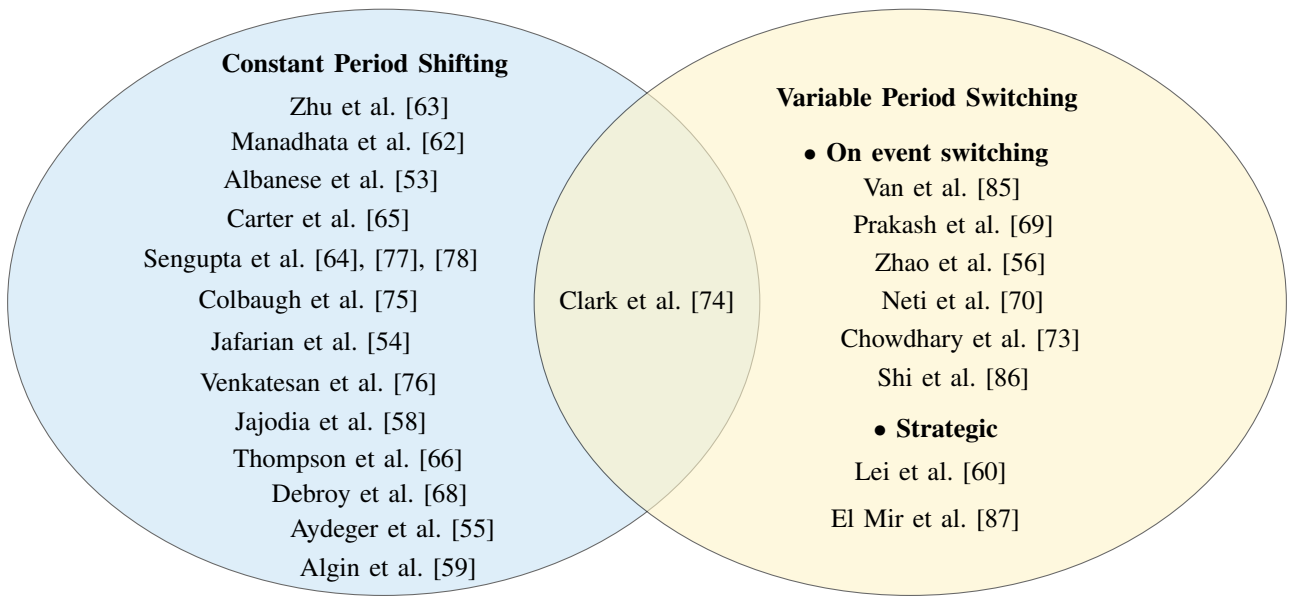
\begin{figure*}
\centering
\begin{tikzpicture}
\begin{scope}[ opacity=0.6]
    \fill[CornflowerBlue!30!white,draw=black] (180:3.5) ellipse (5 and 3.9);
    %\fill[CornflowerBlue!30!white] (180:3) ellipse (6 & 3.6);
    \fill[Goldenrod!30!white,draw=black]  (360:3.5) ellipse (5 and 3.9);
    
 \end{scope}    
     %Refs Constant period switching
    \node at (140:4.9) {\textbf{Constant Period Shifting}};
    \node at (145:4.5) {Zhu et al.~\cite{zhu2013game}};
    \node at (152:4.5) {Manadhata et al.~\cite{manadhata2013game}};
    \node at (159:4.5) {Albanese et al.~\cite{albanese2013moving}};
    \node at (166:4.5) {Carter et al.~\cite{carter2014game}};
    \node at (173:4.5) {Sengupta et al.~\cite{sengupta2017game,sengupta2018moving,senguptageneral}};
    \node at (180:4.5) {Colbaugh et al.~\cite{colbaugh2012predictability}};
    \node at (187:4.5) {Jafarian et al.~\cite{jafarian2012openflow}};
    \node at (194:4.5) {Venkatesan et al.~\cite{venkatesan2016moving}};
    \node at (201:4.5) {Jajodia et al.~\cite{jajodia2018share}};
    \node at (208:4.5) {Thompson et al.~\cite{thompson2014multiple}};
    \node at (215:4.5) {Debroy et al.~\cite{debroy2016frequency}};
    \node at (222:4.55) {Aydeger et al.~\cite{aydeger2016mitigating}};
    \node at (226:4.95) {Algin et al.~\cite{algin2017mitigating}};

    % Hybrid Period Switching
    \node at (0:0) {Clark et al.~\cite{clark2015game}};

    % Ref Variable period switching
    \node at (36:4.6) {\textbf{Variable Period Switching}};
    \node at (28:4.2) {\textbf{$\bullet$ On event switching}};
    \node at (21:4.2) {Van et al.~\cite{van2013flipit}};
    \node at (14:4.2) {Prakash et al.~\cite{prakash2015empirical}};
    \node at (7:4.2) {Zhao et al.~\cite{zhao2017sdn}};
    \node at (0:4.2) {Neti et al.~\cite{neti2012software}};
    \node at (353:4.2) {Chowdhary et al.~\cite{chowdhary2017dynamic}};
    \node at (346:4.2) {Shi et al.~\cite{shi2017chaos}};
    \node at (336:4.2) {\textbf{$\bullet$ Strategic}};
    \node at (329:4.4) {Lei et al.~\cite{lei2017optimal}};
    \node at (322:4.7) {El Mir et al.~\cite{el2016software}};
\end{tikzpicture}
\caption{The time period between two move events is either a fixed interval or decided based on some form of reasoning in the various Moving Target Defenses proposed in the literature.}
\label{fig:when}
% \vspace{-1.6em}
\end{figure*}

\subsubsection{\textbf{Prevention Surface Shifting}}\label{prev}
The goal of an MTD in shifting the Prevention Surface is to make the attacker uncertain about the defense mechanism that is in place; this forces the attacker to spend more resources and come up with sophisticated methods to exfiltrate the data.  It also adds a layer of reasoning on the adversaries part; for example, distinguishing between whether their attack was undetected and went through to the actual system, or was detected and their behavior is presently being monitored becomes difficult.

Investigation of MTD techniques for shifting the prevention surface has been scarce, especially in the context of computer networks. We think this is mostly because an administrator can only use these defenses when they can identify an attack with high accuracy, which is often too strong an assumption. Yet, in some cases, authors make this assumption-- in \cite{chowdhary2017dynamic}, an MTD mechanism that modifies the bandwidth of the network in response to malicious activity is proposed. The configuration space $C$ consists of infinite actions as the bandwidth can take any real value. Similarly, researchers have also considered shifting the latency period when replying to a query that is, with high probability, malicious \cite{clark2015game}. The latter work also considers the deployment of decoy nodes and switches among them to hide from an adversary seeking to actively uncover the prevention surface, i.e. the decoy nodes.

\subsubsection{\textbf{Multi-Surface Shifting}}

By choosing the shift multiple technologies that belong to different surfaces of a system, one can design an MTD that moves more than one surface. Yet, research on hybrid MTDs has been rare. In this regard, we discuss two works that try to shift both the exploration and the attack surface of a system.

In \cite{venkatesan2016moving}, authors show that constructing a simple proxy-based switching approach is ineffective by introducing a new attack known as the proxy harvesting attack that collects a set of IPs over time and then, performs a DDoS attack against all of them. To protect against such attacks, they propose an MTD approach that replaces entire proxies followed by the task of remapping the clients to these new proxies. This renders the exploration effort of the attacker useless and at the same time, reduced the attack efficiency of the attacker. As opposed to looking at a particular network-motivated setting, authors in \cite{zhuang2014towards} and \cite{lei2017optimal} formalize the concept of MTD in which they highlight that the set $C = ES \times AS$, i.e. the elements represents configuration tuples that have technologies from both the attack and the exploration surface.
%This level of abstraction is either too low for generalization or too high for realization.

An interesting future direction could be to leverage existing security research on amalgamation of multiple surfaces and extend it to setting where MTD can be used. For example, NICE combines the techniques involving the detection of attacks to countermeasure (or prevention surface) selection in a single framework \cite{chung2013nice}. This work can be the first step to propose MTD solutions that move both the detection and prevention surface. 

We believe the implementation of systems that integrate multiple surface shifting mechanisms under a single MTD mechanism has several challenges. For example, some of the surfaces might lend themselves easy to management by a centralized controller like SDN, while others are better suited for movement in a decentralized fashion. Characterizing these challenges that deter researchers from considering multi-surface shifting and suggesting methods to overcome that is an interesting research direction.

\subsection{The Timing Function $T$-- When to switch?}

Having defined the possible configurations that a defender can switch between, the next question to ask is {\em at what point in time does a defender choose to execute a switch action that changes their present configuration $c$ to another configuration $c'$}? To answer this question, we divide the works on MTD systems into two broad categories based on how the time-period between multiple switches is determined-— \emph{Constant Period Switching} (CPS) and \emph{Variable Period Switching} (VPS). We first describe the primary characteristics of these categories and then categorize the different MTDs (see Fig \ref{fig:when}).

In CPS, the timing function enforces an MTD system to move, in a randomized fashion, after a constant time-period. This time is dependent on the composition of the set $C$ because attacks on different compositions have different investments in regards to time. On the contrary, in VPS, the timing period between one switch action can vary from the time of another switch. The works in this category can be further subdivided into two distinct categories based on whether the timing function $T$ is reflexive or strategic based on opponent modeling. While the two classes have the exact opposite definition, there is no restriction to not use both in certain scenarios. For example, when multiple surfaces being shifted, it is reasonable to have one surface shift using CPS while the other is shifted via VPS. We conclude his section with a brief discussion in this regard and later, elaborate on it in Section~\ref{research}.

\subsubsection{\textbf{Constant Period Switching (CPS)}}

As mentioned, the key idea of these methods is to move the MTD system after a constant amount of time. In existing works, this is stated as an implicit design choice (for example, in frameworks that model the MTD as a game, the actions of the defender are played after a constant time-period unless they explicitly discretize the time and consider it in the state information), or an explicit one (for example, to ensure security, we move the MTD system after every 100 seconds).

A lot of the research works in MTD literature such as \cite{chowdhary2018mtd,sengupta2017game,sengupta2018moving,manadhata2013game,jajodia2018share}, do not explicitly bring up the topic of a time-period but inherently assume that the system moves after a constant time.
In these works, authors (1) model the problem as a single-step game and generalize the solution of this game to multiple stages (not necessarily repeated games), and (2) showcase the problem of \emph{when to switch} as a different one than \emph{how to switch} and only address the latter, forcing the system admin (or defender) to consider a constant time-period when implementing MTD solutions.

Some research works explicitly state that the time-period is uniform and the equilibrium strategy is played at the start of each stage \cite{zhu2013game, carter2014game,algin2017mitigating}. Note that, some of these works, such as \cite{algin2017mitigating,albanese2013moving}, experiment different timing functions for switching, i.e. see the effectiveness of an MTD if the constant time-period was set to different predefined values. This helps them empirically figure out the constant timing function that works best.
Some other works that employ a constant timing function, use different terminology. For example, authors in \cite{venkatesan2016moving} address DDoS attacks use the notion of a pre-defined fixed frequency of switching while research in \cite{jafarian2012openflow} refers to $T$ as a constant mutation interval. Other works that have syntactic generalizability (by using notations such as $t_i$ that allow a user to use different time intervals for configuration $i$) default to a constant timing function.

An interesting question is {\em how long should this constant time-period be in practice}? In \cite{thompson2014multiple}, the authors vary the time-period from $60-300$ seconds and evaluate the effectiveness of OS rotation in a network-system based on (1) the likelihood of thwarting a successful exploit, (2) the magnitude of impact for an exploit, and (3)  the availability of applications. They show that a smaller $t =60s$ was often good enough to thwart Network Mapping (Nmap)~\cite{lyon2009nmap} attacks, but may allow accurate OS fingerprinting when $t=60s$. The latter does not help the attacker because, during the execution of the attack, the fingerprinted OS might simply have shifted. For $t=300s$, these results look less promising against automated attack systems , but work well when evaluated against human experts. To allow for faster rotations, the authors set up machines with different OS and after every time interval, simply redirect traffic to a new OS. While this makes faster switching less painful, it incurs added cost to maintain multiple machines. To reduce this, we believe that having at least two systems is necessary to allow for such small switching periods, especially when shifting between OSs. With two systems, one VMs sets up an environment while the other handles traffic and the role switches in the next time step.

In \cite{debroy2016frequency}, researchers use the knowledge obtained from historical attack data to obtain a cyber attack inter-arrival (CAIA) rate. Then, they setup an optimization problem to maximize the time-period of switching, constrained upon the fact that the constant timing function has a value less than CAIA. Along similar lines, authors in \cite{aydeger2016mitigating} have looked at \emph{traceroute}~\cite{padmanabhan2003secure} data between possible source-destination pairs to decide on a reasonable time-period for obfuscating links or mutating routes of ICMP packets on the network. This helps to mitigate crossfire attacks.

An interesting case arises in \cite{al2012random} because the states of this MTD represent a bipartite mapping between hosts and virtual IPs (vIP) and the authors let each edge in the mapping have a separate, but constant, mutation time. Some hosts belong to a set of High-Frequency Mutation (HFM) set while others belong to a Low-Frequency Mutation (LFM) set. To ensure that availability of a host is not impacted, the hosts that are in the HFM set map to two vIPs (one that it was mapped to in the previous round and one to which it is mapped in the present round) for a small time duration compared to the time-period of switching. In \cite{albanese2013moving}, the authors also have a local timer maintained by each host; its expiry triggers the change in their vIP. They use {\em ad hoc} messaging to communicate this change to other nodes in the network to facilitate routing.

\subsubsection{\textbf{Variable Period Switching (VPS)}}

The idea behind this switching strategy, as evident from its name, is to have a way to come up with a timing function that varies the switching time based on the present condition of the system. For example, if the system transitions through a series of configurations $\langle \dots, c, c'\dots \rangle$ and the time spent in $c$ and $c'$ is denoted by $t$ and $t'$ respectively, then $t \neq t'$. We believe that doing a VPS along with an MTD mechanism is similar to doing a two-layer MTD where the first layer deals with a meta MTD for shifting the time-surface and then, the second layer MTD is responsible for executing the actual cyber-surface switching. We will now categorize the MTD works under the following sub-classes.

\begin{figure*}
\centering
\begin{tikzpicture}
\begin{scope}[opacity=0.6]
    \fill[Salmon!30!white,draw=black] (175:3.8) ellipse (5 and 3);
    \fill[SpringGreen!30!white,draw=black]  (5:3.8) ellipse (5 and 3);
    
 \end{scope}    
     %Refs Single Stage Games
   \node at (141:4.4)    {\textbf{Single-Stage Modeling}};
   \node at (146:4.2) {Al Mir et al.~\cite{al2012random}};
   \node at (152:4.2) {Jafarian et al.~\cite{jafarian2012openflow}};
   \node at (158:4.2) {Chowdhary et al.~\cite{chowdhary2016sdn,chowdhary2018mtd}};
   \node at (164:4.2) {Zhu et al.~\cite{zhu2013game}};
   \node at (170:4.2) {Manadhata et al.~\cite{manadhata2013game}};
   \node at (176:4.2) {Thompson et al.~\cite{thompson2014multiple}};
   \node at (182:4.2) {Aydeger et al.~\cite{aydeger2016mitigating}}; 
   \node at (188:4.2) {Carter et al.~\cite{carter2014game}};
   \node at (194:4.4) {Sengupta et al.~\cite{vadlamudi2016moving,sengupta2018moving}};
   \node at (200:4.5) {Colbaugh et al.~\cite{colbaugh2012predictability}};
   \node at (205:4.6) {Venkatesan et al.~\cite{venkatesan2016moving}};
   \node at (212:4.4) {Algin et al.~\cite{algin2017mitigating}};

   %Ref Multi-stage game
   
    \node at (36:4.6)    {\textbf{Multi-Stage Modeling}};
    \node at (30:4.3) {Chowdhary et al.~\cite{chowdhary2017dynamic, chowdhary2018adaptive,chowdhary2018markov}};
    \node at (23:4.2) {Manadhata et al.~\cite{manadhata2013game}};
    \node at (16:4.2) {Colbaugh et al.~\cite{colbaugh2012predictability}};
    \node at (9:4.2) {Zhao et al.~\cite{zhao2017sdn}};
    \node at (2:4.2) {Maleki et al.~\cite{maleki2016markov}};
    \node at (355:4.2) {Lei et al.~\cite{lei2017optimal}};
    \node at (348:4.2) {Miehling et al.~\cite{miehling2015optimal}};
    \node at (341:4.2) {Nguyen et al.~\cite{nguyen2018multistage}};
    \node at (334:4.2) {Sengupta et al.~\cite{sengupta2017game,senguptageneral}};
    \node at (327:4.3) {Clark et al.~\cite{clark2015game}};
       
  \end{tikzpicture}
\caption{Game-theoretic modeling of Moving Target Defenses (MTDs), that is necessary for obtaining the movement strategy $M$, can be categorized into either single-stage or multi-stage game modeling. This choice reflects the type of threat model being considered and the characteristic of the particular system for which the MTD is implemented.}
\label{fig:how}
% \vspace{-1.6em}
\end{figure*}
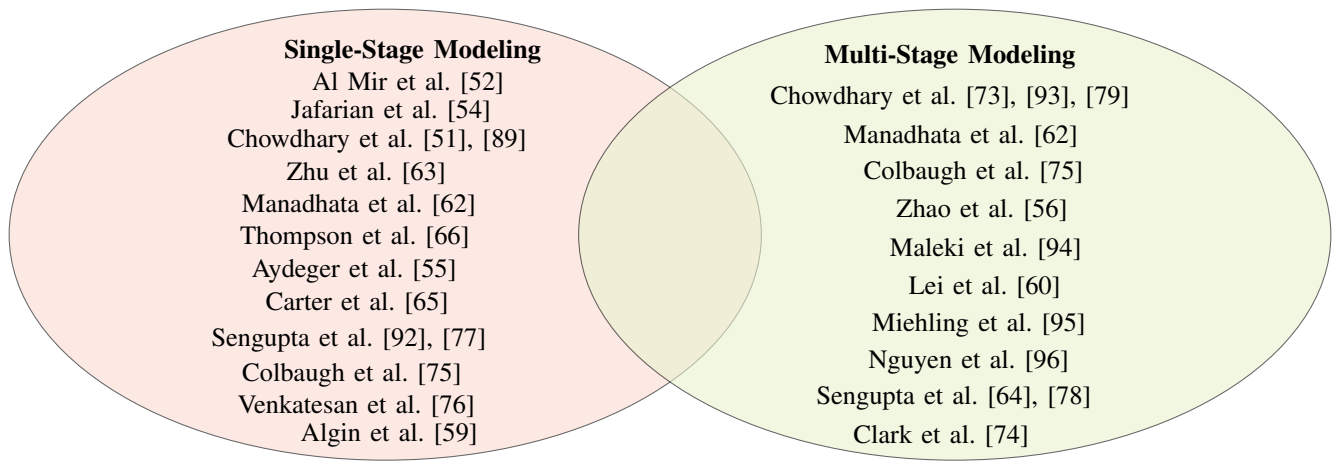

\paragraph{On-event Switching}
Most works that have a general timing function, which isn't constant, fall under this sub-category. The main idea, evident from the nomenclature, is that when a particular event occurs, such as detection of an attack, unavailability of a link or a server, the timing function specifies a time-period to switch. In most cases, the switch action is triggered immediately. 

A well-known case study is that of the \emph{FlipIt} games \cite{van2013flipit}. The true state of the system is represented using a boolean variable $1/0$ that indicates whether the defender or the attacker has control of an underlying software system. The value is unknown to both the players, and a defender strategy is based on its belief that the true value is $0$ (which represents the attacker having access to the server). Thus, the timing function is dependent on this belief.
On similar lines, in \cite{zhao2017sdn}, authors update the belief about the sender type (are they malicious?) upon detection of suspicious packets on the network. This belief, in turn, influences the timing function $T$. Authors in \cite{neti2012software} have a belief variable indicating whether a vulnerability was exploited or not. The timing function orders a switch action if the belief value exceeds a certain threshold.

Other works argue that obtaining belief parameters (such as \textit{with what probability is the sender an attacker vs. a normal user?} or \textit{with what probability is the system compromised?}) that are accurate is difficult in cybersecurity settings because they are based on analysis of historical datasets that are both scarce and might represent a different distribution. These works seek to use the intuitive knowledge of security experts in designing the timing function. In \cite{chowdhary2017dynamic}, if the admin detects a spike in bandwidth usage of a particular link or sub-net, they change the maximum bandwidth value allocated to that link or the network. Authors in \cite{shi2017chaos} scan open connections routinely and upon detection of unexpected connections, move between MTD configurations to protect the system against port-scanning attacks. 

\paragraph{Strategic}
To understand these methods, we design a timing function that produced integer values in the range $[0, t_{\max}]$. These represent discrete time intervals and a value $t$ represents the time at which a move occurs.\\
\indent In \cite{lei2017optimal}, the authors model the MTD setting as a game-theoretic setting. While we discuss this in much more detail in the upcoming section on the design of the movement function $M$, we briefly highlight it here. The basic idea involves representing a game-state that considers the currently deployed configuration $c$ and an integral-valued time variable $t$. In each state that can be represented by the tuple $(c,t)$, it recommends between the actions {\em switch} or {\em stay}.
The stay action takes them to the state $(c, t+1)$, while the switch actions move them to the state $(c',0)$. Note that the stochastic switching or movement policy described later may, with some non-zero probability predict, $c' = c$.

In general, the idea of incorporating the time variable as a part of the state can result in an explosion in the number of states and thus, computing a policy for all states time intensive.
Authors in \cite{el2016software} show that they can model the problem similarly but resort to a simple strategy for defining the timing function. While it may be sub-optimal, scalability is no longer a concern. The timing function considers the impact of (known) vulnerabilities in the currently deployed configuration and based on it, recommends a switching time $t \in \{0,1,\dots,\infty\}$. If there are no vulnerabilities whose impact is greater than a predefined threshold, then the system remains static until a new vulnerability is discovered.

\paragraph{Hybrid Switching}
Although the idea of having both a CPS and VPS in an MTD mechanism seems counter-intuitive at first, authors in \cite{clark2015game} look at MTD mechanisms where one layer shifts the time window of replying to a strategic adversary using an event-based timing function while the other layer that moves the decoy nodes to hide from an active adversary is done using a constant timing function. This acts as a stepping stone for an investigation on the effectiveness of CPS {\em vs.} VPS for a particular surface and finally when a combination of both should be considered.

\subsection{The Movement Function $M$ -- How to switch?}

In this section, we look at how different MTDs come up with a policy for movement. While some surveys (eg. \cite{roy2010survey}) have been solely devoted to analyzing how this question, we propose a unifying framework that maps any MTD system to a Markov Game. This gives us a better sense of the implicit assumptions made by the various MTDs when coming up with $M$. Before getting into the categorization shown in Figure~\ref{fig:how}, we first briefly introduce Markov games \cite{shapley1953stochastic}.

An MTD system can be mapped to a Markov game defined by the tuple $\langle S, A^\mathcal{D}, A^\mathcal{A}, T, R^\mathcal{D}, R^\mathcal{A} \rangle$ where $S$ denotes the state of the MTD system (often represented by deployed configuration $c$), $A^\mathcal{D}$ denotes the set of pure strategies that map to the set of configuration set $C$ and the mixed strategy over this set maps to the movement strategy $M$. The other variables are specific to the domain for which the MTD system is designed and help in coming up with good $M$; $A^\mathcal{A}$ represents the set of attack actions, $T$ represents the transition from one configuration to another depending on the joint action taken by the defender and the attacker, and $R^i$ represents the rewards for the player $i$ (i.e. the defender $\mathcal{D}$ or the attacker $\mathcal{A}$). We will use the superscripts $\mathcal{D}$ and $\mathcal{A}$ to represent the functions for the defender and the attacker respectively.

\subsubsection{\textbf{Single-Stage Modeling}}

In this section, we investigate works that have modeled the interaction between an attacker and a defender as a single-stage game. In these settings, the goal is to come up with a single mixed-strategy that can be played in all configurations (or states). Thus, the movement function $M: H\rightarrow C$ is independent of any history or $H = \emptyset$. As we shall see, the different works consider various notions of equilibrium to come up with this mixed-strategy.

In \cite{manadhata2013game}, the configuration of the system is modified by a pair of actions, one by the defender and one by the attacker. The defender's action, in this case, maps a value of system features represented by the set $F$ to one of the possible actions in the set $\{enabled,~disabled,~modified,~unchanged\}$. The attacker's action set $A^\mathcal{A}$ is comprised of conceptually opposite actions that can {\em dis-enable} a port, {\em disable} a functionality, {\em etc.} and the rewards, $R^\mathcal{A}$ and $R^\mathcal{D}$ for the attacker and the defender, are given by weighing the change in system's features $\Delta F$, i.e. the change in the attack surface $\Delta AS$ and the attack surface measurement $\Delta ASM$, as follows.
\[
R^\mathcal{D}(s,a^\mathcal{D},a^\mathcal{A}) = B_1(\Delta F) + B_2(\Delta AS) - D_1(\Delta ASM)
\]
\[
R^\mathcal{A}(s,a^\mathcal{D},a^\mathcal{A}) = B_3(\Delta ASM) - D_2(\Delta AS)
\]
where the $B_i$ and $D_j$ are weights defined by security experts of the system. Note that the \emph{cost of a defense} action ($D_1$) and the cost of attack action ($D_2$) are incorporated in the reward. Thus, solving for an equilibrium of this game results in strategies that account for the cost-reward trade-off. The authors use the notion of a Subgame Perfect Equilibrium (SPE).\footnote{In sequential games, a strategy profile is an SPE if it is a Nash equilibrium in every subgame of the original game \cite{osborne2004introduction}.}

In \cite{al2012random} and \cite{jafarian2012openflow}, the defender played a predefined strategy that is either uniform random (choose $k$ random vIPs at random) or based on intuitive heuristics (select top-k scanned vIPs). The latter is based on the assumption that the most scanned vIPs will not be scanned as frequently in the new state. The chosen vIPs are then assigned to the $k$ hosts using a purely random matching function. 

In \cite{chowdhary2016sdn,chowdhary2018mtd}, the authors explain the game modeling based on a real-world example and compare the effectiveness of a reactive switching strategy to the Uniform Random Strategy (URS). URS, which is found to be more effective, picks all pure strategies $a \in A^\mathcal{D}$ with equal probability |$1/A^\mathcal{D}|$. An array of works that simply use URS as the defender's policy are \cite{thompson2014multiple,aydeger2016mitigating,algin2017mitigating}. This makes an implicit assumption that the game is a constant sum and the normal form game matrix has symmetric rewards. When no information is available about the attacks and the defender has no preference over the MTD configurations, this may be a reasonable assumption. Fortunately, defenders often have an idea about an attacker's threat model and thus, such assumptions (resulting in URS) result in highly sub-optimal movement strategies.
Works such as \cite{thompson2014multiple} closely mirror URS strategies but make a minute modification-- given the present deployed OS configuration $c$, they consider a URS over remaining OS configurations, i.e. $C \setminus \{c\}$ as the movement function $M$. In \cite{aydeger2016mitigating}, given all networking paths from source to destination, one is picked at random while in \cite{algin2017mitigating}, the authors use to Fisher-Yates shuffling to select a transmission schedule at the start of every round.

In \cite{carter2014game,sengupta2018moving,vadlamudi2016moving,venkatesan2016moving,zhu2013game}, and \cite{wang2019moving} the authors design a single-stage normal form game. Thus, if the defender chooses to deploy a particular configuration $c \in A^\mathcal{D}$ and the attacker chooses a particular exploit $e \in A^\mathcal{A}$, then the rewards for the players can be read from the normal-form game matrix. At equilibrium, a mixed strategy over the defender's actions turns out to be the optimal movement policy. Thus, $M$ is a stochastic function based on the mixed strategy, chooses the next configuration at the end of each switching period. In \cite{wang2019moving}, the authors consider an MTD for the Internet of Things (IoT) and show that Zero-Determinant strategy results in a dominant strategy for the single-stage game.
In \cite{zhu2013game}, the authors assume (1) a zero-sum reward structure, and (2) a threat model in which the attacker is irrational. The opponent model for the attacker is determined based on historical traces of the attacker's response to defense strategies and eventually updates the utilities of the game to reflect the rationality of the attacker. After each update, the resulting Nash equilibrium (NE) is played. In \cite{carter2014game}, the authors assume that if the present system is working with the operating system $c$ and the system is made to switch to the operating system $c'$, lower the similarity between $c$ and $c'$, more secure the move action. To model this, the defender's rewards are defined w.r.t. switch actions and, over time, fine-tuned by simulating the behavior of an adversary. Similar to \cite{zhu2013game}, the NE strategy is chosen to be the defender's policy.

The other works assume that an attacker will eventually, with reconnaissance on its side, figure out the mixed strategy of the defender (using maximum likelihood estimates). Thus, authors of \cite{vadlamudi2016moving,sengupta2018moving} concentrate on finding the Strong Stackelberg equilibrium (SSE) strategy for the defender. The rewards for their game follow a general-sum structure and are obtained using the CVSS metrics of known attacks that can successfully exploit one of the defender's actions, i.e. the MTD configurations that can be deployed. On similar lines, authors in \cite{venkatesan2016moving} use the notion of SE to obtain defender's strategy to thwart DDoS attacks.

\subsubsection{\textbf{Multi-Stage Modeling}}

Works in these sections reason about (1) the history, i.e. paths that the system has taken, which may be a result of the actions taken by the players, to reach the present state, (2) the future, i.e. how the decision in the present state affects the rewards in the future or both.

In \cite{chowdhary2017dynamic}, similar to the work by \cite{manadhata2013game} in single-stage modeling, the authors discretize the continuous action space of the defender, which represents bandwidth values, to two levels-- high and low. Then they find a meaningful defender strategy over them by ensuring that in a repeated game setting, the advantage that an attacker gained by packet flooding attacks (at some stage in the past) is neutralized by reducing their bandwidth to the low state for $1 \leq x < \infty$ number of game stages.
On similar lines, authors in \cite{colbaugh2012predictability} consider a repeated game setting and analyze the defender's policy against self-learning attackers. They infer that in their case, the optimal policy converges to the URS. Other works such as \cite{zhao2017sdn} also update their belief about an attacker based on observations drawn from multiple plays of the game. This belief is then used to come up with an optimal strategy for the defender.

A set of works in MTD leverage the attack graph of an underlying system to better reason about the sequential process of an attack. In \cite{miehling2015optimal}, the authors model the problem of network intrusion via a Bayesian Attack Graph. The action set for the attacker $A^\mathcal{A}$ includes the (probabilistic) paths an attacker can take in this graph. Then, the authors map these paths with the defender's imperfect sensing capabilities to form a Partially Observable Markov Decision Process (POMDP)~\cite{russell2016artificial}. Thus, the state and transition of this POMDP are designed to model the attacker's behavior and the optimal policy becomes the defender's strategy in a particular belief state. A shortcoming of this work is that this strategy may be sub-optimal if the \emph{attacker} deviates away (intentionally or not) from the assumptions that informs the POMDP modeling. On the other hand, modeling the scenario as a Partially Observable Stochastic Game (POSG)~\cite{hansen2004dynamic} results in scalability concerns for any real-world system.

Authors in \cite{maleki2016markov,chowdhary2018markov,valizadeh2019toward}, and \cite{lei2017optimal} relax the assumption about partial observability and formalize MTD in a Markov Game framework. In \cite{maleki2016markov}, authors consider policies over the defender's action set that comprises of either single or multiple IP hops. Each action results in the defender uniformly selecting the next state given that an attacker samples actions randomly from $A^\mathcal{A}$. They can show that multi-element IP hopping actions increase the defender's reward by a greater magnitude compared to static defense strategies. As we will discuss later, the authors do not model the cost of performing a hop action or the downtime associated with switches in the defender's reward function. Thus, the optimal defender policy might be sub-optimal for performance in real-world multi-layered network attack scenarios.

In contrast, the works \cite{chowdhary2018markov,senguptageneral,lei2017optimal} incorporate this trade-off in the rewards ($R^\mathcal{D}$ and $R^\mathcal{A}$) of the Markov Game, and can thus generate policies for the defender that trade-off security corresponding to an attacker's actions and performance at each step of the game. In \cite{lei2017optimal}, the inclusion of time variables and attack strategies in the state allows it to formulate the problem as an MDP but, becomes vulnerable when an attacker is aware of this modeling. It also makes it less scalable for large networks. In all these works, the impact on performance $C^\mathcal{D}$ is a part of $R^\mathcal{D}$ and often, just a \emph{heuristic idea} as opposed to simulation in a real system that informs these values. The more accurate these measures become, the better is the strategy.
Authors in \cite{sengupta2017game} model the performance and security concerns in a different way than the above methods. The performance cost represents the switching cost while the rewards of the state represent the security costs of deploying a particular configuration. Then, they formulate an optimization problem that produces a defender strategy that reduces the cost of switching over two steps (one switch) and maximizes the security over a single step. Although the authors in \cite{sengupta2017game} do not discuss how sub-optimal their switching strategies are, recent work by \cite{li2020spatial} extends their approach to a Markov Game setting with the Stackelberg Equilibrium concept.

Lastly, authors in \cite{clark2015game} look at an MTD defense mechanism for deception (obfuscation of the links to send attacker to decoy nodes) and model attackers who are actively trying to uncover this deception over a repeated stage interaction. For one problem, they use the Nash equilibrium (NE) while for the other (identifying decoy nodes) they consider the Stackelberg equilibrium (SE) as the defender's strategy. We believe that it is necessary to highlight a shortcoming of the different modeling choices, especially in light of multi-stage attacks and APT scenarios, however optimizing the model and ensuring the scalability is a difficult task, thus we highlight these as possible research opportunities in the Section~\ref{research}.
\section{Implementation of Moving Target Defenses}
\label{sec:mtd_impl}

In this section, we discuss how the various Moving Target Defenses (MTDs) have been implemented using research test-beds and industrial products. First, we briefly discuss the tools for MTD implementation. In this regard, we highlight that traditional networking technologies like middleboxes have a set of disadvantages. To overcome this, users can leverage the advancement in networking technologies such as Software Defined Networking (SDN) and Network Function Virtualization (NFV). We briefly emphasize their role in making MTD a pragmatic solution for real-world systems. Second, we highlight how the existing MTDs, discussed in the survey as shown in table \ref{tab:adel}, have been evaluated. We select a subset of works and conduct a detailed case-study to highlight how movement strategies can be implemented in practice. Third, the sub-section~\ref{testbed} provides details of the test-beds used for academic research and industry products.

\subsection{Middleboxes for enabling Moving Target Defenses}

Middleboxes are the devices used by network operators to perform network functions along the packets data-path from source to destination, e.g., Web Proxy, Firewall, IDS. This provides a decentralized framework that can be used alongside existing network technology to implement strategies for MTDs. Furthermore, researchers have focused significant effort on several issues associated with middleboxes, such as ease of use, ease of management and deployment, the design of general-purpose middleboxes for different network functions {\em etc.} This makes them seem like a good choice for enabling the practical implementation of MTDs.

\begin{table}[t!]
\centering
\begin{tabular}{|l|c|c|c|}
\hline
\textbf{Middlebox} & \textbf{Misconfiguration} & \textbf{Overload} & \textbf{Physical/Electric}  \\
\hline
Firewall & 67.3\% & 16.3\% & 16.3\% \\
\hline
Proxies & 63.2\% & 15.7\% & 21.1\% \\
\hline
IDS & 54.5\% & 11.4\% & 34\% \\
\hline
\end{tabular}
\caption{Common causes of middlebox failures. Misconfiguration is a dominant cause of failure because of different underlay and overlay network.}

 \label{tab:1}
  \vspace{-1.2em}
\end{table}

%When the traditional security tools detect some security issues (vulnerabilities, weak security configurations) with the deployed middleboxes in a data center network, the natural question comes to mind, \emph{why are those issues not fixed in a production environment right away?}
Unfortunately, a survey of various middlebox deployments conducted by Sherry \textit{et al.}~\cite{sherry2012survey} reveals some drawbacks such as increased operating costs caused by misconfiguration and overloads that affect their normal functioning. As shown in the Table~\ref{tab:1}, based on the results in the survey \cite{sherry2012survey} of 57 enterprise network administrators from \emph{NANOG} network operators group, the misconfigured and overloaded middleboxes are the major reasons for middlebox failure. Furthermore, about 9\% of administrators reported about six and ten hours per week dealing with middlebox failures. Also, the adoption of new and secured middlebox technologies is slow in the industry based on the survey results. In the median case, enterprises update their middleboxes every four years. The use of traditional networks for incorporation of MTD defense can increase chances of network misconfiguraion and outages.

Moreover, this static nature of the middleboxes themselves, in contrast to the dynamic nature of the system they enable, provides an assymetric advantage to the attackers as noted by Zhuang \textit{et al.}~\cite{zhuang2014towards}. The attackers can perform necessary network reconnaissance, identify the services and configuration of the applications, and operating systems by leveraging the middleboxes. This information helps the attacker in choosing best-fit attack methods against the static configuration and select the best time to attack the system. The attackers can use the compromised resource to maintain the foothold in the network for a long period of time and try to exploit other high-value targets in the same network.

\begin{table*}[!htp]
	\centering
		\begin{tabular}{|p{15mm}|p{15mm}|p{35mm}|p{55mm}|p{25mm}|}
			\hline
			\textbf{Research Work} &\textbf{SDN/NFV}  & \textbf{Implementation Layer} & \textbf{Technologies/ Testbed} & \textbf{Maturity Level} \\
			\hline
			\hline
            ~\cite{chowdhary2018markov}, 2019 & \checkmark & network and application & Sample Use-Case & simulation \\
            \hline
			~\cite{steinberger2018ddos}, 2018 & \checkmark & network and application & ONOS~\cite{berde2014onos} & emulation \\
            \hline
            ~\cite{schlenker2018deceiving}, 2018 &  & network & Personalized use cases & simulation \\
            \hline
            ~\cite{jajodia2018share}, 2018 &  & abstract & Personalized use cases & simulation \\
            \hline
            ~\cite{chowdhary2018adaptive}, 2018 & \checkmark & network and application & Personalized use-cases & simulation \\
            \hline
            ~\cite{chowdhary2018mtd}, 2018 & \checkmark & network & Personalized test-bed with VMs & emulation \\
            \hline
            ~\cite{polyverse}, 2018 &  & application & Polyverse & commercial \\
            \hline
            ~\cite{shu2018ensuring}, 2018 & & network and application & Personalized Test-bed with VMs & simulation \\
            \hline
            ~\cite{nguyen2018multistage}, 2018 &  & network & personalized use-cases & simulation \\
            \hline
            ~\cite{trapx-deceptiongrid}, 2018 &  & network and application & Deception Grid~\cite{trapx-deceptiongrid}, Crypto Trap~\cite{cryptotrap} & commercial \\
            \hline
		    ~\cite{sengupta2018moving}, 2018 &  & network and application & Mininet & both  \\
            \hline
		    ~\cite{hong2017optimal}, 2017 & \checkmark & network  & Opendaylight Helium~\cite{medved2014opendaylight} & simulation \\
		    \hline
            ~\cite{wang2017u}, 2017 & \checkmark & network  & Ryu SDN~\cite{ryusdn} & simulation 
            \\
            \hline
            ~\cite{homescu2017large}, 2017 & & application & LLVM compiler~\cite{lattner2004llvm} & emulation \\
            \hline
            ~\cite{sengupta2017game}, 2017 &  & application & personalized use-cases & simulation \\
            \hline
            ~\cite{han2017evaluation}, 2017 & & application & CMS application~\cite{baxter2002content} & simulation \\
            \hline 
            ~\cite{connell2017performance}, 2017 &  & application & Simpy \cite{simpytestbed} & simulation \\
            \hline
            ~\cite{chowdhary2017dynamic}, 2017 & \checkmark & network & network simulator + ODL controller & emulation \\
            \hline
            ~\cite{lei2017optimal}, 2017 & & abstract & Personalized Test-bed with VMs & simulation \\
            \hline
            ~\cite{zhao2017sdn}, 2017 & \checkmark & application & Mininet with POX controller & emulation \\
            \hline
            ~\cite{shi2017chaos}, 2017 & \checkmark & network & CloudLab~\cite{ricci2014introducing} & emulation \\
            \hline
            
            ~\cite{pappa2016moving}, 2016 & & network &
            Smart grid, energy management system (EMS) & emulation \\
            
            \hline
            ~\cite{sengupta2017game}, 2017 &  & application & personalized test-bed & simulation \\
            \hline
            ~\cite{morphisec}, 2017 &  & network and application & Morphisec & commercial \\
            \hline
            ~\cite{algin2017mitigating}, 2017 &  & network & Personalized Simulation Environment & simulation \\
            \hline
            ~\cite{aydeger2016mitigating}, 2016 & \checkmark & network & Mininet with Floodlight controller & emulation \\
            \hline
            ~\cite{maleki2016markov}, 2016 &  & network & Personalized use-cases & simulation \\
            \hline
            ~\cite{chowdhary2016sdn}, 2016 & \checkmark & abstract & OpenStack \cite{sefraoui2012openstack} with ODL ~\cite{medved2014opendaylight} controller & emulation \\
            \hline
             ~\cite{ahmed2016mayflies}, 2016 & & network & OpenStack~\cite{sefraoui2012openstack} & emulation \\
            \hline
            ~\cite{debroy2016frequency}, 2016 & \checkmark & network & GENI~\cite{berman2014geni} & emulation \\
            \hline
            ~\cite{venkatesan2016moving}, 2016 &  & network & Rocketfuel Dataset & simulation \\
            \hline
            ~\cite{taylor2016automated}, 2016 & & network & ARCSYNE \cite{yackoski2013applying} \& SDNA ~\cite{yackoski2013mission,yackoski2011self} & emulation \\
            \hline
            ~\cite{el2016software}, 2016 & \checkmark & network and application & Mininet with POX controller & simulation \\
            \hline
            ~\cite{chadha2016cybervan}, 2016 & \checkmark & network & CyberVAN & commercial \\
            \hline
            ~\cite{debroy2016frequency}, 2016 & \checkmark & application& GENI~\cite{berman2014geni} & emulation \\
            \hline
            ~\cite{prakash2015empirical}, 2015 & & abstract & Personalized simulation environment & simulation \\
            \hline
            ~\cite{zaffarano2015quantitative}, 2015 & & network and application & Cyber Quantification Framework  & emulation \\
            \hline
            ~\cite{clark2015game}, 2015 &  & network & matlab & simulation \\
            \hline
            ~\cite{miehling2015optimal}, 2015 &  & network & personalized use-cases & simulation \\
            \hline
            ~\cite{luo2015rpah}, 2015 &  & network and application & RPAH framework & simulation \\
            \hline
            ~\cite{thompson2014multiple}, 2014 &  & network and application & CORE Impact Pro & commercial \\
            \hline
            ~\cite{carter2014game}, 2014 &  & application & personalized test-bed & simulation \\
            \hline
            ~\cite{kampanakis2014sdn}, 2014 & \checkmark & network & Personalized Cloud System & emulation \\
            \hline
            ~\cite{jia2013motag}, 2013 & & network and application & PlanetLab~\cite{chun2003planetlab} & emulation \\
            \hline
            ~\cite{chung2013nice}, 2013 &  & network and application & Personalized virtual cloud system & emulation \\
            \hline
            ~\cite{van2013flipit}, 2013 &  & abstract & Personalized test-bed & simulation \\
            \hline
            ~\cite{zhu2013game}, 2013 &  & application & game-based theoretical framework & analytic \\
            \hline
            ~\cite{jafarian2012openflow}, 2012 & \checkmark & network & NOX in Mininet & emulation \\
            \hline
            ~\cite{neti2012software} , 2012 &  & abstract & Proposes to use a modified CTF environment & analytic \\
            \hline
            ~\cite{crouse2012improving} , 2012 &  & network and application & Personalized test-bed & emulation \\
            \hline
            ~\cite{colbaugh2012predictability}, 2012 &  & network & KDD dataset~\cite{dhanabal2015study} & simulation \\
            \hline
            ~\cite{al2012random}, 2012 & \checkmark & network &Personalized use cases & simulation \\
            \hline
           ~\cite{dunlop2011mt6d}, 2011 &   & network and application & SLAAC~\cite{thomson2007ipv6} & simulation\\
            \hline
            ~\cite{al2011toward}, 2011 &  & network &  MUTE & simulation \\
            \hline

		\end{tabular}
		\caption{A taxonomy of MTD Implementation categorized defenses based on the use of SDN/NFV, the layer of network protocol stack they protect, key technologies they use, and the level of maturity at which they have been implemented.}
		\label{tab:adel}
		\vspace{-1.6em}
	\end{table*}

\subsection{SDN and NFV for enabling Moving Target Defenses}

\emph{Software Defined Networking} (SDN)~\cite{kreutz2015software} is defined as a networking paradigm that decouples the control plane from the data plane, allowing a global view of the network, and centralized control capabilities. SDN deals with packet headers, from layers 2-4 of OSI model and other protocols such as MPLS~\cite{cox2017advancing}. SDN and NFV have emerged as a state-of-the-art network architecture for data centers and backbone networks. Google's \textit{B4 project}~\cite{jain2013b4} shows the feasibility of SDN for handling real-world network challenges such as traffic engineering and Quality of Service (QoS).

The decoupling allows a network architecture where switches act as forwarding devices, maintaining flow tables populated with flow rules. The new architecture allows considerably more flexible and effective network management solutions, and a unified programmable interface that can be utilized by software developers for application deployment~\cite{ali2015survey}. 
   The SDN architecture can be vertically split into three layers described below: 
  \begin{itemize}
    \item \textbf{Application Plane:} It consists of end-user business applications that consume SDN communication and network services. The examples include network security or virtualization applications.  
    \item \textbf{Control Plane:} The control plane consists of SDN controllers such as ONOS~\cite{berde2014onos}, OpenDaylight~\cite{medved2014opendaylight} providing open Application program interfaces (APIs) to monitor network forwarding behavior. The communication interface between the application and control plane is known as northbound interface. The interface between control and data plane is known as southbound interface.  
    \item \textbf{Data Plane:} The forwarding elements such as OpenFlow switches are present in the data plane. The forwarding devices can be physical or virtual switches. Control plane installs flow rules in order to govern the forwarding behavior of data plane devices. 
    \end{itemize}

  Network Functions Virtualization (NFV)~\cite{han2015network} has emerged as a technology that provides a virtualized implementation of hardware-based equipment such as firewall, routers, and Intrusion Detection Systems (IDS). Virtual Network Functions (VNFs) can be realized through virtual machines (VMs) or containers running on top of the physical server of cloud computing infrastructure. SDN acts as an enabling technology for NFV and help in centralized management, making it easier to implement MTD. Despite the great benefits offered by SDN and NFV, its use to implement MTDs has been limited to a few research works (as can be seen in Table \ref{tab:adel}).

\subsection{SDN-based MTD Applications and Case Study}

We categorize the MTD industrial, and research implementations from the perspective of networking technologies used, e.g., traditional network or SDN/NFV as shown in the Table~\ref{tab:adel}, as highlighted in column 2. It is noteworthy that SDN/NFV has been a dominant technology for MTD research. Around 34\% of the research works in Table~\ref{tab:adel}, utilize SDN/NFV for implementation of MTD or cyber-deception. Column 3 describes the layer in network protocol stack where the describe MTD solution has been implemented. As on may notice, most research works are implemented at the network or the application layer. In the fourth column of the table, we identify the key technologies used by these research works; if they utilize a particular implementation test-bed (for example, GENI~\cite{berman2014geni}, then the name of the test-bed is mentioned in this column. Column 5 of the table shows the level of \emph{maturity} of the MTD research work. We categorized the research works into four levels - \emph{analytic} - if only numerical results or thought-based experiments have been presented in the said paper. We put the research work under the category \emph{simulation} - if the research work describes some simulated network, e.g., \emph{Mininet}. The \emph{emulation} category consists research works that use close to real world networks, e.g. multiple VMs deployed in a cloud testbed, e.g., GENI~\cite{berman2014geni}. Lastly, if the research work has been deployed in some commercial product dealing with live attacks in a production grade network, we consider these research works under category \emph{commercial}. 

We observed that more than 50\% of the research works use simulated networks or applications when experimenting with MTD techniques, whereas $\sim$ 34\% of research works have used emulated environments for performing MTD based analysis research. Only few MTD solutions e.g., $\sim$ 13\% have implemented MTD at commercial level (in production grade networks for dealing with live attacks). This shows that though MTD has been well accepted in research community, and benefits of MTD can help in dealing with different threat models, the industry adoption of MTD is rather slow. The key reason behind this is the adverse impact that MTD can induce on Quality of Service (QoS), and the additional cost-overhead associated with deployment of MTD. We discuss these factors under \emph{Qualitative} and \emph{Quantitative} metrics in Section~\ref{effect}.

\begin{figure}[!t]
    \centering

    \begin{tikzpicture}[
            state/.style={rectangle},
            node distance=2.3cm,
            ]
            \centering
            \node[state] (router) [text width=1cm, align=center] {{\Huge \faWifi}\\{\footnotesize Router}};
            
            \node[state] (ofTop) [above left of=router, xshift=-0.7cm, yshift=0cm, text width=1.2cm, align=center] {{\faChevronRight\faChevronLeft}\\{\footnotesize Open-flow Switch}};
            \node[state] (ofBottom) [below left of=router, xshift=-0.7cm, yshift=-0.5cm, text width=1.2cm, align=center] {{\faChevronRight\faChevronLeft}\\{\footnotesize Open-flow Switch}};
            
            \node[state] (dbTop) [above left of=ofTop, xshift=-1cm, yshift=-0.85cm, text width=1.45cm, align=center] {{\Huge \faDatabase}\\{\footnotesize Database}};
            \node[state] (wsTop) [below left of=ofTop, xshift=-1cm, yshift=0.85cm, text width=1.45cm, align=center] {{\Huge \faServer}\\{\footnotesize Web Server}};
            
            \node[state] (dbBottom) [above left of=ofBottom, xshift=-1cm, yshift=-0.2cm, text width=1.45cm, align=center] {{\Huge \faDatabase}\\{\footnotesize Database}};
            \node[state] (authSer) [left of=ofBottom, xshift=-0.32cm, text width=1.45cm, align=center] {{\Huge \faSignIn}\\{\footnotesize Auth Server}};
            \node[state] (wsBottom) [below left of=ofBottom, xshift=-1cm, yshift=0.2cm, text width=1.45cm, align=center] {{\Huge \faServer}\\{\footnotesize Web Server}};
            
            \node[state] (internet) [right of=router, xshift=0.2cm, text width=1.2cm, align=center] {{\Huge \faGlobe}\\{\footnotesize Internet}};
            
            \node[state] (nox) [below of=router, text width=1.2cm, xshift=0.7cm, yshift=0.15cm, align=center] {{\Huge \faServer}\\{\footnotesize NOX Controller}};
            
            \path[black, -]
                (dbTop) edge (ofTop)
                (wsTop) edge (ofTop)
                
                (dbBottom) edge (ofBottom)
                (authSer) edge (ofBottom)
                (wsBottom) edge (ofBottom)
                
                (ofTop) edge (router)
                (ofBottom) edge (router)
                
                (router) edge (nox)
                (router) edge (internet)
                ;
                
\end{tikzpicture}
    
    \caption{OpenFlow-based Random Host Mutation mutates the virtual IPs of hosts based on a pool of available IP addresses.}
    \label{fig:7}

\end{figure}
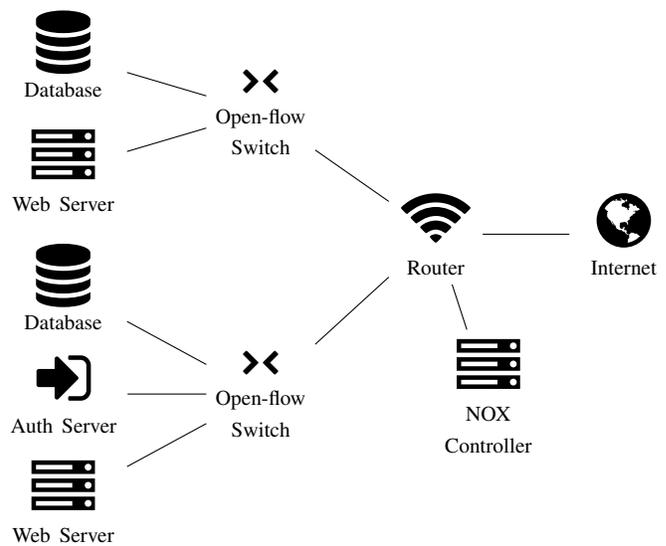

\subsubsection{SDN-based Network Mapping and Reconnaissance Protection}
The first step of the Cyber Kill Chain (CKC) is the identification of vulnerable software and OS versions. Most scanning tools make use of ICMP, TCP or UDP scans to identify the connectivity and reachability of the potential targets. The replies to the scans can also reveal the firewall configuration details, i.e., what traffic is allowed or denied. The time to live (TTL) information can also help in the identification of a number of hops to the attack target~\cite{kampanakis2014sdn}.

SDN-enabled devices can help in delaying the network attack propagation by hiding the real response and replying back with a random response in order to confuse the attacker. As a result, the attacker will believe that random ports are open in the target environment. The attack cost will be increased since the attacker will need to distinguish the real reply from the fake reply. SDN-enabled devices can also introduce random delays in TCP handshake request that will disrupt the reconnaissance process utilized by the attacker for identification of TCP services. The cost-benefit analysis of MTD adaptations against network mapping attempts has been discussed by Kampankis \textit{et al.}~\cite{kampanakis2014sdn}. The survey considers cost-benefit aspects of MTD in Section~\ref{effect}, under the quantitative metrics of MTD.

\subsubsection{Service Version and OS Hiding} The attacker needs to identify the version of OS or vulnerable service in order to mount an attack. For instance, the attacker can send \emph{HTTP GET} request to Apache Web Server, and the response can help in identification of vulnerability associated with a particular version of the Apache software. If the attacker gets a reply 404 Not Found, he can identify some obfuscation happening at the target software. A careful attacker can thus change the attack vector to exploit the vulnerability at the target.

An SDN-enabled solution can override the actual service version with a bogus version of the web server. Some application proxies leverage this technique to prevent service discovery attempts by a scanning tool. Another attack method is known as Operating System (OS) Fingerprinting, where the attacker tries to discover the version of the OS which is vulnerable. Although modern OS can generate a random response to TCP and UDP requests, the way in which TCP sequence numbers are generated can help an attacker in the identification of OS version.

In an SDN-enabled solution, the OS version can be obfuscated by the generation of a random response to the probes from a scanning tool. SDN can introduce a layer of true randomization for the transit traffic to the target. The SDN controller can manage a list of OS profiles and send a reply resembling TCP sequence of a bogus OS, thus misguiding the attacker.

\begin{figure}[!tp]
    \centering
    \includegraphics[trim=80 5 80 5,clip,width=0.5\textwidth]{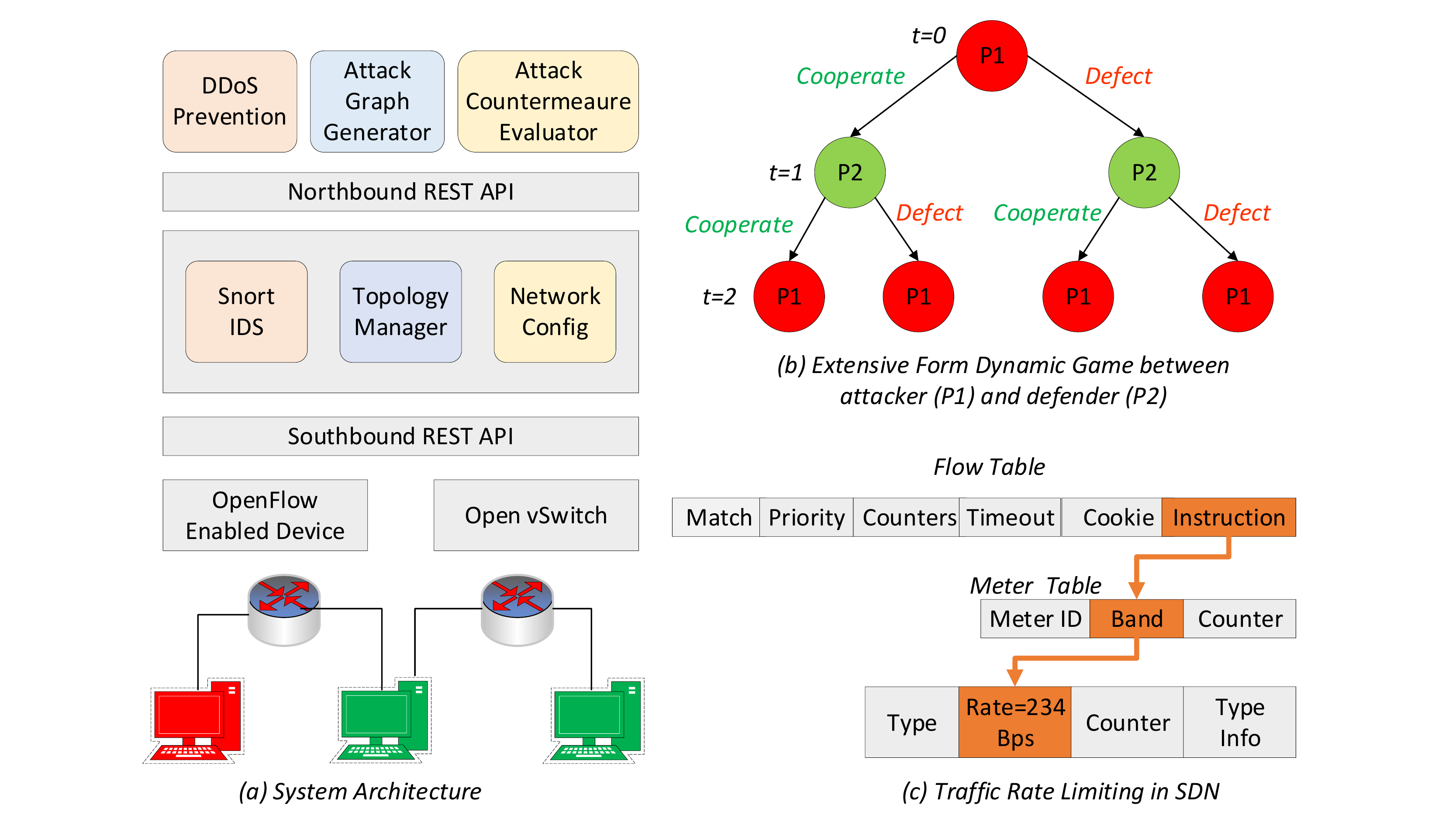}
    \caption{SDN/NFV based MTD that models the interaction between an attacker and the defender as a game to select an optimal countermeasure strategy. The work imposes limiting on bandwidth and views the MTD as an adaptation problem.}
    \label{fig:case-study3}
    % \vspace{-1.6em}
\end{figure}

\subsubsection{Protection against multi-stage attacks and service disruption} Once the attackers have obtained necessary information, they proceed towards targeted attacks, with the aim of stealing information - SQL Injection or service disruption - Distributed Denial of Service (DDoS). The SDN-based MTD can introduce various countermeasures at network-level such as network shuffling~\cite{hong2017optimal}, route modification~\cite{steinberger2018ddos}, IP, and port obfuscation as discussed by Wang \textit{et al.}~\cite{shi2017chaos}.

We now discuss some case studies which show use of SDN/NFV capabilities for implementation of MTD techniques - what, when, and how to switch? Jafarian \textit{et al.}~\cite{jafarian2012openflow} use OpenFlow based random host mutation technique to switch virtual IP (what to switch) targeted by reconnaisance attempts. Debroy \textit{et al.}~\cite{debroy2016frequency} use SDN framework to identify optimal rate of migration (when to switch) and ideal migration location for VMs under DoS attack. Chowdhary \textit{et al.}~\cite{chowdhary2017dynamic} use SDN-environment to create a analyze the hosts mounting DDoS attacks on critical network services. The SDN-controller downgrades network bandwidth (how to switch) using a Nash-Folk theorem based punishment mechanism. 

\noindent $\RHD$ \textbf{Case Study (What to Switch)-- OpenFlow Random Host Mutation.}
SDN makes use of OpenFlow protocol for control plane traffic. Jafarian \textit{et al.}~\cite{jafarian2012openflow} propose OpenFlow enabled MTD architecture as shown in Figure~\ref{fig:7}, can be used to mutate IP address with a high degree of unpredictability while keeping a stable network configuration and minimal operational overhead.

The mutated IP address is transparent to the end host. The actual IP address of the host called real IP (rIP), which is kept unchanged, but it is linked with a short-lived virtual IP address (vIP) at regular interval. The vIP is translated before the host. The translation of rIP-vIP happens at the gateway of the network, and a centralized SDN controller performs the mutation across the network. A Constraint Satisfaction Problem (CSP) is formulated in order to maintain mutation rate and unpredictability constraints. The CSP is solved using Satisfiability Modulo Theories (SMT) solver.

Sensitive hosts have a higher mutation rate compared to the regular hosts in this scheme. The OF-RHM is implemented using Mininet network simulator and NOX SDN controller. OF-RHM is able to thwart about 99\% of information gathering and external scanning attempts. The framework is also highly effective against worms and zero-day exploits.

\begin{figure}[!t]
    \centering

    \begin{tikzpicture}[
            state/.style={rectangle},
            node distance=2.3cm,
            ]
            \centering
            \node[state] (controller) [text width=1.3cm, align=center] {{\Huge \faServer}\\{\footnotesize Open-flow Controller}};
            
            \node[state] (attacker) [above left of=controller, xshift=-0.5cm, text width=1.2cm, align=center] {{\Huge \faUserSecret}\\{\footnotesize Attacker}};
            \node[state] (user) [below left of=controller, xshift=-0.5cm, text width=1cm, align=center] {{\Huge \faUser}\\{\footnotesize User}};
            
            \node[state] (auth) [above of=controller, yshift=-0.1cm, text width=1.5cm, align=center] {{\Huge \faSignIn}\\{\footnotesize Auth Server}};
            
            \node[state] (cloud) [right of=controller, text width=1.5cm, align=center] {{\Huge \faCloud}\\{\footnotesize Cloud env.}};
            
             \node[state] (mig) [below right of=cloud, xshift=1cm, text width=1.7cm, align=center] {{\Huge \faFilesO}\\{\footnotesize Migration Destination VM}};
             \node[state] (target) [above right of=cloud, xshift=1cm, text width=1.7cm, align=center] {{\Huge \faFilesO}\\{\footnotesize Target Application VM}};
            
            \path[BrickRed, ->, dashed]
                (attacker) edge (controller)
                (controller) edge (cloud)
                (cloud) edge (mig);
            
            \path[ForestGreen, ->]
                (user) edge (controller)
                (controller.-23) edge (cloud.200)
                (cloud) edge (target);
                
            \path[black, <->, dotted]
                (auth) edge (controller)
                (controller.23) edge (cloud.160);
                
            \path[black, ->, dotted]
                (cloud.50) edge (target.195)
                (target) edge (mig);
                
\end{tikzpicture}
    
    \caption{Virtual Machine migration using frequency minimal MTD seeks to migrate applications to new VMs and redirect an attacker's DDoS attacks without effecting availability to regular users. The goal of this work was to find an optimal timing function $T$, termed as the migration frequency.}
    \label{fig:8}
    % \vspace{-1.6em}
\end{figure}
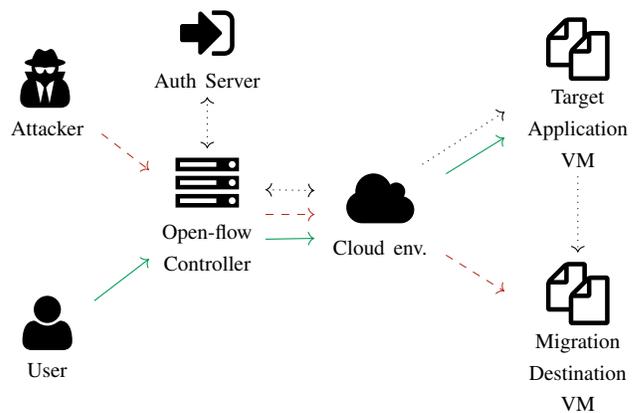

\noindent $\RHD$ \textbf{Case Study (How to Switch)-- Dynamic Game-based Security in SDN-enabled Cloud Networks}.
DDoS attacks are a major security threat affecting networks. Attacker's leverage sophisticated bots to generate a huge volume of network traffic, and overwhelm critical network services as discussed by Chowdhary \textit{et al.}~\cite{chowdhary2017dynamic}. The case study presented in this research work, focused on the MTD decision {\em how to move?} The framework presented in Figure~\ref{fig:case-study3} (a) to communicate with OpenFlow devices using southbound REST API, whereas any application plane decision and network analytic is performed using northbound REST API.

The Snort IDS~\cite{roesch1999snort} detects malicious DDoS patterns, and uses Nash Folk Theorem~\cite{fudenberg2009folk} to analyze the behavior of a malicious node, e.g., red color VM shown in Figure~\ref{fig:case-study3} (a). The scenario is represented as an extensive-form game spanning multiple rounds. If the node sending network traffic P2, the Defect - Figure~\ref{fig:case-study3} (b), the normal bandwidth B is reduced to $\frac{B}{2}$. The bandwidth is reduced in each subsequent round by network admin P1, till average bandwidth over a period of time $t=\{0,1,2,..\} \in T$ is B.

The SDN controller utilizes Instruction field present inside the flow table in order to change the band rate of the node flagged as malicious, as shown in Figure~\ref{fig:case-study3} (c). Thus SDN-based rate-limiting serves as an effective countermeasure against flooding/loss of availability attacks like DDoS.

\noindent $\RHD$ \textbf{Case Study (When to Switch): Frequency Minimal MTD using SDN.}
\emph{Frequency minimal MTD}~\cite{debroy2016frequency} approach considers resource availability, QoS and exploits probability for performing MTD in an SDN-enabled cloud infrastructure. The design goals of this research work are as follows:
\begin{enumerate}
    \item Identification of optimal frequency of VM migration, ensuring minimal wastage of cloud resources.
    \item Finding the preferred location of VM migration without impacting the application performance. 
\end{enumerate}

As shown in Figure~\ref{fig:8}, the normal clients can access the services hosted in the cloud network via regular path, and the attack path represents the path along which the attacker tries to exploit the target application. The VMs periodically share their resource information such as storage and compute capacity with the controller along the control path. Depending upon the level of threat, the migration of VM can be proactive or reactive. The real IP address of the VM hosting cloud application is hidden from the clients. The data-path shows the path along which VM application and its back-end database are migrated. VM migration is based on the following factors:

\begin{itemize}
    \item \textbf{VM Capacity:} This parameter considers the capacity of migration target in terms of computing resources available to store files and database of current and future clients.
    \item \textbf{Network Bandwidth:} The lower the network bandwidth between the source and target VM, the slower will be the migration process and the longer will be the exposure period (in case of active attacks) for the VM being migrated. This parameter considers bandwidth between source and target while performing VM migration.
    \item \textbf{VM Reputation:} This is the objective indicator of VM robustness to deter future cyber attacks. It is the history of VM in terms of cyber attacks launched against the VM. This parameter is considered in order to ensure VM’s suitability for migration.
\end{itemize}
   
This research work estimates the optimal migration frequency and ideal migration location based on the parameters described above. The VM migration mechanism is highly effective in dealing with denial-of-service (DoS) attacks. 

%\subsection{Comparitive Analysis of SDN-based MTD}

 %\subsection{Security Modeling  MTD}
\subsection{MTD Testbeds: Research and Prototyping}\label{testbed}
 The practicality of MTD can be established by the deployment of MTD techniques and tactics over an underlying network. In this section, we identify some platforms which can assist MTD researchers in conducting experiments or serve as a guideline when creating MTD case studies.

\subsubsection{GENI Platform} {Global Environment for Networking Innovation} (GENI)~\cite{berman2014geni} provides a distributed virtual environment for at-scale networking and security experimentation. Individual experiments can be conducted in isolation using network sliceability (ability to support virtualization and network isolation). Each experimental slice is provided with network and computing resources. The users can request resources for an allotted time period, in which they can conduct experiments and release the resources once they have finished the experiment.

Debroy \textit{et al.}~\cite{debroy2016frequency} utilize GENI framework for implementation of a {VM migration} MTD solution. The authors utilized the {InstaGENI} platform at the Illinois site to host a news feed application targeted by DoS attacks. The setup also involved four non-malicious users, one remote SDN controller, and attacking VMs, all hosted at physically distributed locations. The attacker VMs were utilized for sending a large number of HTTP GET requests to news feed site in order to achieve a DoS attack. Proactive and reactive frequency minimal (FM) MTD countermeasures were deployed in response to targeted DoS attacks. The research work shows the capability of GENI platform to host similar MTD experiments where users can study the impact on network bandwidth, VM performance, attack success rate when MTD security countermeasures are implemented at scale on a cloud platform.  

 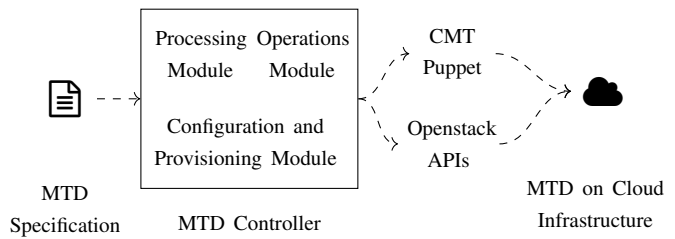
\begin{figure}[!tp]
    \centering

    \begin{tikzpicture}[
            scale=0.8,
            state/.style={rectangle},
            node distance=2cm,
            ]
            \centering
            \node[state] (spec) [text width=0.6cm, align=center] {{\Large \faFileTextO}};
            \node[state] (spec_text) [below of=spec, yshift=0.5cm, text width=1.5cm, align=center] {{\footnotesize MTD Specification}};
            
            \node[state] (mtd_cnt) [draw=black, right of=spec, xshift=0.45cm, text width=2.65cm, text height=2.15cm, align=center] {};
            \node[state] (process) [right of=spec, xshift=-0.2cm, yshift=0.6cm, text width=1.2cm, align=center] {\footnotesize Processing Module};
            \node[state] (operation) [right of=process, xshift=-0.65cm, text width=1.2cm, align=center] {\footnotesize Operations Module};
            \node[state] (cnp) [right of=process, xshift=-1.4cm, yshift=-1.2cm, text width=2.5cm, align=center] {\footnotesize Configuration and Provisioning Module};
            \node[state] (mtd_cnt_text) [below of=mtd_cnt, yshift=0.35cm, text width=2.5cm, align=center] {{\footnotesize MTD Controller}};
            
            \node[state] (cmt) [right of=operation, text width=0.9cm, align=center] {\footnotesize CMT Puppet};
            \node[state] (openstack) [right of=cnp, xshift=0.7cm, text width=1.1cm, align=center] {\footnotesize Openstack APIs}; 
            
            \node[state] (cloud) [right of=cmt, yshift=-0.5cm, text width=0.6cm, align=center] {{\Large \faCloud}}; 
            \node[state] (cloud_text) [below of=cloud, xshift=-0.1cm, yshift=0.5cm, text width=1.9cm, align=center] {{\footnotesize MTD on Cloud Infrastructure}};

            \path[black, ->, dashed]
                (spec) edge (mtd_cnt)
                
                (mtd_cnt) edge[out=0,in=180] (cmt)
                (mtd_cnt) edge[out=0,in=180] (openstack)
                
                (cmt) edge[out=0,in=180] (cloud)
                (openstack) edge[out=0,in=180] (cloud)
                ;
                
            % \path[NavyBlue, -]
            %     (es_s) edge node[below] {Exploration Surface} (es_e);
                
            % \path[PineGreen, -]
            %     (dps_s) edge node[below] {Detection and Prevention Surface} (dps_e);
                
    \end{tikzpicture}
    
    \caption{A platform that takes an abstract specification of a cloud system as input and outputs the corresponding system on a cloud infrastructure. Advantages of cloud automation using Automated Enterprise Network Compiler (ANCOR) include performing live instance migration.}
    \label{fig:6}
    % \vspace{-1.6em}
\end{figure}  
 
\subsubsection{OpenStack Cloud} {OpenStack}~\cite{sefraoui2012openstack} is a cloud operating system that consists of compute, storage, and networking resources. The users can log in through GUI to provision isolated virtual networks or utilize OpenStack APIs to create and configure the network. OpenStack is compatible with the existing virtual solutions such as {VMWare ESX},   {Microsoft's Hyper-V}, {KVM}, {LXC}, {QEMU}, {UML}, {Xen}, and {XenServer}. 

\textbf{Mayflies}~\cite{ahmed2016mayflies} utilized OpenStack for designing a {fault-tolerant} MTD system. The research work utilizes VM {introspection} to checkpoint the current state of the live node/VM, and {reincarnation} - node {commission/decommission} based on attacks against the introspected node. The strategy allows the Mayflies framework to deal with attacks in a short interval of time, avoiding the attack progress. Zhuang \textit{et al.}~\cite{zhuang2013investigating} conduct MTD experiments to simulate a {pure-random MTD} strategy and an intelligent MTD approach based on intelligent attack indicators. The experiments were conducted using \emph{NeSSi2}~\cite{kosowski2011nessi2}, an {open-source}, discrete event-based network security {simulator}. The authors proposed implementation on an OpenStack based MTD testbed as future work.

\textbf{MTD CBITS} MTD-platform for \textit{Cloud-Based IT Systems} (MTD CBITS)~\cite{bardas2017mtd} as shown in  Figure~\ref{fig:6} evaluates the practicality and performs detailed analysis of security benefits when performing MTD on a cloud system such as OpenStack.

 The platform makes {automated changes} to the IT systems running on the cloud infrastructure. One adaptation performed in MTD CBITS is {replacing} the {running components} of the system with new ones. The system parameters are stored in the operational model and can be viewed using the MTD system inventory - {CMT (Puppet) APIs}. Any MTD adaptation is also recorded in an operational model for future reference. {OpenStack API} utilizes Puppet agents to communicate with the live instances of the cloud system. MTD controller communicates with agents over a private network. 
 
%  {|p{45mm}|p{50mm}|p{13mm}|p{13mm}|p{15mm}|p{13mm}|}

\subsubsection{CyberVAN Testbed}~\cite{chadha2016cybervan} provides a testing and experimentation platform for cybersecurity research. The platform supports high fidelity network, by representing the network in a discrete event simulator. CyberVAN allows hosts represented by VMs to communicate over the simulated network. The testbed utilizes network simulator \emph{ns-3}~\cite{henderson2008network} to simulate network effects such as end-to-end network latency, link capacities, routing protocols, {\em etc.} The platform also supports {wireless networks} by modeling {mobility}, {interference}, {propagation effects}, as well as the details of different {waveforms}. The {Scenario Management GUI} allows the experimenter to access and manage the elements of an attack scenario, including network traffic and logging, running analytic tools on the VM, saving results of the experiment, pausing and restarting the experiments, {\em etc.}

\subsubsection{MTD SCADA Testbed for Smart Grid Network}
 \begin{figure}[!tp]
    \centering
    \includegraphics[trim=10 50 10 20,clip,width=0.5\textwidth]{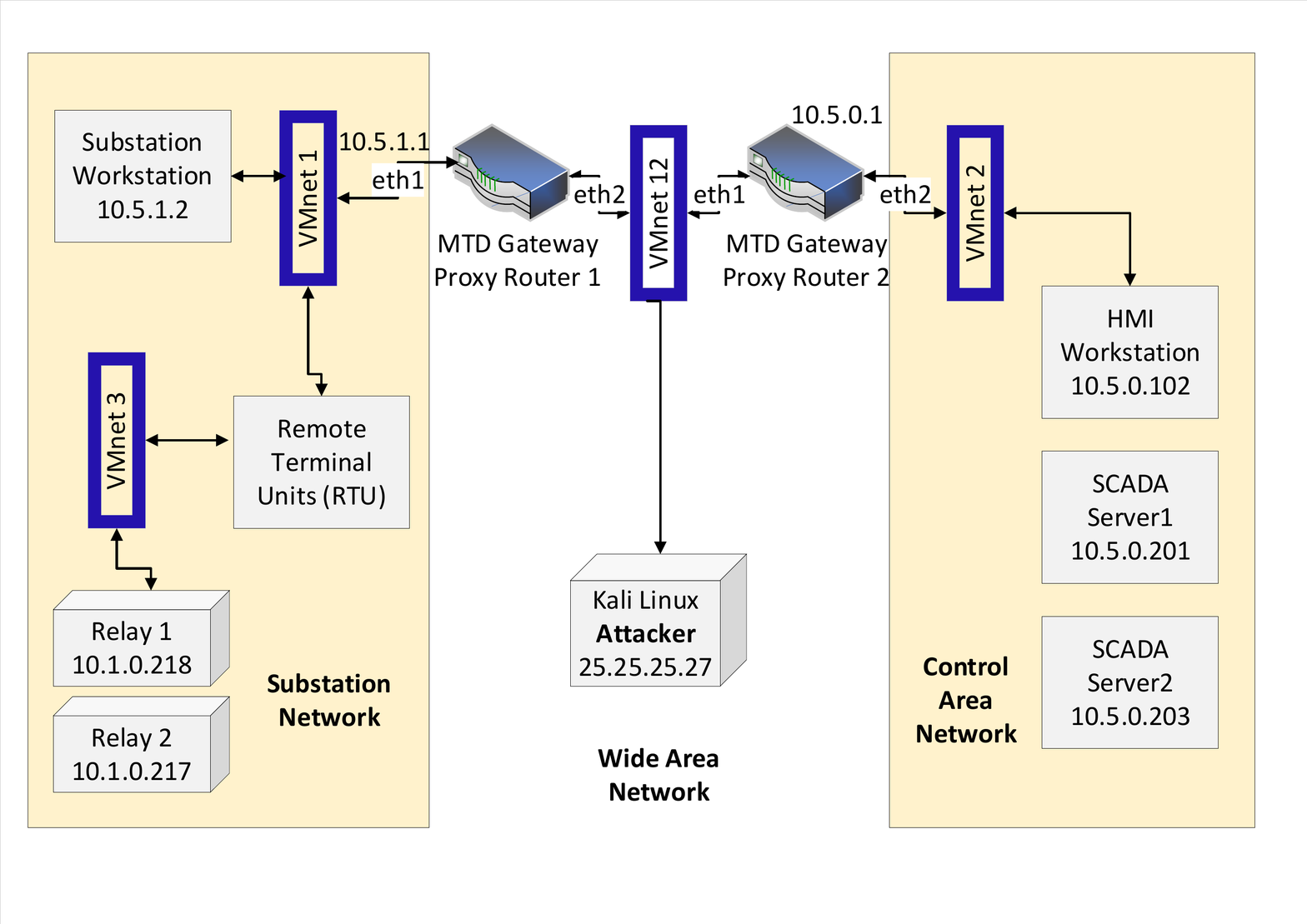}
    \caption{An MTD-based architectural platform for IP hopping. The solution has been targeted at Supervisory Control and Data Access Systems (SCADA). The MTD gateway routers act as dynamic proxies, periodically mutating the IP address of the external interfaces exposed on the public wide-area network.}
    \label{fig:6-1}
    % \vspace{-1.6em}
\end{figure} 

In the context of smart-grid environments, Pappa \textit{et. al.} has proposed an MTD for IP hopping~\cite{pappa2016moving} (see Figure~\ref{fig:6-1}). The network consists of two SCADA servers running Siemens Spectrum Power TG Energy Management Systems (EMS) software. The software periodically polls Remote Terminal Units (RTUs) for network traffic measurements. The substation networks are connected to the Wireless Area Network (WAN) using MTD gateway routers. The remote attacker can utilize the publicly exposed service over the WAN to identify   security vulnerabilities. The services can be exploited using traditional means or using APT techniques. The gateway routes in the network act as dynamic proxies, mutating IP address of the external interfaces, while providing end-to-end SCADA communication. The IP generation algorithm employs a random initial seed to initiate the IP generation at the network boot up. Both routers are assigned set $(j,k)$ of $n$ random IP addresses at the start, in combination of initial seed vector. The initial seed vector ensures order between the two gateway routers is synchronized; this helps to prevent IP collision and network outages. Additionally, the MTD algorithm used creates a dynamic network technology which makes the job of network mapping difficult for the attacker.

%TrapX and other companies using smart honeypots*

\subsubsection{MTD Commercial Domain Solutions}
\textbf{TrapX:} The production-grade decoy network technologies such as \emph{DeceptionGrid}~\cite{trapx-deceptiongrid} and \emph{CryptoTrap}~\cite{cryptotrap} deceive the attackers by imitating the true assets. DeceptionGrid provides {agentless} visibility and control appliance, that dynamically identifies and evaluates {network endpoints} and {applications} when they connect to the main network. {Assuta Medical Center} incorporated DeceptionGrid as a part of their network security suite. The use of this framework helps to counter not only known attacks but also APTs and {zero-day} attack scenarios. The DeceptionGrid created a network of {traps} (decoys) that camouflaged real medical devices such as {MRI} \& {CT scanners}, {X-ray machines}, and {ultrasound equipment} (PACs). The solution has been deployed on many {VLANs} across {Assuta Medical Center} and provided better visibility into the lateral movement of the attackers. 

\textit{CryptoTrap} on the other hand is utilized for {countering} the {ransomware} early in their exploitation lifecycle. This helps in countering malware propagation while protecting valuable network assets. The traps (decoys) are masked in the form of {SMB network shares} across the network. The fake data of the company is replicated across the traps. Once the attacker touches these traps, the targeted computer is disconnected from the network and security administrator is alerted about the incident. In effect, the attacker is held captive in the trap and attacker actions are logged for further {threat intelligence}. 

\textbf{Polyverse:} \emph{Zero-day} vulnerabilities ,e.g., \emph{Heartbleed}~\cite{carvalho2014heartbleed} (vulnerability discovered in  \textit{OpenSSL} software in 2014), can cause a significant damage in an underlying network. There is no known attack signature to identify these vulnerabilities, hence they can {bypass} security {monitoring software} unnoticed.
The attackers trying to target software or OS {memory-based zero-day} vulnerabilities start with the assumption that the gadgets they are trying to access are located at a certain address or within a specific offset from the absolute base address. \emph{Polyverse}~\cite{polyverse} employs MTD strategy to defeat this assumption of the attackers. The polymorphic version of Linux is utilized by Polyverse in order to create a high-level of {entropy} in the software system in such a way that the entire {memory} structure is {diversified}. With the polymorphic versions of Linux, the entire Linux software stack is roughly randomized. The resulting program is semantically identical to the original program (functionally equivalent), however, nearly every machine instruction is changed. 

\textbf{MorphiSec:} Some threat vectors classified as {Advanced Evasive Attacks} {cloak} the {malicious intent} in order to deceive the security monitoring tools. Some of these techniques include {Polymorphism} (changing malware signature), {Metamorphism} (changing malware code on the fly), {Obfuscation} (hiding malicious activity), {behavior changes} (waiting for normal user activity before executing). \emph{Morphisec}~\cite{morphisec} uses MTD at {network-level} (route changes, random addresses and ports), firewall level (policy changes), {host-level} (OS version, memory structure changes), and {application level} (randomizing storage format addresses, application migration, multilingual code generation) in order to deceive and detect attacks using evasive attack techniques.

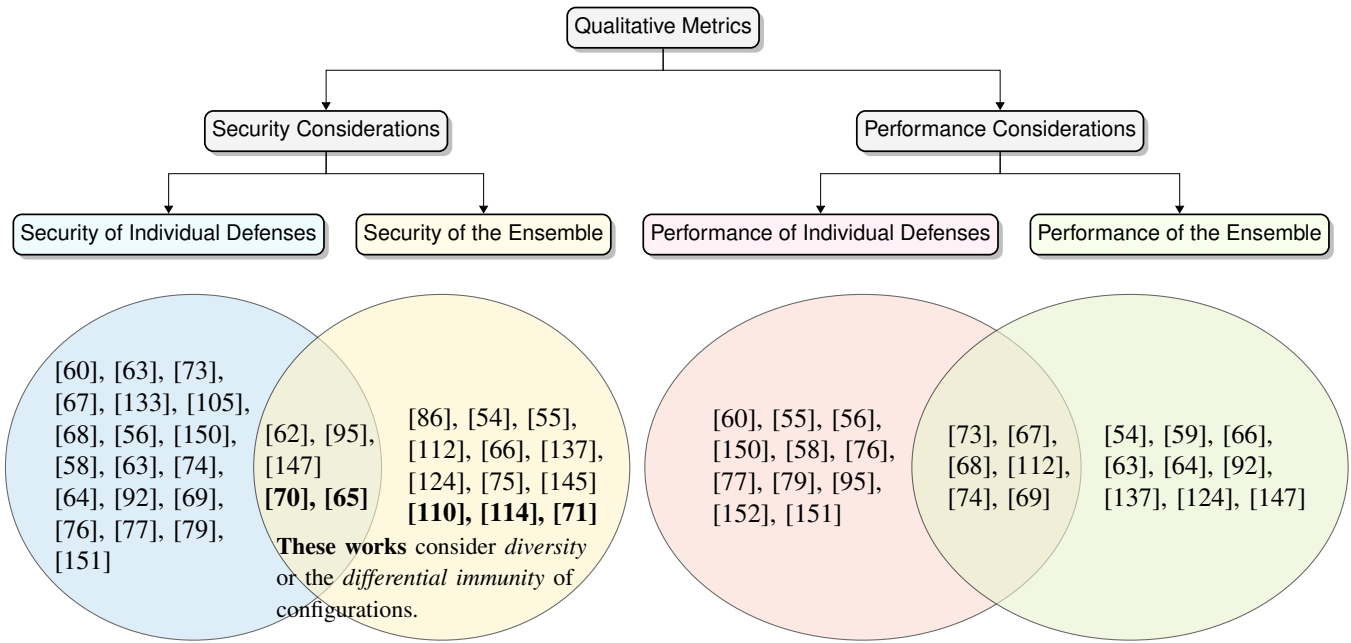
\begin{figure*}[t]

\resizebox{2.0\columnwidth}{!} {
\centering
    \begin{forest}
        for tree={
          %ellip,
          %square,
          draw,
          thick,
          rounded corners,
          font=\sffamily,
          inner color=Gray!9,%JungleGreen!30,
          outer color=Gray!9,%JungleGreen!30,
          align=center,
          child anchor=north,
          parent anchor=south,
          drop shadow,
          l sep=1cm,
          s sep=0.5cm,
          edge path={
                \noexpand\path[
                    >={Triangle[]},
                    ->,
                    \forestoption{edge}]
                (!u.parent anchor) -- +(0pt,-10pt) -|
                (.child anchor)\forestoption{edge label};
            }
        }
    [Qualitative Metrics
        [Security Considerations
            [Security of Individual Defenses, inner color=CornflowerBlue!10!white, outer color=CornflowerBlue!10!white]
            [Security of the Ensemble, inner color=Goldenrod!10!white, outer color=Goldenrod!10!white]
        ]
        [Performance Considerations
            [Performance of Individual Defenses, inner color=Salmon!10!white, outer color=Salmon!10!white]
            [Performance of the Ensemble, inner color=SpringGreen!10!white, outer color=SpringGreen!10!white]
        ]
    ]
    \end{forest}
}

\begin{tikzpicture}
\begin{scope}[opacity=0.6]
    \fill[CornflowerBlue!30!white,draw=black] (180:4.15) ellipse (2.5 and 2.3);
    \fill[Goldenrod!30!white,draw=black]  (180:0.85) ellipse (2.5 and 2.3);
    \fill[Salmon!30!white,draw=black] (0:4.75) ellipse (2.9 and 2.3);
    \fill[SpringGreen!30!white,draw=black]  (0:8.3) ellipse (2.9 and 2.3);
 \end{scope}
    \node at (90:2.7) {};
    % Is aware of the security of individual defender actions
    \node[text width=3cm] at (180:4.5) {\cite{lei2017optimal,zhu2013game,chowdhary2017dynamic,el2016softwarel,kampanakis2014sdn,steinberger2018ddos,debroy2016frequency,zhao2017sdn,meier2018nethide,jajodia2018share,zhu2013game,clark2015game,sengupta2017game,vadlamudi2016moving,prakash2015empirical,venkatesan2016moving,sengupta2018moving,chowdhary2018markov,rass2017defending}};
    
    % Is aware of the security of individual defender actions and the ensemble
    \node[text width=1.5cm] at (180:2.45) {\cite{manadhata2013game,miehling2015optimal,bardas2017mtd}\\
    \textbf{\cite{neti2012software,carter2014game}}};
    
    % Is aware of the security of the ensemble
    \node[text width=3cm] at (0:0.2) {\cite{shi2017chaos,jafarian2012openflow,aydeger2016mitigating,wang2017u,thompson2014multiple,dunlop2011mt6d,ahmed2016mayflies,colbaugh2012predictability,zhuang2013investigating}\\\textbf{\cite{hong2017optimal,homescu2017large,crouse2012improving}}};

    \node at (245:1.65)[text width=4.7cm]   {\small \textbf{These works} consider \textit{diversity} or the \textit{differential immunity} of configurations.};
    
    \node[text width=3cm] at (0:4.25) {\cite{lei2017optimal,aydeger2016mitigating,zhao2017sdn,meier2018nethide,jajodia2018share,venkatesan2016moving,sengupta2018moving,chowdhary2018markov,miehling2015optimal,nguyen2008survey,rass2017defending}};
    
    \node[text width=1.75cm] at (0:6.75) {\cite{chowdhary2017dynamic,el2016softwarel,debroy2016frequency,wang2017u,clark2015game,prakash2015empirical}};
    
    \node[text width=3cm] at (0:9.45) {\cite{jafarian2012openflow,algin2017mitigating,thompson2014multiple,zhu2013game,sengupta2017game, vadlamudi2016moving,dunlop2011mt6d,ahmed2016mayflies,bardas2017mtd}};

\end{tikzpicture}
 
\caption{A Moving Target Defense is more effective if it considers the security and performance impacts of the system configurations, both in unison and also when used as part of an ensemble that the MTD leverages. We categorize the works which explicitly consider these metrics either while modeling the defense or during its evaluation. Note that there is no work which considers all four kinds of metrics, i.e. falls in the intersection zone both on the left and on the right.}
\label{fig:qual}
% \vspace{-1.6em}
\end{figure*}

\section{Evaluating the Effectiveness of\\Moving Target Defenses}\label{effect}

In this section, we present a list of qualitative and quantitative metrics that can help in determining the effectiveness of MTD. First, we discuss the quality of the defender actions $C$, i.e. the various system configurations it can choose to deploy, in terms of performance and security. This discussion helps in identifying a set of qualitative metrics. As we will see, most works either consider this metric implicitly when choosing the configuration set or simply ignore it.
Second, we devote a sub-section on qualitative evaluation metrics. These are mathematical functions to help a defender capture measures such as the cost-benefit, the risk analysis, {\em etc.} of an MTD. As we shall see, they can be measured by simulation on test-beds or by using metrics based on security domain knowledge such as CVSS, attack graphs representations {\em etc.}

\subsection{Qualitative Evaluation}

When using a Moving Target Defense (MTD) that moves between multiple system configurations, we would want to believe that it increases the security of the deployed system without negatively impacting the performance for legitimate users. Thus, we look at how different MTDs consider these aspects either in the modeling of the defense or its evaluation. In particular, we look at two major aspects-- {security considerations} and {performance considerations}-- to evaluate the quality of an MTD. Under each of these sub-headings, we consider the quality of each individual defense and then, the {overall ensemble} of constituent defenses. This helps us identify {non-trivial heuristics} to understand when an MTD might succeed or fail. For example, an MTD that lacks {\em differential immunity}, i.e. all the constituent system configurations ($\in C$) are all vulnerable to the same attack, can never be secure regardless how well it is implemented in practice.

In regards to quality metrics, we categorize the various MTDs in Figure \ref{fig:qual}. An MTD is part of a set if they explicitly consider the particular qualitative metric when designing either the configuration set $C$ (i.e. {\em what to move?}), the timing function $T$ (i.e. {\em when to move?}), or the movement function $M$ (i.e. {\em how to move?}). Note that only a subset of works discussed under security considerations (i.e. shown in the {\color{CornflowerBlue!80!black} blue} and/or {\color{Goldenrod!90!black}yellow} boxes on the left of Fig. \ref{fig:qual}) are featured under {performance considerations} (i.e. shown in the {\color{Salmon!90!black} pink} and/or {\color{SpringGreen!90!black}green} boxes on the right of Fig. \ref{fig:qual}). We discovered that majority of the defenses proposed only consider (and therefore highlight) the improvement with regards to security and ignore the impact on performance. We will now discuss the details about how the different works capture the various qualitative aspects that we put forth in this survey.

\subsubsection{Security Considerations}
We first look at works that reason about the security risks of individual actions followed by works that reason about the security risks associated with an ensemble and lastly, discuss works that consider both.

\paragraph{Considers only Security of Individual Defenses} Most MTDs that consider some form of game-theoretic modeling, consider the security of individual defenses in the ensemble by representing them as a part of the defender's utility value \cite{lei2017optimal,sengupta2018moving,jajodia2018share,vadlamudi2016moving,carter2014game,sengupta2018moving,rass2017defending,chowdhary2018markov}. For most of these works, the security metrics considered for each action are obtained using scoring metrics designed by experts such as the Common Vulnerability Scoring Services (CVSS) discussed earlier. These works are able to come up with an intelligent movement strategy $M$ based on reasoning over known vulnerabilities (more specifically, common vulnerabilities and exposures (CVE)) for which CVSS scores are readily available. This can result in highly sub-optimal behavior when faced with zero-day vulnerabilities. In that regard, authors in \cite{carter2014game} model the security of a configuration as inversely proportional to the probability with which an adversary can come up with a new attack given the attacks it performed in the earlier time steps. In \cite{rass2017defending}, the authors try to model zero-day attacks by asking security experts to annotate how effective they think a particular {countermeasure} or defense action will be against zero-day vulnerabilities. As the annotations might be inaccurate (due to the lack of actual black-hat hackers who invent zero-day attacks in the set of security experts who annotate the defense methods), the effectiveness of their MTD cannot be accurately determined. As to how, if possible, utility values can capture the security risks associated with zero-day vulnerabilities is an interesting question and remains to be explored by future works.

Other works that also only capture the security of individual constituent system configurations are more domain or problem specific in nature. In \cite{chowdhary2017dynamic}, each defense action, before being played, is weighed based on the penalty it imposes on an attacker (who tried to do a DDoS attack) over a repeated game setting. In \cite{el2016softwarel}, authors choose an action (to migrate a VM or keep running) after reasoning about the security risks associated with the present vulnerable state. In \cite{debroy2016frequency}, the security risk is based on the reputation of the current state, which in turn looks at the type and number of attacks that were done in previous time steps when the particular system was deployed. On similar lines, researchers in \cite{zhu2013game} find a more compact way to model and continuously update the risk associated with deploying a particular defense action. Some works that consider the longitudinal effects of a movement strategy $M$ model the epistemic effects of movement on an attacker's knowledge about the network. For example, \cite{zhao2017sdn} considers the fingerprint of the overall {network} when a particular defense action is selected, while \cite{meier2018nethide} reasons about the topology information a particular defense action leaks to an attacker. In contrast to all these works, \cite{kampanakis2014sdn} models the security risk associated with a particular defense action as inversely proportional to the cost (it believes) an attacker would spend to compromising it.

\paragraph{Considers only Security of the Ensemble} Works that measure only the security of an ensemble as, opposed to security of actions, mostly showcase the security benefits of the MTD ensemble by comparing them to the security benefits provided by a static system configuration \cite{shi2017chaos,jafarian2012openflow,aydeger2016mitigating,thompson2014multiple,zhuang2013investigating}. In essence, these works consider the static system as the control case but unfortunately, do not ensure that this static system is the most secure constituent defense configuration in the ensemble. This way, they might overestimate and thus, over-promise the security guarantees of the designed MTD.

On the other hand, works that consider the security metrics associated with the ensemble (as a whole) at modeling time look at metrics that are similar to {\em entropy} or {\em diversity} of the ensemble. In \cite{dunlop2011mt6d}, the authors try to use the large address-space of IPv6 to their advantage and are able to create more uncertainty for an attacker who is trying to pinpoint the address to use for a successful attack. On similar lines, authors in \cite{ahmed2016mayflies} argue that fast reincarnations of a functional system (i.e. bringing the service down and starting it up on a new location/container, {\em etc.}) increases the entropy and makes it more costly for an attacker to continuously keep attacking such a system. On a different note, researchers in \cite{hong2017optimal} try to select a defense configuration from the ensemble based on how diverse it is with respect to the current configuration that might have been attacked. In order to do this, they use a topological distance measure, which in their case is the {symmetric difference} between the edge sets of the {current} and the {consecutive defense} configuration. Although they do not explicitly recognize it as a diversity metric like \cite{homescu2017large,crouse2012improving}, they bring to light an interesting issue that most MTD papers seem to either miss or assume by default. If there was an attack, that with extremely little modification, could exploit all the defender's configuration that is a part of the MTD, an MTD would not be an effective defense strategy. For example, MTD work that randomly selects a classifier for malware detection assumes by default that each classifier is not vulnerable to the same attack \cite{colbaugh2012predictability} which seems to be an incorrect assumption to make given current research works \cite{szegedy2013intriguing,sengupta2018mtdeep}. We believe that it would be easier to convince practitioners about the effectiveness of an MTD if researchers can show that diverse defender actions can indeed be generated at a low cost. This is the goal of \cite{homescu2017large}, who create an approach at the compiler level to {increase software diversity} and \cite{crouse2012improving}, who use a genetic algorithm to draw a pool of configurations that maximize diversity. 

\paragraph{Considers both}
Only a few works take a holistic view in regards to security and consider both the security of each individual defense and the ensemble as a whole. In \cite{manadhata2013game}, the general level at which they define the {utility values}, one can capture the security risks associated with an {individual defense} action in terms of {attack surface features} and {attack surface measurements} while the security risks associated with an ensemble can be captured via the utility variable for attack surface shifting. In \cite{bardas2017mtd}, the security for individual defense actions is trivial for their settings, where defending a node that is not a stepping stone is not beneficial at all for the defender. The security risks associated with an ensemble are evaluated through experimentation with a static defense.  Particularly, the try to increase the number of interruptions for an attacker to start from one point in the network and reach a goal node. In \cite{miehling2015optimal}, the selection of the {entire ensemble} is based on the fact that the set of {possible defenses} have a {global property} of {covering} each of the {leaves} that an attacker might want to reach and each individual defense action has some utility in terms of security associated with it. Lastly, the works \cite{carter2014game} and \cite{neti2012software} model both the {diversity} of {constituent defenses} present in the MTD ensemble and also model the security of each individual defense. In \cite{neti2012software}, each {configuration} is mapped to a set of {vulnerabilities} and thus, a diverse ensemble is composed of system configurations that do not have a lot of overlapping vulnerabilities. In \cite{carter2014game}, since they use Linux based operating systems, the diversity is modeled in terms of lines of code that is {different} between two {Operating Systems}.

\subsubsection{Performance Considerations}

As mentioned above, a large set of works for developing MTDs focus on showcasing the security benefits and sweep under the rug the performance costs. Note that in the case of MTDs, the impact on performance may arise due to a variety of reasons. First, each system configuration (individual defense action), that is a part of the MTD ensemble, has a performance cost associated with it and moving to a high-cost configuration {impacts} the performance. These concerns are termed as performance considerations for individual defense. Second, the switching from one configuration to another may need a defender to deal with (1) downtime, (2) legitimate requests on the wire that were meant for the previous configuration in a graceful manner and/or (3) keep all the different configurations running (at least two of them) to facilitate a faster switch. All these costs can be termed as shuffling costs and are categorized as performance considerations of the ensemble because they only arise when there is an MTD ensemble.

We will now describe how the various MTD systems consider the performance costs associated with each individual configuration followed by the performance cost associated with the ensemble. We also discuss works that reason about both these factors together and analyze the affect on the overall {Quality of Service} (QoS) when such defenses are is deployed.

\paragraph{Considers only performance of individual defense actions}
Similar to the case of security metrics, when we look at performance considerations for an individual defense, most game-theoretic works model these as a part of the defender's utility function \cite{lei2017optimal,venkatesan2016moving,sengupta2018moving,chowdhary2018markov,nguyen2018multistage}. Various equilibrium concepts in these games yield movement strategies for the defender that gives {priority} to constituent {defenses} that have {low performance costs} while ensuring that the security is not impacted by a lot. In \cite{lei2017optimal} and \cite{nguyen2018multistage}, the reward functions are defined at an abstract level and the authors point out that they can be used to consider the performance cost of constituent defenses. In \cite{venkatesan2016moving} and \cite{sengupta2018moving}, the authors consider the impact of placing {Network-based Intrusion Detection Systems} (NIDS) on the {latency} of the network and use {centrality} based measures as {heuristic} guidance for it.

A large number of works also look at problem-specific instances. Authors in \cite{aydeger2016mitigating} perform experiments to evaluate their MTD against crossfire attacks and find that the average time for packet transfer from one point to another increases because a particular path selected at random can be highly sub-optimal. On similar lines, in \cite{zhao2017sdn}, the authors consider the performance cost of doing a {\em defend} action on legitimate user traffic. On a different note, \cite{meier2018nethide} tries to solve a multi-faceted problem where the MTD tries to {obfuscate} the {network topology} to an attacker and, at the same time, ensures that it {does not negatively impact} a {defender's ability} to debug network issues. This is done by leveraging the knowledge asymmetry about the network topology that a defender and an attacker has. In \cite{jajodia2018share}, the {performance costs} of a  \emph{honeynet} configuration, which is the defense action for this MTD, represents the number and \emph{quality of resources} (or honey) necessary for developing a credible honeynet that fools an attacker. Lastly, authors in \cite{rass2017defending} combine both the costs of setting up a good defense and the usability of that defense for legitimate traffic as the performance cost for a particular placement of countermeasures.

\begin{figure}[t]
    \centering
    \resizebox{0.90\columnwidth}{!} {
\begin{forest}
  for tree={
    child anchor=west,
    parent anchor=east,
    grow=east,
    draw,
    text width=3.2cm,
    anchor=west,
     font=\sffamily,
     inner color=Gray!9,
     outer color=Gray!9,
    edge path={
      \noexpand\path[\forestoption{edge}]
        (.child anchor) -| +(-4.5pt,0) -- +(-4.5pt,0) |-
        (!u.parent anchor)\forestoption{edge label};
    },
  }
  [Quantitative Metrics
    [Usability Metrics,
      inner color=SpringGreen!10!white,
      outer color=SpringGreen!10!white,
      [\textit{QoS Metrics}\\\cite{jafarian2012openflow,crouse2015probabilistic,carter2014game,el2016softwarel, algin2017mitigating, venkatesan2016moving,sengupta2018moving, shi2017chaos, hong2017optimal},
      inner color=SpringGreen!20!white,
      outer color=SpringGreen!20!white,
      ]
      [\textit{Cost Metrics}\\\cite{wang2016towards,lei2016moving,taylor2016automated, sengupta2018moving, albanese2012time, alavizadeh2018evaluation, han2017evaluation},
      inner color=SpringGreen!20!white,
      outer color=SpringGreen!20!white,
      ]
    ]
    [Security Metrics,  outer color=Salmon!10!white
, inner color=Salmon!10!white
,
     [\textit{Policy 
      Conflict Analysis}
      \\\cite{chowdhary2016sdn,pisharody2017brew,pisharody2016security,hamed2006taxonomy,khurshid2013veriflow, chowdhary2018markov}, outer color=Salmon!25!white
, inner color=Salmon!25!white
,
      ]
      [\textit{Attack Graph/Tree}\\\cite{chung2013nice,sheyner2003tools,schneier1999attack,chowdhary2016sdn,miehling2015optimal, nguyen2018multistage, wang2008attack,roy2012attack}, outer color=Salmon!25!white
, inner color=Salmon!25!white,
      ]
       [\textit{Risk Metrics}\\\cite{hong2016assessing,hong2013scalable,schiffman2005complete,zhu2013game, chowdhary2018markov, chowdhary2018mtd, chung2013nice}, outer color=Salmon!25!white
, inner color=Salmon!25!white,
      ]
      [{\em CIA Metrics}\\\cite{zaffarano2015quantitative, chowdhary2016sdn,connell2017performance,chowdhary2018mtd, thompson2014multiple,crouse2012improving}, outer color=Salmon!25!white
, inner color=Salmon!25!white, 
      ]
    ]
  ]
\end{forest}
}
    \caption{Categorization of MTDs based on the fine-grained quantitative metrics, under the broader umbrella of security and usability metrics, they use for evaluation.}
    \label{fig:quantitative}
    % \vspace{-1.6em}
\end{figure}
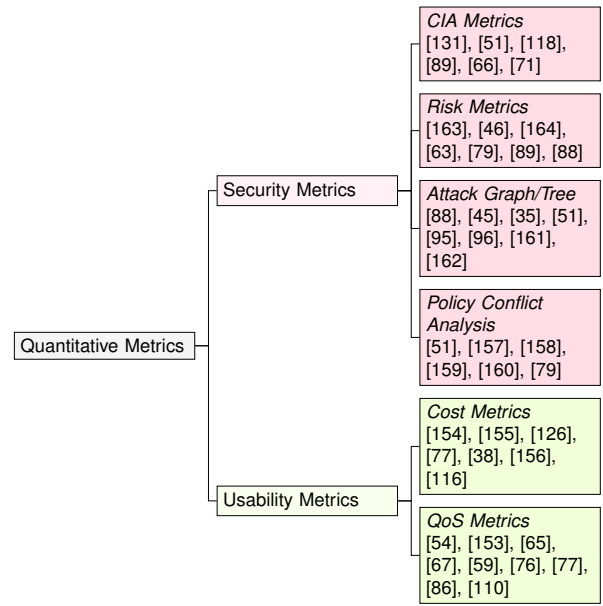

\paragraph{Considers only performance of the ensemble}
Multiple works try to capture the performance costs that result because of the dynamics involved is shuffling between the different configuration but do not look at the  \emph{performance impacts} resulting because of a particular \emph{bad constituent configuration}. In \cite{zhu2013game} and \cite{sengupta2017game}, the authors consider the \emph{one-step cost} of switching from one defense action to another one and seek to find a strategy that minimizes this. 

Other works, instead of accounting for the performance impact of the ensemble, compare an MTD defense to a static system configuration by measuring usability metrics such as latency, availability to legitimate users, {\em etc.} In \cite{dunlop2011mt6d}, authors notice an overhead of 40 bytes for the IPv6 header and latency of 12ms during address change as opposed to 3ms when MTD is not implemented on the network packets. They point out that a more efficient implementation might help in reducing this gap. On similar lines, authors in \cite{ahmed2016mayflies} highlight that creating an entire file system replica takes two more minutes in the case of an MTD system in comparison to a non-MTD enabled system refresh. On the contrary, \cite{bardas2017mtd} does not notice negligible performance overhead for instance replacements in a $14$ node system (where one is a controller node and the others are compute nodes). Although, they do notice a larger number of {HTTP error packets} on the wire when using the MTD. To ensure that the usability to legitimate users is not impacted at all, authors in \cite{thompson2014multiple} keep multiple systems running with the different configurations. At every switch, they simply pick the system that serves the request. On the downside, they incur the cost of maintaining multiple services (at least two). An interesting side effect of using MTD in \cite{algin2017mitigating}, mainly done to thwart selective jamming attacks, is the reduction of delay in transmitting packets over the network. Existing approaches that schedule packet delivery in a deterministic way land up in packet collision scenarios. The random schedule selection in each round is shown to reduce the number of collisions, improving end-to-end (ETE) packet delivery time.

\paragraph{Considers both} A small section of works either consider or evaluate their MTD in regards to both {performance} of {single constituent} defenses and the {ensemble}. In \cite{chowdhary2017dynamic}, the authors model the performance {costs} associated with {each defense action} as a part of the utility values and consider the {shuffle cost} as a part of the rewards the attacker and a legitimate user gets in a repeated game setting. On similar lines, in the empirical analysis of MTDs in the context of {Flip-it} games \cite{prakash2015empirical}, the authors model the cost of each defender action as a part of {defender's utility} value while the state of the system after a move action considers the number of {servers} being {controlled} by the {defender}-- an indirect way of measuring system performance. Authors in \cite{el2016softwarel} consider the availability of each defense action over a \emph{bounded time horizon} and for \emph{shuffling costs}, consider the downtime or unavailability that results when the system is migrating from one system configuration to another. On similar lines, the authors of \cite{debroy2016frequency} model the \emph{host capacity} and \emph{network bandwidth} associated with each system configuration and also determine the next configuration based on the performance of other defenses in the previous time steps. For some works, the performance costs associated with the ensemble do not arise due to the shuffle but represent the cost of ensuring exact same performance across defense configurations \cite{wang2017u} or cost of implementing a system that can support the various configuration \cite{clark2015game}. More specifically, authors in \cite{wang2017u} create a system that ensures that the logical virtual identifier distances between any two nodes remain the same, while in \cite{clark2015game}, the authors assume an extra cost of creating an ensemble that is not necessary when not using MTD solution.

\subsection{Quantitative Metrics}
 
%\hl{References to be added}
Although we realize that the quality of MTD is key is determining the quality of defense being offered, quantifiable measures about these qualitative terms are essential for effective evaluation.
For example, shuffling \emph{x \%} number of hosts leads to \emph{y \%} reduction in attack success probability, and \emph{z \%} increase in overhead/quality of service (QoS) for the normal user. In this sub-section, we present the quantitative analysis of MTD research works discussed in the survey. We categorize quantitative analysis into \emph{Security Metrics} and \emph{Usability Metrics} as shown in Figure~\ref{fig:quantitative}.

%\begin{itemize}
%    \item \textbf{Security Metrics} as shown in the Figure~\ref{fig:quantitative}, covers attack quantification using CvSS metrics - \emph{Confidentiality}, \emph{Integrity}, and \emph{Availability} (CIA). This metric also considers \emph{attack representation methods} (ARMs) Attack Graphs and Attack Trees. We discuss MTD research works from the perspective of security metrics in next sub-section.  
    
    %security metrics also cover the \emph{risk analysis} 
    
    %\item \textbf{Risk Analysis} - This aspect of qualitative analysis considers \emph{cost-benefit} modeling of MTD research works. MTD research works that use \emph{probabilistic modeling} or \emph{game theoretic modeling} mechanisms are discussed in this subsection. We consider \emph{reward/utility} modeling for MTD research works in particular, under this category.
    
%    \item \textbf{Usability Metrics} - This category as shown in the Figure~\ref{fig:quantitative}, analyzes MTD research work from the aspects such as \emph{QoS} (network bandwidth, delay), \emph{impact} on existing mission metrics and the \emph{cost} of deploying MTD defense.
    
%\end{itemize}

%\begin{itemize}
%    \item \textbf{Security Metrics:} 
%    \item \textbf{Risk Analysis:}
%    \item \textbf{Utility Metrics:} 
%\end{itemize}
% \usepackage{multirow}

\begin{figure}[t]
\centering
\begin{tikzpicture}
    \pie [polar, rotate=90, scale font, explode=0.2, style = drop shadow,
            color = {CornflowerBlue!30!white, SpringGreen!30!white, Goldenrod!30!white, BurntOrange!30!white, Salmon!30!white, NavyBlue!30!white}]
    {32/,%Reconnaissance,
     22/,%Vulnerability Exploitation,
     11/,%DoS or DDoS,
     11/,%Multi-stage attacks,
     10/,%APT \& Data Exfiltration,
     14/}%Others}
\end{tikzpicture} 
\vspace{-0.7em}
\end{figure}
\begin{table}[t]
% \small
\centering
    % \resizebox{\linewidth}{!}{%
\begin{tabular}{p{1.8cm}p{6cm}}
\rowcolor{CornflowerBlue!30!white} \textbf{Reconnaissance} & \cite{jafarian2012openflow} \cite{al2012random} \cite{connell2017performance}  \cite{farris2015quantification} \cite{cai2016moving} \cite{zhao2017sdn} \cite{wang2016towards}   \cite{schlenker2018deceiving}  \cite{manadhata2013game}     \cite{prakash2015empirical} \cite{crouse2012improving} 
\cite{shi2017chaos}  \cite{chowdhary2018mtd}   \cite{luo2015rpah}
\cite{al2011toward} \cite{maleki2016markov}   \cite{miehling2015optimal}   \cite{nguyen2018multistage}  
\cite{kampanakis2014sdn} \cite{hong2017optimal} \cite{wang2017u}  \cite{dunlop2011mt6d}                               \\
\rowcolor{SpringGreen!30!white} \textbf{Vulnerability Exploitation }         &                       \cite{vadlamudi2016moving} \cite{sengupta2017game}      \cite{carter2014game} \cite{el2016softwarel} \cite{neti2012software} 
\cite{thompson2014multiple} \cite{bohara2017moving}  \cite{nguyen2018multistage} 
\cite{christodorescu2011end}  \cite{han2017evaluation} \cite{zhuang2012simulation} \cite{ahmed2016mayflies}     \cite{bardas2017mtd}  
\cite{xu2014comparing} \cite{homescu2017large}                          \\
\rowcolor{Goldenrod!30!white}
\textbf{(D)DoS Attack}         &          \cite{steinberger2018ddos}   \cite{aydeger2016mitigating}   \cite{cai2016moving}  \cite{debroy2016frequency}       \cite{chowdhary2017dynamic} 
\cite{luo2015rpah}   \cite{al2011toward}  \cite{wang2016towards}       \\
\rowcolor{BurntOrange!30!white} \textbf{Multi-Stage Attack}                                    &  \cite{wang2008attack} \cite{chowdhary2018markov} \cite{chowdhary2016sdn}   \cite{lei2017optimal}    \cite{chowdhary2018adaptive}
\cite{zhuang2012simulation}  \cite{hong2014scalable}  \cite{zhuang2013investigating}                           \\
\rowcolor{Salmon!30!white} \textbf{APT and Data Exfiltration}                                                  &      \cite{van2013flipit} \cite{taylor2016automated} \cite{shu2018ensuring} \cite{rass2016gadapt} \cite{rass2017defending}    \cite{taylor2016automated} \cite{venkatesan2016moving}                   \\
\rowcolor{NavyBlue!30!white} \textbf{Other types of Attack}                                 &         \cite{clark2015game}    
\cite{chowdhary2018mtd} \cite{chung2013nice} \cite{colbaugh2012predictability}  \cite{sengupta2018moving}  
\cite{taylor2016automated} \cite{venkatesan2016moving}  \cite{sengupta2018mtdeep} \cite {albanese2013moving}                   \\
\end{tabular}
% }
\caption{Threat Models considered in the MTDs proposed in the surveyed papers.}
\label{tab:6}
% \vspace{-1.8em}
\end{table}

\subsubsection{\textbf{Security Metrics}}
An important aspect of network defense is representation and visualization of network attacks. Enterprise networks are becoming large and complex with different network overlay and underlay technologies. The adage \emph{what can't be measured cannot be effectively managed} applies aptly here. Security metrics such as the ones shown in the Figure~\ref{fig:quantitative}, covers attack quantification using CVSS metrics-- Confidentiality, Integrity, and Availability (CIA). This metric also considers attack representation methods (ARMs)-- Attack Graphs and Attack Trees. We now discuss the MTD research works from this perspective of security metrics.

We surveyed $\sim 70$ MTD research articles and analyzed them from the perspective of different threats models they targeted. A summary of the findings has been provided in Table~\ref{tab:6}. The key observations are as follows a) $32 \%$ of research works focus on defense against {reconnaissance} attempts, b) $22 \%$ of these research works target {vulnerability exploitation}, c) $11 \%$ research works use MTD defense against {DoS/DDoS} attacks. There has been a limited focus on using MTD defense for dealing with {stealthy attacks} like {APT} and {data-exfiltration} (7/68 $\sim 10 \%$ research works). Also, some papers used other types of attacks such as timing-based attacks \cite{clark2015game}, and network intrusions \cite{chowdhary2018mtd,chung2013nice,colbaugh2012predictability, sengupta2018moving}.

%\begin{figure}[h]
%    \centering
%\begin{tikzpicture}
%\pie [rotate = 180, text=inside]
%    { 50/ recon~\cite{jafarian2012openflow, connell2017performance, zhao2017sdn, schlenker2018deceiving, manadhata2013game}\\
%    ~\cite{prakash2015empirical, crouse2012improving, chowdhary2018mtd, shi2017chaos, miehling2015optimal}
%    , 12/ ddos~\cite{aydeger2016mitigating, el2016softwarel, debroy2016frequency, chowdhary2017dynamic}, 
%     38/ multi-stage and APT\\ ~\cite{taylor2016automated, shu2018ensuring, chowdhary2016sdn, lei2017optimal, venkatesan2016moving, chowdhary2018markov}\\
%     ~\cite{zhu2013game, van2013flipit,chowdhary2018adaptive}
%     }
%\end{tikzpicture}
%\caption{Quantitative Overview of Threat Vectors Targeted by MTD Research works}
%    \label{fig:quant2}
%\end{figure}

\textbf{CIA Metrics:} \emph{Confidentiality}, \emph{Integrity}, and \emph{Availability} (CIA) are used as quantitative metrics for measurement of impact on system under attack. Zaffarano \textit{et al.}~\cite{zaffarano2015quantitative} use confidentiality metric for measuring information exposed by modeling tasks. The mission \emph{M} confidentiality valuation \emph{v} is expressed as, \emph{Conf (M, v)} = $\frac{1}{|T|} \sum_{t \in T}(t, unexposed)$. As the measurement equation suggests, the goal is to maintain information as {unexposed} over time ($t \in T$). Similarly \emph{Integrity} valuation \emph{v} for mission $M$ is quantified as $Int (M, v) = \frac{1}{|T|} \sum_{t \in T}(t, intact)$. High integrity is ensured by keeping information intact over time. 

Conell et al. in~\cite{connell2017performance} consider {availability} as an important metric for analyzing impact of MTD countermeasure. The system reconfiguration rate $\alpha$ is modeled as a function of system resources using {Continuous Time Markov Chain} (CTMC) modeling. The analysis of the effect of reconfiguration on the availability is considered for fine-tuning MTD decision. MASON~\cite{chowdhary2018mtd} framework utilizes {Intrusion Detection System} (IDS) alerts and vulnerability score \emph{CVSS} calculated on the basis of CIA values to identify critical services in the network. A \emph{Page Rank} based threat scoring mechanism is utilized for combining static and dynamic information, and prioritizing network nodes for MTD countermeasure {port hopping}. It is noteworthy, that port-hopping for $40-50\%$ services can help reduce overall threats in the network by $97\%$. This research work, however, does not consider the usability impact induced by the MTD countermeasure. 

{\em MORE} \cite{thompson2014multiple} framework shows reduction in reconnaissance attempts and exploits targeting software integrity violation on frequent rotation of Operating Systems (OS) of the network under attack. Experimental analysis shows that a rotation window of $60$s makes \emph{nmap fingerprinting} attempts ineffective. {Temporal} and {spatial} diversity have been used to introduce genetic algorithm based MTD by Crouse \textit{et al.}~\cite{crouse2012improving}. The experiments on {average vulnerability} of different configurations show decaying vulnerability rates for evolved configurations selected from the {chromosome pool} of VM configurations.

\textbf{Attack Graph/ Attack Tree:} CVSS present only a piece of quantitative information about the vulnerabilities such as complexity of performing network attack, impact on confidentiality or integrity of the system if the attack is successful, {\em etc.} This information alone is not sufficient for taking MTD decisions. {Attack representation methods} (ARMs) such as \emph{Attack Graph}~\cite{sheyner2003tools} and \emph{Attack Tree}~\cite{schneier1999attack} answer questions such as (a) What are the possible {attack paths} in the system (b) What attack paths can be taken by the attacker to reach a specific target node in the network. SDN based scalable MTD solution~\cite{chowdhary2016sdn} makes use of attack graph-based approach to perform the {security assessment} of a large scale network. Based on the security state of the cloud network, MTD countermeasures are selected. 

%\begin{figure*}
%    \centering
%    \includegraphics[width=\textwidth]{figures/system-modules.pdf}
%    \caption{System modules and operating layers of an MTD-based solution using SDN. The system comprises of quantitative metric based on vulnerabilities, attack graph at overlay network and policy conflict detection at the logical network layer.}
%    \label{fig:10}
%\end{figure*}

%The Figure~\ref{fig:10} shows system modules and operating layers, which are the part of SDN based MTD framework. The {overlay network} is responsible for vulnerability analysis, attack graph generation. The {physical network} consists of Open vSwitch (OVS), running on top of the physical server. The SDN controller interacts with OVS using {OpenFlow APIs}. 

{Bayesian attack graphs} have been used by Miehling \textit{et al.}~\cite{miehling2015optimal} for defending the network against vulnerability exploitation attempts. The defender's problem is formulated as a {Partially Observable Markov Decision Process} (POMDP) and {optimal defense} policy for selecting countermeasures is identified as a solution for the {POMDP}. 

The security analysis of a large-scale cloud network in real time is a challenging problem \cite{sabur2019s3}. Attack Graphs help in identification of possible attack scenarios that can lead to exploitation of vulnerabilities in the cloud network. The attack graphs, however, suffer from scalability issues, beyond a few hundred nodes as we discussed in Section~\ref{sec:background}. 

\textbf{Risk Metrics:} 
Like any other system, MTD systems have an associated risk once an organization considers deploying whole or part of the MTD technique. According to the \emph{National Institute of Standards and Technology} (NIST), there are several attacks, service disruptions, and errors caused by human or machines that may lead to breaking benefits and critical assets at the organization or national level. Risks assessment is a critical measure and has many ways to deploy and use. In this survey, we identify and highlight research work that has adopted and took into consideration risks associated with deploying MTD solution. Specifically, we emphasize on the research work that evaluates the {cost} of the adopted MTD solution, since system administrators need to take into account the cost of using the MTD solution. 

Risk assessment in the MTD has a direct and strong {relationship} with the {effectiveness} of the deployed MTD technique. According to \cite{hong2016assessing}, the system administrator can determine how good the MTD solution is by examining the associated risk. Therefore, the authors \cite{hong2016assessing} studied the effect of deploying each MTD technique alone (i.e. shuffle, diversity, and redundancy) by inspecting the Hierarchical Attack Representation Method (HARM). {HARM} is basically a \textit{ARM} at the upper layer such as attack graph (AG), and another ARM in the lower layer such as attack tree (AT), where these to ARM have a one-to-one mapping between them. Moreover, the authors also studied the associated risk by computing the {instance measure} (IM) which uses vulnerability's {base score}, {impact score}, etc~\cite{hong2013scalable}.

Feedback-driven multi-stage MTD has been proposed by Zhu \textit{et al.}~\cite{zhu2013game} for dealing with multi-stage attacks like Stuxnet~\cite{langner2011stuxnet}. Author's quantify the damage or cost caused by an attacker at different stages of the network. The game between attacker-defender is modeled as a {finite zero-sum} matrix game with a bounded {cost function}, and a {mixed-strategy} {Saddle Point Equilibrium} (SPE). Players utilize cost-function learned online to update MTD strategies. The numerical results show that {feedback mechanism} allows network defense to respond to unexpected (exogenous) events and reduce unusual peak of {risk} imposed by different vulnerabilities.

In~\cite{taylor2016automated}, the authors provide {metrics} for MTD evaluation and risk analysis. For risk metrics, they proposed statistical metrics to study the effect of how the attacker can quickly conduct and succeed in adversarial attacks. The authors assumed the system will always have a running task that can be measured. The validity of the metrics was studied by simulating the APT attack scenario, where they assumed the APT will always have some sort of overhead that can be measured and detected. Finally, the {performance} of the proposed metrics was measured also by examining the {CPU utilization} of the designed system.

Cheng \textit{et al.} \cite{lei2017optimal} consider the game theoretical formulation of MTD systems; specifically, they model it as a {Markov Game}. The authors provided a theorem, subject to probabilistic constraints, to calculate the revenue for the defensive and offensive approaches in MTD systems. Their work depended on testing different defensive and offensive strategies and was tested using a networking setup that includes vulnerable services and a {firewall} component as well.

Another work considered the {statistical approach} to evaluate the likelihood of a successful attack is the work conducted by \cite{venkatesan2016moving}. The authors proposed an approach to determine the minimum effort required a system to detect stealthy botnets. Moreover, the {entropy} was measured to determine how close an adversary is to the detection point, where {high entropy} indicates the {attacker} is far in {distance} from the detector. The evaluation of the proposed approach was conducted on a real ISP network obtained from \cite{spring2002measuring}. The results of the proposed algorithm show that the detection strategy has a {complexity} of $O(N^3)$ although theoretical complexity analysis indicates the algorithm is $\sim$ $O(N^6)$. 

Chung \textit{et al.} \cite{chung2013nice} provide a detailed and comprehensive evaluation for the optimal countermeasure selection over a set of vulnerable attack paths in the attack graph. By evaluating the CVSS score of vulnerabilities, the authors determined the countermeasure selection option, taking into account the ration of the {Return of Investment} (ROI). Countermeasure option that produces the smallest ROI is considered the optimal one. The system performance of \textit{NICE} \cite{chung2013nice} is proven to be efficient in terms of network delay, CPU utilization, and the traffic load. 

To study the effect of a number of intrusions on the system's threat, Chowdhary \textit{et al.}~\cite{chowdhary2016sdn} use a {statistical approach} also to evaluate how the number of intrusion in the system, with the number of vulnerabilities, affect the threat score. The threat scoring algorithm is similar to \textit{Page Rank} algorithm. To evaluate the proposed work, two experiments were conducted. One experiment is to test the threat scoring engine on {software vulnerabilities} and {IDS alerts}. The second experiments study the {effect} of {port hopping} attack. The results show that as the number of services in the system increase, the service risk value remains unchanged between a specified interval. However, the first few services show an increase in the number of risk value. Finally, the port hopping attack showed a reduction in threat score. 
%This quantitative metric considers \emph{cost-benefit} modeling of MTD research works. MTD research works that use \emph{probabilistic modeling} or \emph{game theoretic modeling} mechanisms are discussed in this subsection. We consider \emph{reward/utility} modeling for MTD research works in particular, under this category.

\textbf{Policy Conflict Analysis:} The MTD countermeasures such as network address switching can dynamically and rapidly insert new type of traffic, or new {flow rules} (environment managed by SDN). Pisharody \textit{et al.}~\cite{pisharody2017brew, pisharody2016security} show how different MTD countermeasures such as {network address} change, {load-balancing}, and {intrusion detection} can cause {security policy violations}. The research works discuss {SDN-based MTD}, but the {policy conflicts} can cause {security violations}~\cite{hamed2006taxonomy}, {loops}, and {blackholes} in the network as discussed by Khursid \textit{et al.}~\cite{khurshid2013veriflow}. The violations of {network-wide} in-variants and {security policies} must be analyzed before deployment of MTD countermeasure. This research challenge is an important {quantitative metric}, which has not been considered by a lot of research works. We discuss this as a possible {research oppurtunity} in Section~\ref{research}.

%The change in checks the possible conflicts in the OpenFlow rules after MTD countermeasure deployment. Six types of security conflicts, i.e., (i) Redundancy (ii) Generalization (iii) Correlation (iv) Shadowing (v) Overlap and (vi) Imbrication are identified based on overlap in match and action fields of flow rules. The flow rule extraction, conflict detection, and resolution happen at the logical layer in the cloud network. The conflicted flow rules are identified and corrected in order to provide security compliance in the SDN environment post-MTD countermeasure.

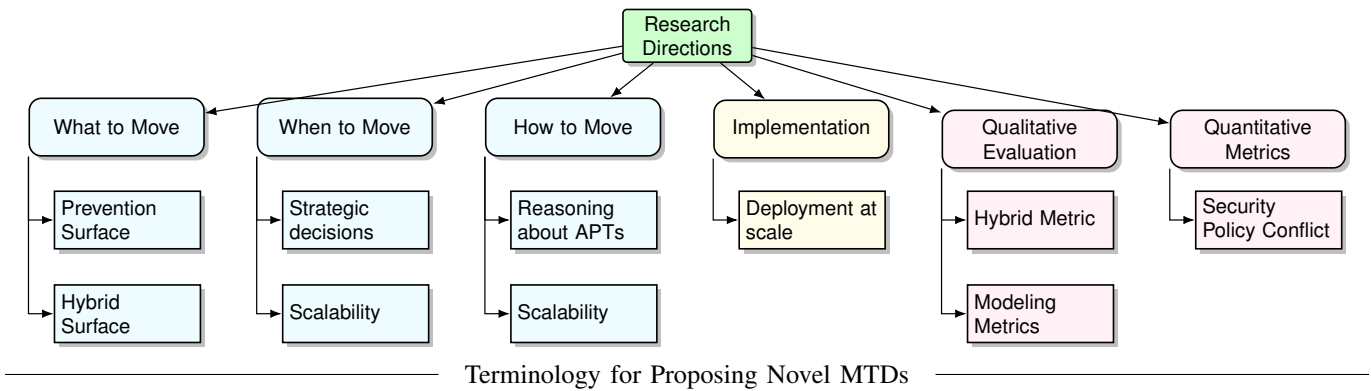
\begin{figure*}[htp]
\centering
\resizebox{2.0\columnwidth}{!} {
    \begin{forest}
        for tree={%
            thick,
            drop shadow,
            l sep=0.6cm,
            s sep=0.9cm,
            node options={draw,font=\sffamily},
            edge={semithick,-Latex},
            where level=0{parent}{},
            where level=1{
                minimum height=1cm,
                child,
                parent anchor=south west,
                tier=p,
                l sep=0.1cm,
                for descendants={%
                    grandchild,
                    minimum height=1cm,
                    anchor=110,
                    edge path={
                        \noexpand\path[\forestoption{edge}]
                        (!to tier=p.parent anchor) |-(.child anchor)\forestoption{edge label};
                    },
                }
            }{},
        }
    [Research Directions,
        [What to Move
          , inner color=CornflowerBlue!10!white
          , outer color=CornflowerBlue!10!white
            [Prevention Surface
            , inner color=CornflowerBlue!10!white
            , outer color=CornflowerBlue!10!white
                [Hybrid Surface
                , inner color=CornflowerBlue!10!white
                , outer color=CornflowerBlue!10!white
                ]
            ]
        ]
        [When to Move
        , inner color=CornflowerBlue!10!white
        , outer color=CornflowerBlue!10!white
            [Strategic decisions
            , inner color=CornflowerBlue!10!white
            , outer color=CornflowerBlue!10!white
                [Scalability
                , inner color=CornflowerBlue!10!white
                , outer color=CornflowerBlue!10!white
                ]
            ]
        ]
        [How to Move
        , inner color=CornflowerBlue!10!white
        , outer color=CornflowerBlue!10!white
            [Reasoning about APTs
            , inner color=CornflowerBlue!10!white
            , outer color=CornflowerBlue!10!white
                [Scalability
                , inner color=CornflowerBlue!10!white
                , outer color=CornflowerBlue!10!white
                ]
            ]
        ]
        [Implementation
            , inner color=Goldenrod!10!white
            , outer color=Goldenrod!10!white
            [Deployment at scale
            , inner color=Goldenrod!10!white
            , outer color=Goldenrod!10!white
            ]
        ]
        [Qualitative Evaluation
        , inner color=Salmon!10!white
        , outer color=Salmon!10!white
          [Hybrid Metric
          , inner color=Salmon!10!white
          , outer color=Salmon!10!white
            [Modeling Metrics 
            , inner color=Salmon!10!white
            , outer color=Salmon!10!white
            ]
          ]
        ]
        [Quantitative Metrics
          , inner color=Salmon!10!white
          , outer color=Salmon!10!white
              [Security Policy Conflict
              , inner color=Salmon!10!white
              , outer color=Salmon!10!white
              ]
        ]
    ]
    \end{forest}
}
\headline{Terminology for Proposing Novel MTDs}
 
\caption{Our categorizations help in identifying various directions that have been under explored during the development, implementation and evaluation of Moving Target Defenses (MTDs). In addition, it also provides a terminology that can help to underlying identify assumptions made by existing work and describe future MTDs.}
\label{fig:fw}
% \vspace{-1.6em}
\end{figure*}

\subsubsection{\textbf{Usability Metrics}} 
This category as shown in the Figure~\ref{fig:quantitative}, analyzes MTD research work from the aspects such as {QoS} (network bandwidth, delay), {impact} on existing mission metrics and the {cost} of deploying MTD defense.

\textbf{QoS Metrics:} MTD can induce some performance cost on the existing system resources. Jafarian \textit{et al.}~\cite{jafarian2012openflow} identify the {virtual IP} (vIP) {mutation}, {range allocation}, and {range distribution} constraints in order to minimize the QoS impact which can be induced by {vIP} collisions, as well as, maintain optimal-level of unpredictability.
Probabilistic performance analysis of MTD reconnaissance defenses has been conducted by Crouse \textit{et al.}~\cite{crouse2015probabilistic}. The research work analyzes quantifiable MTD metrics such as {reconnaissance, deception} performance, \emph{attack success probability} vs \emph{connection drop probability}, attacker's success probability under different conditions such as {network-size}, {number of vulnerable computers}. 

Taylor et al. in~\cite{taylor2016automated} use mission and attack metrics for analyzing the effectiveness of a network defense. The research work analyzes dynamic defenses such as {\em Active Re-positioning in Cyberspace for Synchronized Evasion} (ARCSYNE) and {\em Self-shielding Dynamic Network Architecture} (SDNA) using mission and adversary activity set. Mission \emph{Success}, i.e., the rate at which mission tasks are completed, and Mission \emph{Productivity}, i.e., how often are mission tasks successful are used as {QoS} measurement metrics are used for evaluations.

A statistical analysis of static {\em vs.} dynamic attacks against different MTD strategies-- uniform, random, diversity-based, evolution-based, and optimal-- has been conducted by Carter \textit{et al.}~\cite{carter2014game}. Experimental results on performance {\em vs.} adaptability shows that diversity-based MTD is the optimal strategy against most attack scenarios. They also show that uncertainty about the adversary type-- slow adversary or fast-evolving adversary-- can adversely impact the effectiveness of an MTD.

El-Mir~\cite{el2016softwarel} models {performance parameters} such as {availability}, {downtime}, and {downtime cost} using a {Continuous Time Markov Chain} (CTMC) model. The experimental results showed that cost-effective {VM migration} can be performed in an SDN-based network with {limited impact} on {network performance}. The research work utilizes normalized CVSS score as a key metric for initiating VM migration. 

Sengupta \textit{et al.}~\cite{sengupta2018moving} analyze the performance impact of placing IDS (NIDS and HIDS) at all possible enforcement points in a cloud network. It is noteworthy that the placement of more than $15$ detection agents in their simulated network fails to provide any additional intrusion detection benefit, whereas the network throughput decreases drastically from $16$ Gbps in the case of a single detection agent to $\sim 6$ Gbps when $15$ detection agents are placed. 

CHAOS~\cite{shi2017chaos} analyzes how the delay intentionally introduced by MTD impacts the packet count in a SDN-managed network. The SDN controller utilizes host-mutation and {decoy-severs} to deceive an adversary. The obfuscated network increases the {cost} and {difficulty} for an adversary targeting the network. The percentage of {information disclosure} reduces from $90\%$ to $10\%$ in a CHAOS protected network, with slight impact on packet delay ($1$s to $1.5$s for $1800$ packets). 

\textbf{Cost Metrics:} Protection against Distributed Denial of Service (DDoS) attacks is one of the important priorities for many cyber systems. Wang \textit{et al.} \cite{wang2016towards} present a cost-effective MTD solution against DDoS and {\em Covert Channel} attacks. Through MTD adaptation, their work \cite{wang2016towards} aims to answer two main questions: 1) what is the adaptation cost?, and 2) what is the cost incurred by a defender if an attacker succeeds in exploiting a particular vulnerability?. This solution does not rely on IDS-generated alerts while making the adaptation. The adaptation cost includes any cost related to purchasing required software or hardware that helps in the adaptation process. Lei \textit{et al.}~\cite{lei2016moving} utilize change-point analysis method for evaluation MTD cost-benefit for a multi-layer network resource graph. The proposed method analyzes mission productivity ($\Delta M$), and attack success productivity ($\Delta A$) on dynamic network address translation (DNAT). The evaluation results show reduced attack success probability using DNAT over a network under observation. The path enumeration mechanism used in this research work can, however, suffers from scalability challenges because of frequent path probability calculation and update operations. The cost and effectiveness evaluation of reactive and proactive network defense strategies, has been conducted by Taylor \emph{et al.}~\cite{taylor2016automated} using Measurement of Effectiveness (MOE) metrics. The research work considers hop-delay for different attack success rates, and static defense policies. They show that an attacker's productivity, i.e., how quickly attacker can perform adversarial tasks increases against static defense, whereas attacker's confidentiality, i.e., ability to remain undetected is same for both the static and the dynamic defense case. 

\section{Research Opportunities}
\label{research}

In this section, we highlight some lessons learnt and the under explored aspects of Moving Target Defenses (MTDs). These lead to a discussion on promising directions for future research (a brief summary is highlighted in Figure \ref{fig:fw}).

\subsection{The Configuration Set (What to move?)}

Works in MTD have concentrated mostly on the movement of the exploration, the detection and the attack surface. While moving the exploration and the attack surface helps a defender take away the advantage of reconnaissance that an attacker has, the goal of moving the detection surface is to improve the scalability and Quality of Service (QoS).

The movement of the prevention surface, which comprises of security modules such as Firewalls, IPS {\em etc.} has only been investigated by a couple of works (see discussion in Section~\ref{prev}). An MTD research can consider exploring {Next-Generation Firewall} (NGFW) architectures that combine security modules such as {firewall}, {content filter}, {anti-virus} {\em etc.} to provide a multi-layered defense-in-depth solution. Some current implementations of NGFW, that can be leveraged for testing the effectiveness of these defenses, are \emph{Cisco ASA}~\cite{frahim2014cisco} and \emph{PAN Firewall}~\cite{panfirewall}. 
{
Further, with the rise of mobile technologies, MTDs can prove to be effective defenses.
Although we discuss a few works on thwarting jamming attacks \cite{algin2017mitigating} and identity shifting in mobile ad-hoc networks \cite{albanese2013moving}, MTDs for different surfaces of in mobile infrastructure networks can be a promising direction for future research.}

As previously stated, the movement of different surfaces in a single framework, although challenging, can provide greater security benefits than the movement of a single surface. Investigation in this area would require one to identify sets of configurations across the various surfaces that are compatible (in terms of performance) with one another. This prevents the number of strategies from multiplying uncontrollably by leveraging the expertise of system designers.

The categorization of the various software surfaces opens up the possibility of considering other logical surface level distinctions and thus, MTDs that shift these surfaces. For example, \textit{Microsegmentation}~\cite{mammela2016towards} is a method of creating {secure zones} in data centers and cloud deployments to isolate workloads from one another and secure them individually. With MTD formalism, one can test the existing hypothesis and develop new ones for microsegmentation. We believe that formal modeling, in line with \cite{chowdhary2018markov}, one might discover that advanced services will be more effective when applied at a {granular-level} (as close to the application as possible in a distributed manner).

% ====
% The content of the following sentence is discussed in the hybrid surface shifting paragraph. Thus omitting
% =====
%An MTD mechanism which explores \emph{dynamic} and \emph{flexible} security policy composition based on the current situation of the cloud network can be considered as a potential research direction for \emph{prevention surface shifting}.

\subsection{The timing function (When to Move?)}

{
The timing problem has mostly been ignored in previous works on MTDs. While some works perform empirical studies to test the best (constant) time-period for switching, they can be highly specific to threat model and the elements being shifted. For example, while 15 second time-periods are shown to be reasonable when protecting against jamming attacks \cite{algin2017mitigating}, 60 seconds time periods are needed to defend against Network Mapping attacks \cite{lyon2009nmap} attacks. Also, the complexity of changing the virtual IP address vs. the underlying virtual machine impose different constraints on the lower bound on feasible time-periods.
}

{
Note that the handful of works that empirically determine time-periods are by no means complete. We need an extensive study just to come up with reasonable time periods for particular surfaces, how it effects that attack model and provide guidelines on efficient implementation methods.
}

On the other hand, a few existing approaches that do address the timing problem theoretically, suffer from scalability issues. In \cite{lei2017optimal}, the authors land up increasing number of states in their Markov Game by including time as a parameter. Inferring the defender's strategy in such Markov Games cripples the MTD to work beyond small networks. Effective solutions or improved modeling could both be interesting research directions for the future.

\subsection{The Movement Function (How to Move?)}

Randomization in the movement is a necessary part of effective MTDs. Given the extensive use of randomization in cryptography, MTD defenses tend to use a Uniform Random Strategy (URS) an the movement strategy \cite{zhuang2014towards}. Several works have argued that often the defender has performance information about their system and known attacks that can be used to exploit their system. In such cases, a game-theoretic modeling can result in better strategies that offer higher gains in terms of both security and performance metrics. Unfortunately, the latter works suffer from scalability issues, encouraging practitioners to default to URS. Works that can leverage existing knowledge without suffering from scalability issues can fill an important gap in MTD research.

Recently, it has been shown that attacks on networks are better characterized by persistent approaches such as APTs. In such scenarios, attackers are known to demonstrate sequential attack behavior spread over a long time.
A relatively new line of work investigates modeling MTDs to come up with strategies that are effective against APTs. Although these approaches leverage the use of attack graphs to bootstrap the modeling process \cite{rass2016gadapt, rass2017defending}, they do not scale well. While real-world cloud services host thousands of nodes, researchers have only considered small-scale settings \cite{chowdhary2018markov,senguptageneral}. Furthermore, works such as \cite{miehling2015optimal} and \cite{nguyen2018multistage} that try to consider partial observability limit the scalability even further. Study on the design of approximation approaches and their effectiveness when compared to scalable baselines such as URS will be key for future research.

Current research often makes strong assumptions about the threat model. This makes results regarding the effects of such defenses questionable. In the future, we hope to see more studies that try to figure out realistic attack scenarios. Figuring out an attacker's behavioral model might also lead the modeling community to relax assumptions often made by rationality of an attacker.

\subsection{Qualitative and Quantitative Evaluation}

{
As seen in Figure \ref{fig:qual}, none of the existing works demonstrate either empirically or model the impact of a proposed MTD on all four types of metrics. The lack of testing against real-world attackers also makes it difficult to prioritize which metrics a defender needs to care about and to what extent.
}

% Although ignoring certain metrics for evaluation may result in degradation of performance or security of an actual deployed MTD, hardly any papers deploy their system against real-world attackers . Moreover, coming up with an MTD that incorporates all these metrics when coming up with effective movement strategies should not be difficult because one can survey existing works to realize what metrics to incorporate.

% \begin{figure}[t]
%     \centering
%     \includegraphics[width=\linewidth]{figures/figure11.pdf}
%     \caption{Motivating scenario for addressing policy conflicts in a SDN-managed cloud environment that may arise due to the movement of Moving Target Defenses.}
%     \label{fig:11}
% \end{figure}

%\subsection{Quantitative Evaluation}
 {
Beyond human studies to understand attack behavior, use of MTDs may introduce a new attack surface in cyber-systems.}
For example, firewall filtering rules need to be carefully analyzed to prevent conflicting security policies that may arise at the movement time because such conflicts might result in either dropping legitimate user packets or introducing new attack points. Although some works such as \cite{chowdhary2016sdn} have tried to address this issue of identifying security policy conflict for an SDN-managed cloud network, it is not immediately clear how it can be adapted in the context of other MTDs. A clear idea of when such scenarios arise and finding ways to address them would constitute an important line of research in the future.

A key idea is to have a continuous feedback cycle that verifies the security policy in place post MTD-countermeasure deployment. This can be done by ensuring end-to-end integration and regression test for the various use-cases pertaining to network traffic. Another solution could be to incorporate the policy conflicts that might arise into the modeling of the MTD. This would produce {\em safe} movement policies that forsee the use of MTD as a new attack surface and avoid policy conflicts.

\iffalse
We now highlight an example to motivate such a scenario and then briefly talk about a few works that can be.

In the data center network shown in Figure~\ref{fig:11}, we have \emph{Tenant A} hosting a web farm. Traffic on \emph{TCP port 443} is allowed into the IP addresses for the web servers. When an attack directed against \emph{host A2} is detected, the MTD application responds with countermeasures and takes two actions: 
\begin{itemize}
    \item a new webserver ({host A3}) is spawned to handle the load of host A2; and 
    \item the IP for host A2 is migrated to the Honeypot network and assigned to \textit{host Z}.
\end{itemize}

In order to run forensics and isolate an attacker, the \textit{HoneyPot} network permits all in-bound traffic, but allows no out-bound traffic to other sections of the data center. These actions result in new flow rules being injected into the flow table that 
\begin{itemize}
    \item permits all traffic inbound to the IP that originally belonged to host A2, but now belongs to host Z.
    \item modifies an incoming packet’s destination address from host A2 to host A3 if the source is considered to be a non-adversarial source.
    \item stops all outbound traffic from the IP that originally belonged to host A2 but now belongs to host Z to the rest of the data center.
    \item permits traffic on port 443 to host A3 (not of great importance to our case). The original policy allowing only port 443 to the IP of host A2, and the new policy allowing all traffic to the IP address of host Z are now in conflict.
\end{itemize}
\fi

\subsection{Proposing novel Moving Target Defenses}

An important goal of this survey is to establish a common terminology for MTD researchers. Thus, we try to categorize an array of existing works in this terminology. For example, consider the work \cite{chowdhary2018markov}. This MTD can be categorized as a moving target defense for the movement of detection surfaces with fixed interval switching formulated as a multi-stage game that performs simulation studies of simple use-cases and measures the security and performance of individual defenses in these settings.

An interesting idea would be to turn this goal of ours on its head and explore the design of new MTDs based on the permutations of the various categorization aspects designed in this survey. For example, a hybrid surface shifting MTD that (1) shifts the detection surface and then based on a stochastic environment, shifts the prevention surface, (2) models this problem as a two-step game, (3) considers rewards that incorporate performance of individual actions and security of the ensemble, and (4) showcases experimental results on an emulated testbed is a novel Moving Target Defense.
\section{Conclusion}\label{concl}

In this survey, we looked at various Moving Target Defenses (MTDs) that have been proposed for enhancing network security. First, we categorized them based on {\em what} surfaces these defenses move (the configuration set), {\em when} a move operation occurs (the timing function), and {\em how} they move between the different constituent system configurations (the movement function). In doing so, we highlight how the movement of particular software surfaces is linked to Advanced Persistent Threats; thereby, allowing us to understand how the various MTDs can help thwart real-world sophisticated attacks. The dearth of works that consider the simultaneous movement of different surfaces points to an exciting research direction and possibly, the invention of more effective MTDs. In answering {\em how to move}, we notice that many approaches leverage Artificial Intelligence (AI) methods in general and game-theoretic techniques in particular for crafting intelligent movement strategies.

Second, we discuss how these MTDs can be implemented in practice. We find that the use of centralized technologies like Software Defined Networking (SDN) helps in implementing the MTD countermeasures with limited impact on network performance. We showcase how the surveyed MTDs are implemented in the context of real-world systems ranging from simulation studies to use in commercial products. We highlight the key technologies leveraged by the various MTDs, the layers of the network protocol stack at which an MTD is effective and the level of maturity at which it is implemented. We briefly described a few test-beds that either have been leveraged by existing MTDs and encourage researchers to use them for evaluating the effectiveness of a proposed MTD. We conclude that (1) SDN/NFV is a dominant technology used by MTDs and (2) industrial adoption of MTD solutions is still limited to few application-security products.

Third, we discuss various metrics used for measuring the effectiveness of MTDs. We put forth two categorizations based on whether these metrics (1) consider security and performance impact of the designed system and (2) are used in the modeling phase to generate particular behavior {\em vs.} used for evaluation of the MTD.
%This categorization can serve as a guideline for ensuring seamless adoption and capability maturity
We notice that none of the proposed MTDs consider all the metrics we put forth. One wonders if a defense that models all the proposed metrics can be realized in practice and if so, prove to be better that existing defenses both in terms of performance and security.

Lastly, we highlight areas of network security where the scope of developing MTDs hasn't been investigated. We believe that coming up with defenses which can fill these gaps will improve security and/or performance aspects of various network systems. We conclude by showcasing how our categorization provides a common terminology for researchers to describe existing MTDs and develop future ones.
\section{Acknowledgement}
This research is supported in part by following research grants: Naval Research Lab N00173-15-G017,  AFOSR  grant  FA9550-18-1-0067,  the  NASA  grant NNX17AD06G, ONR grants N00014-16-1-2892, N00014-18-1-2442,  N00014-18-12840,  NSF—US  DGE-1723440, OAC-1642031, SaTC-1528099, 1723440 and NSF—China 61628201 and 61571375 and a JP Morgan AI Research Faculty Award. Sailik Sengupta is supported by the IBM Ph.D. Fellowship. Also, Abdulhakim Sabur is a scholarship recipient from Taibah University through Saudi Arabian Cultural Mission (SACM).
% === Need to find where this goes in === %
   % The application of MTD-based countermeasure to a network can be \textit{random} or \textit{strategic}. The random mutations of current system aim to create an alternate configuration for the current network-setup. OF-RHM~\cite{jafarian2012openflow} for instance, utilizes a random mapping between the \textit{real IP (rIP)} and \textit{virtual IP (vIP)} addresses. The mutations are coordinated across different OpenFlow~\cite{mckeown2008openflow} switches. While, the \textit{rIPs} remain the same, the \textit{vIPs} are updated continuously with high-degree of unpredictability. The end-user remains transparent to the mutations. 
   
  % The strategic MTD, on the other hand checks the security and usability information, such as, vulnerabilities, available system resources (compute power, memory requirements, active connections on the service being changed). Zhuang \textit{et. al.}~\cite{zhuang2013investigating} use \textit{Conservative Attack Graph (CAG)} of the network vulnerabilities to analyze the paths attacker might take for exploiting high-value target nodes in the network. Based on the results of \textit{CAG} the \textit{Adaptation Engine} performs the analysis of MTD strategy, e.g., migration of database from one network host to other within the cloud network. 

	\ifCLASSOPTIONcaptionsoff
	\newpage
	\fi
	
% 	\balance

	\bibliographystyle{IEEEtran}
	\bibliography{main}
% 	\vspace*{-3\baselineskip}
	% --- For IEEE --- %
	
	\begin{IEEEbiography}[{\includegraphics[width=0.8in,height=1.25in,clip,keepaspectratio]{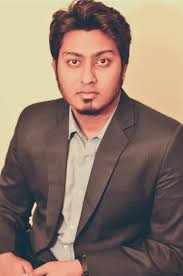}}]{Sailik Sengupta} is a Ph.D. student in Computer Science at Arizona State University. He received his B.E. in Computer Science and Engineering from Jadavpur University (2013) and then worked for Amazon where he became an application-security certifier. Sailik is interested in solving problems that arise in multi-agent settings where the participating agents are non-cooperative or adversarial. He has looked at challenges in cyber-security, adversarial machine learning, and human-AI interaction. Sailik was awarded the IBM Ph.D. fellowship in 2018.
	\end{IEEEbiography}
% \vskip 0pt
% \vspace*{-3\baselineskip}
	\begin{IEEEbiography}[{\includegraphics[width=0.8in,height=1.25in,clip,keepaspectratio]{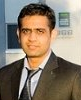}}]{Ankur Chowdhary}
		is a Ph.D. candidate in Computer Science at Arizona State University, Tempe, AZ, USA. He received B.Tech in Information Technology from GGSIPU in 2011 and MS in Computer Science from ASU in 2015. He has worked as Information Security Researcher for Blackberry Ltd., RSG and Application Developer for CSC Pvt. Ltd. His research interests include SDN, Web Security, Network Security, and  application of AI and Machine Learning in the field of Security.
	\end{IEEEbiography}

% \vspace*{-3\baselineskip}	
	\begin{IEEEbiography}[{\includegraphics[width=0.8in,height=1.25in,clip,keepaspectratio]{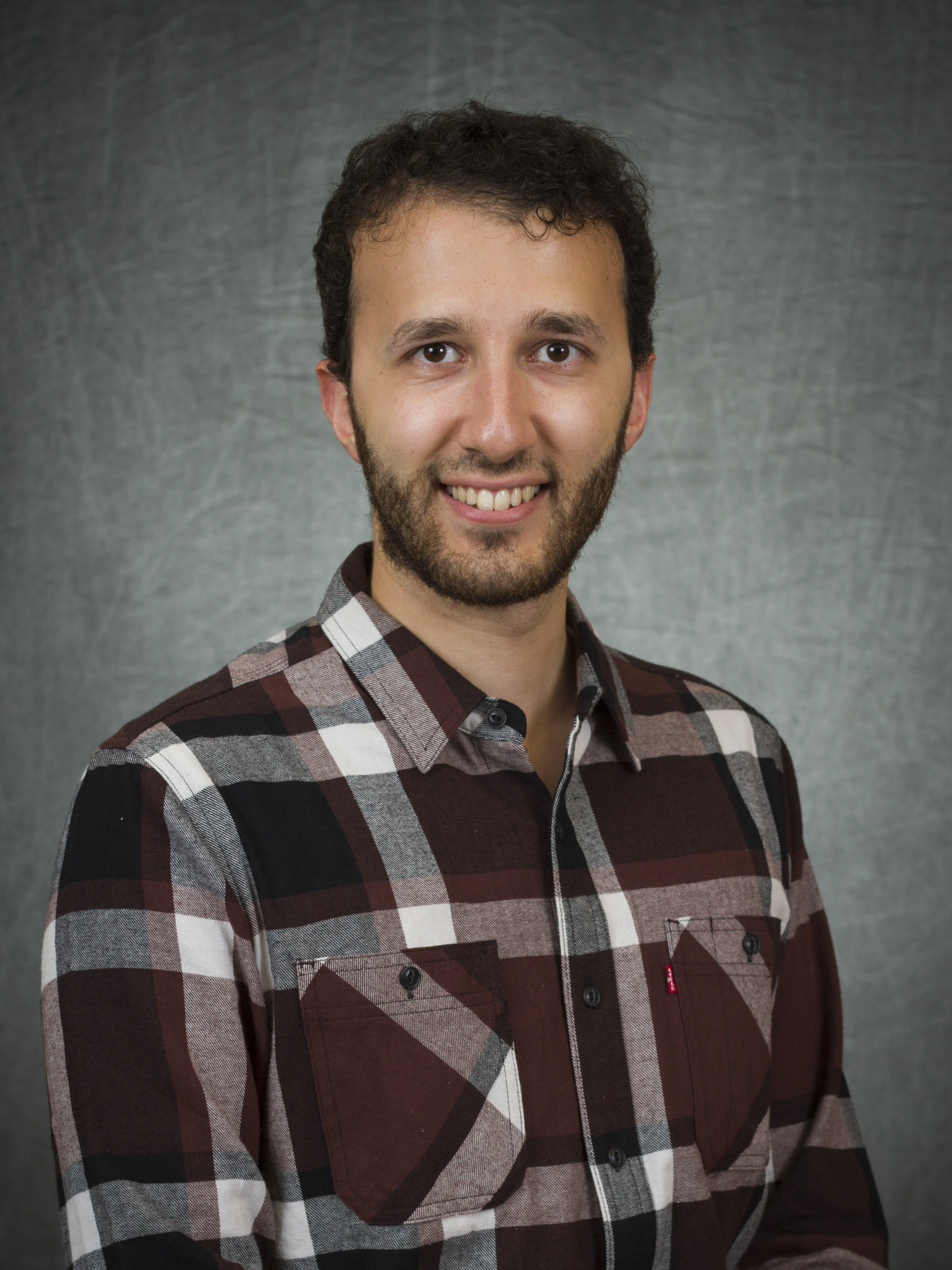}}]{Abdulhakim Sabur}
	is a Ph.D student in Computer Engineering at Arizona state University, Tempe, AZ, USA. He received his B.S. degree (with Honor) in Computer Science and Engineering from King Saud University, Saudi Arabia in 2015 and a Master degree in Computer Engineering from Arizona State University in 2018. He worked as a researcher in King Abdulaziz City for Science and Technology (KACST) and a teaching assistant in Taibah University.  His research interest include Network and information security, vulnerability analysis and management, automated policy and security checking in software defined networking systems.
	\end{IEEEbiography}
% \vspace*{-3\baselineskip}	

% \vspace*{-3\baselineskip}	
	\begin{IEEEbiography}[{\includegraphics[width=0.8in,height=1.25in,clip,keepaspectratio]{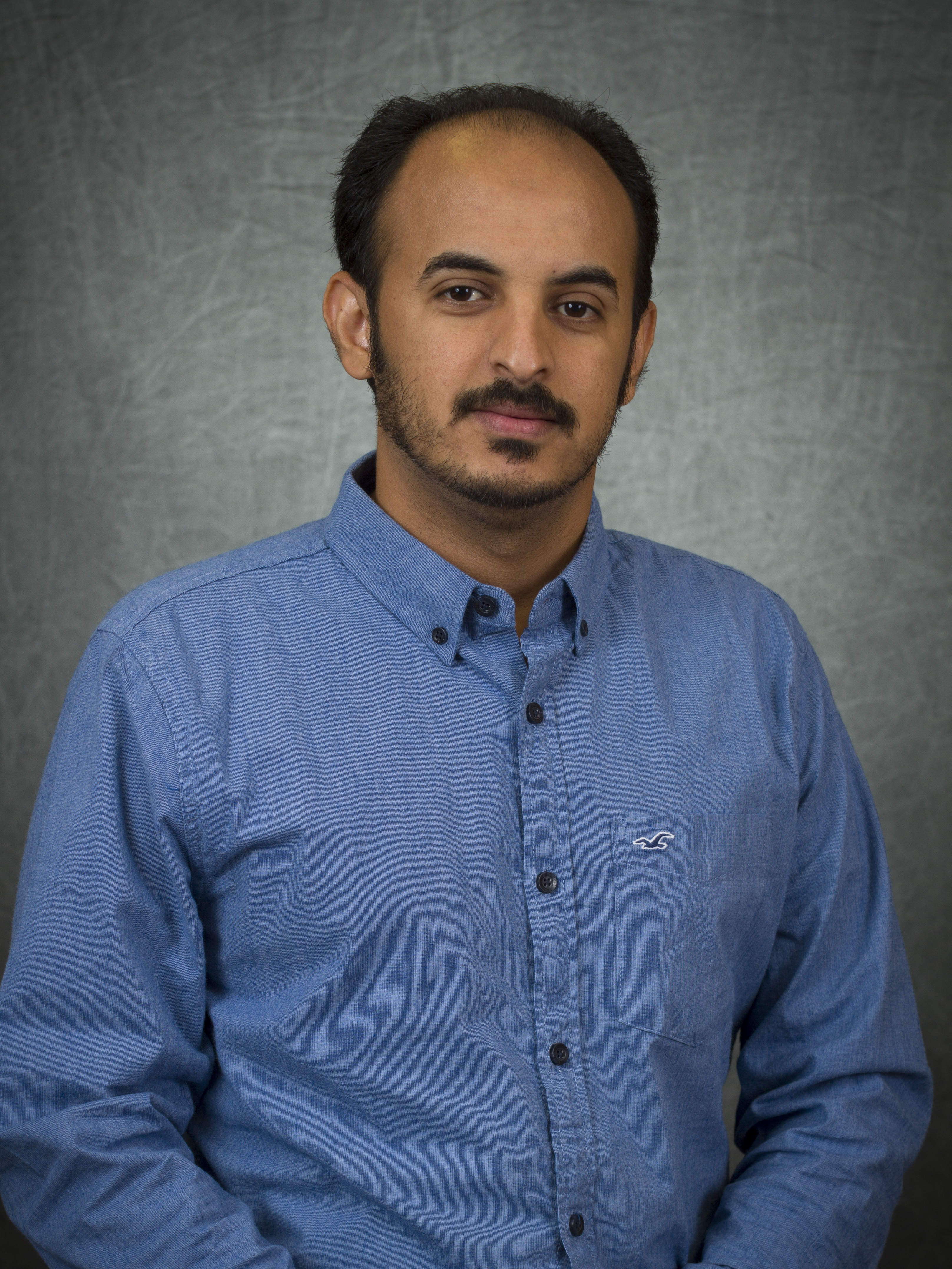}}]{Adel Alshamrani}
		is an assistant professor in department of cybersecurity, College of Computer Science and Engineering at University of Jeddah, Jeddah, Saudi Arabia. He received his B.S. degree in computer science from Umm Al-Qura University, Saudi Arabia in 2007, M.S. degree in computer science from La Trobe University Melbourne, Australia, in 2010, and PhD in computer science from Arizona State University in 2018. He has eight years of work experience in information security, network engineering, and teaching while working in the Faculty of Computing and Information Technology, King Abdul Aziz University, and University of Jeddah. His research interests include information security, intrusion detection, and software defined networking. He is the Chief Information Security Officer (CISO) at the University of Jeddah. 
	\end{IEEEbiography}
% \vskip 0pt
% \vspace*{-3\baselineskip}

    \begin{IEEEbiography}[{\includegraphics[width=0.8in,height=1.25in,clip,keepaspectratio]{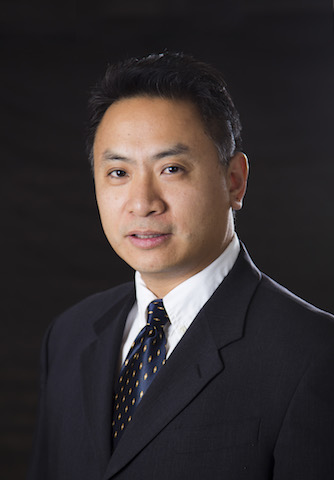}}]{Dijiang Huang}
		received the B.S. degree from Beijing University of Posts and Telecommunications, China, and the M.S. and Ph.D. degrees from the University of Missouri, Kansas in 1995, 2001, and 2004 respectively. He is an Associate Professor with the School of Computing Informatics and Decision System Engineering, Arizona State University. His research interests include computer networking, security, and privacy.	He is an Associate Editor of the Journal of Network and System Management (JNSM) and the IEEE Communications Surveys and Tutorials. He has also served as the chair at multiple international conferences and workshops. His research was supported by the NSF, ONR, ARO, NATO, and Consortium of Embedded System (CES). He was the recipient of the ONR Young Investigator Program	(YIP) Award.
	\end{IEEEbiography}

	\begin{IEEEbiography}[{\includegraphics[width=0.8in,height=1.25in,clip, keepaspectratio]{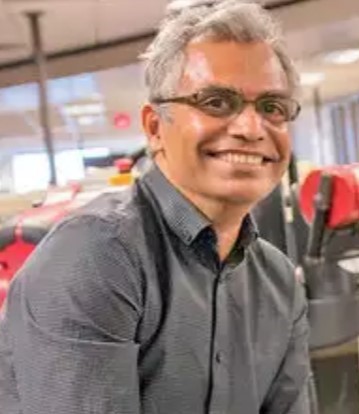}}]{Subbarao Kambhampati (Rao)}
    	 is a professor of Computer Science at Arizona State University. He received his B.Tech. in Electrical Engineering (Electronics) from Indian Institute of Technology, Madras (1983), and M.S.(1985) and Ph.D.(1989) in Computer Science (1985,1989) from University of Maryland, College Park. Kambhampati studies fundamental problems in planning, decision making, and game theory. Kambhampati is a fellow of AAAI, AAAS and ACM, and was an NSF Young Investigator. He received multiple teaching awards, including a university last lecture recognition. Kambhampati is the past president of AAAI and was a trustee of IJCAI. He was the program chair for IJCAI 2016, ICAPS 2013, AAAI 2005 and AIPS 2000 and served on the board of directors of Partnership on AI. Kambhampati's research, as well as his views on the progress and societal impacts of AI, have been featured in multiple national and international media outlets.
	\end{IEEEbiography}
	
\end{document}